# A construction of lattice chiral gauge theories

Rajamani Narayanan and Herbert Neuberger *

*School of Natural Sciences, Institute for Advanced Study,
Olden Lane, Princeton, NJ 08540*

hep-th/9411108  15 Nov 94

**Abstract**

Path integration over Euclidean chiral fermions is replaced by the quantum mechanics of an auxiliary system of non–interacting fermions. Our construction avoids the no–go theorem and faithfully maintains all the known important features of chiral fermions, including the violation of some perturbative conservation laws by gauge field configurations of non–trivial topology.

---

* Permanent Address: Department of Physics and Astronomy, Rutgers University, Piscataway, NJ 08855-0849

# 1. Introduction

A surge of interest [1–6] in the problem of constructing lattice chiral gauge theories has been generated recently by a proposal of David Kaplan [1]. The ultimate goal of this activity is to make a study of chiral gauge theories outside perturbation theory possible. While the simulation of pure gauge theories on the computer is conceptually easy, the inclusion of chiral fermions has been a theoretical stumbling block for years [7]. The nature of the stumbling block is essentially kinematical in that it only involves the fermion dynamics in a fixed gauge background, and this problem only requires to make sense of a quadratic (also infinite and Grassmann) path integral. The fermion dynamics results in an effective action (the logarithm of the chiral determinant) which is a complex functional of the gauge field background. A correct local lattice regularization of the chiral determinant should satisfy two main requirements: Firstly, it should reproduce perturbation theory (for small and smooth gauge fields). Secondly, it should reproduce instanton physics correctly (for large and smooth gauge fields that carry non-zero topological charge). In the present paper we present such a regularization. Having overcome the kinematical problem we hope to render the insufficiently understood dynamics of chiral matter coupled to nonabelian gauge fields amenable to non–perturbative study. There are further obstacles on the road to efficient numerical simulation; we hope to convince the reader that the seemingly insurmountable kinematical obstruction has been removed and therefore it is worth investing efforts into dealing with the more technical problems still ahead.

Our regularization is based on a formal representation of the chiral determinant by the overlap of two states in an auxiliary Hilbert space. The two states are the many body ground states of two Hamiltonians describing non-interacting fermions. After regularization, in a finite Euclidean volume, the states become finite vectors with components labelled by lattice sites, spinorial and internal symmetry indices. The overall phases of the states are defined in a way that reflects several discrete symmetries of the continuum target theory. In [4], the overlap representation was arrived at in a rather convoluted manner: Kaplan[1], started from a Dirac fermion on a $2d+1$ dimensional lattice with a mass term inhomogeneous in one direction creating a transverse domain wall. The Dirac operator has a zero mode with a definite chirality bound to the domain wall. Kaplan's proposal was to couple this zero mode to a gauge field in order to construct a $2d$ dimensional chiral gauge theory. At the same time there was an independent proposal by Frolov and Slavnov [8] to perturbatively regulate an anomaly free chiral gauge theory by using an infinite number of Pauli Villars regulators. It was shown in [3] that the above two proposals can be seen as related once it is recognized that the abstract central idea is to describe a single chiral fermion by an infinite flavor tower of Dirac fermions. The Dirac mass matrix is representable by a finite operator in infinite flavor space whose analytical index is equal to unity times a prefactor of the order of the ultra–violet cutoff. The infinity of the number of fermions can be rigorously controlled and the result is the overlap formula for the chiral determinant [4].

The overlap formula can be motivated directly, without recourse to an infinite number of fermions. In this paper we construct, in complete detail, chiral gauge theories on a lattice starting with the overlap formalism. Since there is no reference to an infinite number of Fermi fields, one does not have to worry about the apparent need to tackle the seemingly impossible task of incorporating them exactly. Neither does one stumble into any of the many pitfalls arising out of truncating them and spoiling the important property of the mass matrix, namely, the non-zero index. The direct motivation of the overlap formulation is elegant, natural and convincing. From this vantage point, the infinite number of fermions appears as an after thought, a possible new interpretation of the overlap. Still, we feel that it is necessary to keep both view points in mind for a complete understanding of our construction. For example, if we insist on a description of our regularization in a Hamiltonian (rather than Euclidean action) framework, we need locality in the physical time direction and are forced to think in terms of an infinite number of fermions.

In this paper we focus on the lattice regularization of the fermionic part of the chiral gauge theory. The gauge part can be regulated on the lattice in standard manner. There are three parts to this paper. In the first part (sections 2–6), we deal with the overlap formula and the formulae for fermion expectation values. In section 2, we present the connection between the chiral determinant and the overlap formula at a formal level. The formal overlap is easier to regulate than the chiral determinant. The latter vanishes for the large class of gauge fields carrying non-zero topological charge. In section 3 we show how the overlap formula can easily reproduce this important property. The phase of the chiral determinant contains all the parity breaking effects induced by the chiral fermions and, in particular, possible anomalous breaking of gauge invariance. This role is taken up by the phase of the overlap which is sensitive to the phases of the two participating states. The real part of the overlap does not depend on these phases and its definition is naturally gauge invariant. Gauge invariance is broken by employing a Wigner–Brilouin phase choice for the many body ground states. In section 4, we show that this phase choice preserves several important formal properties of the imaginary part of the continuum induced action. Having defined the overlap, we proceed to construct representations of fermion expectation values in section 5. In particular, we also show how to define fermion expectation values in an anomaly free gauge theory when the background gauge field carries a non-zero topological charge. In section 6 we describe how to compute the chiral determinant and the fermion expectation values in perturbation theory.

Having discussed the overlap formula and the fermion expectation values, we proceed to construct a lattice regularization of these quantities in the second part (section 7–9) of the paper. In section 7 we show that the two many body Hamiltonians whose many body ground states comprise the overlap can be regulated in a straightforward manner. The regularization so obtained is simpler to understand and easier to work with than that obtained in [4], but both regularizations are conceptually equivalent. Section 8 deals with the vanishing of the overlap in non-trivial gauge backgrounds. An important byproduct is a clean definition of the topological charge for lattice gauge fields, this time based directly on the reaction of fermions. Having constructed the needed ingredients, regularized anomaly free theories are put together in section 9. Vector theories with arbitrary numbers of flavors are also similarly constructed. The fermionic propagators in the vector theories satisfy the basic properties needed to prove the well known mass inequalities involving mesons and baryons. The proof is now in a rigorously regularized theory and holds for an arbitrary number of flavors, both for massless and for massive fermions.

The third part (section 10–11) deals with some tests in two and four dimensions. In section 10 we focus on two dimensional abelian models. We compute the overlap for perturbative gauge fields and show that both the magnitude and the phase agrees with the continuum result. The overlap reproduces the correct topological charge when the background gauge field carries one. We study the overlap in a gauge background made up of a localized instanton and a well separated localized anti-instanton. The quantitative behavior of the overlap as a function of the instanton anti-instanton separation is as expected from continuum considerations. The fermionic two-point function also has the expected continuum behavior. By explicitly looking at a background with a single localized instanton, we extract the zero mode and show that it has the right shape. Further, we study the axial anomaly in a vector theory in the presence of a background gauge field that carries non-zero topological charge. We find that the continuum anomaly still is reproduced, indicating that the definition of the overlap is also valid for non-perturbative gauge fields. All results are in complete agreement with the exact results for the Schwinger model and it is argued that there is a good chance this formalism will work if applied to a chiral rather than vector gauge theory in two dimensions. Four dimensional gauge theories are discussed in section 11. Since the computational effort is larger in four dimensions, fewer tests have been carried out so far. We show that the perturbative anomaly is reproduced and the overlap correctly accounts for the number of zero modes both in abelian and non-abelian backgrounds. The results of these important tests stongly indicate that this formalism should also work for chiral gauge theories in four dimensions. We end this paper with a discussion in section 12 of the expected behavior when the dynamics of the gauge fields are included.

## 2. Chiral determinant as an overlap of Slater determinants

The massless Euclidean Dirac operator,

$$\mathbf{D} = \gamma_\mu(\partial_\mu + iA_\mu(x)), \tag{2.1}$$

in the chiral basis,

$$\gamma_\mu = \begin{pmatrix} 0 & \sigma_\mu \\ \sigma_\mu^\dagger & 0 \end{pmatrix}, \quad \gamma_5 = \begin{pmatrix} 1 & 0 \\ 0 & -1 \end{pmatrix}, \tag{2.2}$$

is of the form

$$\mathbf{D} = \begin{pmatrix} 0 & \mathbf{C} \\ -\mathbf{C}^\dagger & 0 \end{pmatrix}, \tag{2.3}$$

where

$$\mathbf{C} = \sigma_\mu(\partial_\mu + iA_\mu(x)). \tag{2.4}$$

In 4-D

$$\sigma_1 = \begin{pmatrix} 0 & 1 \\ 1 & 0 \end{pmatrix}, \sigma_2 = \begin{pmatrix} 0 & -i \\ i & 0 \end{pmatrix}, \sigma_3 = \begin{pmatrix} 1 & 0 \\ 0 & -1 \end{pmatrix}, \sigma_4 = \begin{pmatrix} i & 0 \\ 0 & i \end{pmatrix}; \tag{2.5}$$

and in 2-D

$$\sigma_1 = 1, \ \sigma_2 = i. \tag{2.6}$$

$A_\mu(x)$ is a finite hermitian matrix for each $x$ and $\mu$ and this includes as special cases all compact gauge groups with the fermion being in any representation.

In a vector theory, integration of the fermions in a fixed background gauge field yields $\det \mathbf{D}$ which by virtue of (2.3) is

$$\det \mathbf{D} = \det \mathbf{C} \det \mathbf{C}^\dagger = \det \mathbf{C}\mathbf{C}^\dagger = \det \mathbf{C}^\dagger \mathbf{C} = |\det \mathbf{C}|^2 \tag{2.7}$$

This shows that the vector determinant is a product of two chiral determinants of opposite chirality. $\mathbf{D}$, $\mathbf{C}\mathbf{C}^\dagger$, $\mathbf{C}^\dagger \mathbf{C}$ are operators that map vectors from one space into itself and therefore one can think of their determinants as products of eigenvalues and proceed to regularize these products. The same is not true of the determinants of $\mathbf{C}$ and $\mathbf{C}^\dagger$ individually. $\mathbf{C}$ is the chiral Dirac operator and is a mapping from $V_L$ to $V_R$ where elements of $V_L$ are decreed to transform as $(\frac{1}{2}, 0)$ under the Euclidean Lorentz group and those of $V_R$ are decreed to transform as $(0, \frac{1}{2})$ under the same Lorentz group. As such $\det \mathbf{C}$ cannot be thought of as a product of eigenvalues.

Let $\{u_i\}$ and $\{v_j\}$ be orthonormal bases of $V_R$ and $V_L$ respectively. These bases were picked without any reference to the gauge potential $A$. A reasonable definition, compatible with (2.7), of $\det \mathbf{C}$ would be

$$\det \mathbf{C} = \det_{ij} < u_i \ \mathbf{C} v_j > . \tag{2.8}$$

This formal definition is also compatible with the known rules for Feynman graphs because if we expanded $\mathbf{C}$ around $\mathbf{C}_0 = \sigma_\mu \partial_\mu$ and chose the bases as plane waves with ordinary spinorial structure we indeed recover unrenormalized perturbation theory. Explicitly, up to normalization factors, (2.8) is an overlap of two many body wave functions*, i.e.,

$$\det \mathbf{C} = \int [d\xi] \left[ \det_{ij}(u_i(\xi_j)) \right]^* \det_{ij}((\mathbf{C} v_i(\xi_j))). \tag{2.9}$$

The two many body wave functions in (2.9) are written as Slater determinants and the variables $\xi$ denote all the continuous and discrete arguments of the single particle wave functions.

We now proceed to describe one approach to obtain the two many body wavefunctions whose overlap is $\det \mathbf{C}$ and where Lorentz invariance is maintained. The idea is to build a larger vector space, $V = V_R \oplus V_L$, and two hermitian operators in it that can be viewed as Hamiltonians, $\mathbf{H}^\pm$:

$$\mathbf{H}^\pm = \begin{pmatrix} \pm m & \mathbf{C} \\ \mathbf{C}^\dagger & \mp m \end{pmatrix} = \gamma_5 [\pm m + \gamma_\mu (\partial_\mu + i A_\mu(x))]; \quad m > 0 \tag{2.10}$$

$\mathbf{H}^\pm$ are Dirac Hamiltonians in $(d+1)$ dimensions in the presence of a static (from the point of view of the $(d+1)$ dimensional world) gauge field whose time component vanishes. Suppose now that the mass squared, $m^2$, is much larger than $||\mathbf{C}\mathbf{C}^\dagger||$ (at this point we are still at the formal level and the fact that $||\mathbf{C}\mathbf{C}^\dagger|| = \infty$ is ignored). The Dirac sea for $\mathbf{H}^+$ ($\mathbf{H}^-$) is given by a Slater determinant made out of one particle wave functions with large bottom (top) components. This would be the ground state for a system of identical fermions (in the grand canonical ensemble and with zero chemical potential) all subjected to the single particle Hamiltonian $\mathbf{H}^+$ ($\mathbf{H}^-$). If $m$ is very large the many body ground states for $\mathbf{H}^+$ and for $\mathbf{H}^-$ will be orthogonal. We cannot therefore ignore $\mathbf{C}$ entirely. Taking into account Fermi statistics we conclude that the overlap between the two ground states is linear in $\det \mathbf{C}$ to leading order in $\frac{||\mathbf{C}\mathbf{C}^\dagger||}{m^2}$. Since both ground states are Lorentz invariant objects we have a Lorentz invariant formula. Here we still refer to the *four* dimensional Lorentz group as the "Lorentz Group".

We now derive the explicit connection between the overlap of the two many body states of $\mathbf{H}^\pm$

---

* This is equivalent to the mathematical definition [9] which regards $\det \mathbf{C} \in (\det V_L)^* \otimes \det V_R$ as the natural element corresponding to $\mathbf{C} \in V_L^* \otimes V_R$. $\det_{ij}(u_i(\xi_j)) \in \det V_R$ and $\det_{ij}(v_i(\xi_j)) \in \det V_L$ are two single element bases while the map $\det \mathbf{C}$ between the two one dimensional spaces acts by $(\det \mathbf{C}) \det_{ij}(v_i(\xi_j)) = \det(\mathbf{C} v_i(\xi_j))$. (2.9) computes the (single) matrix element of $\det \mathbf{C}$ relative to the two bases.

and det **C**. The single particle eigenstates are given by

$$\sum_{y\beta j} \mathbf{H}^{\pm}(x\alpha i, y\beta j; A)\psi_K^{R\pm}(y\beta j; A) = \lambda_K^{R\pm}\psi_K^{R\pm}(x\alpha i; A); \quad \lambda_K^{R\pm} > 0$$
$$\sum_{y\beta j} \mathbf{H}^{\pm}(x\alpha i, y\beta j; A)\psi_K^{L\pm}(y\beta j; A) = \lambda_K^{L\pm}\psi_K^{L\pm}(x\alpha i; A); \quad \lambda_K^{L\pm} < 0$$
(2.11)

$\alpha, \beta$ and $i, j$ denote spin and color indices. Let $|L\pm>_A$ denote the many body states obtained by filling all the $\lambda_K^{L\pm}$ states. We write

$$\psi_K^{L\pm} = \begin{pmatrix} u_K \\ v_K \end{pmatrix}$$
(2.12)

and (2.10) implies

$$\mathbf{C} v_K = (\lambda_K^{L\pm} \mp m) u_K \qquad \mathbf{C}^\dagger u_K = (\lambda_K^{L\pm} \pm m) v_K$$
(2.13)

Operating with **C** and $\mathbf{C}^\dagger$ on the last two equations we get:

$$\mathbf{C}^\dagger \mathbf{C} v_K = \left([\lambda_K^{L\pm}]^2 - m^2\right) v_K \qquad \mathbf{C} \mathbf{C}^\dagger u_K = \left([\lambda_K^{L\pm}]^2 - m^2\right) u_K$$
(2.14)

The eigenvalues $\lambda_K^{L\pm}$ of $\mathbf{H}^\pm$ are simply related to the eigenvalues of $\mathbf{C}^\dagger \mathbf{C}$ and $\mathbf{C} \mathbf{C}^\dagger$ (the spectra of these two operators are identical except for possible zero modes). The manifold of all **C**'s is parameterized by the set of all hermitian matrix valued vector potentials with no restrictions. As such, it is important that we write expressions for the eigenfunctions that are valid for all possible vector potentials, including the cases where $\mathbf{C}^\dagger \mathbf{C}$ and/or $\mathbf{C} \mathbf{C}^\dagger$ have zero modes (in these cases $\mathbf{H}^\pm$ has some eigenvalues equal to $m$ in absolute value). We note that $\lambda_K^{L\pm} < 0$ since the $(d+1)$ dimensional Dirac problems we are interested in always have a nonvanishing gap in the fermion spectrum. For $\mathbf{H}^+$ the single particle wave functions of interest are

$$\psi_K^{L+} = -\frac{1}{\sqrt{2N_{KK}^+}} \begin{pmatrix} \mathbf{C} \frac{1}{\lambda_K^{L+} - m} v_K \\ v_K \end{pmatrix} = -\frac{1}{\sqrt{2N_{KK}^+}} \begin{pmatrix} \mathbf{C} \frac{1}{-\sqrt{\mathbf{C}^\dagger \mathbf{C} + m^2} - m} v_K \\ v_K \end{pmatrix}$$
$$= \frac{1}{\sqrt{2N_{KK}^+}} \begin{pmatrix} \frac{1}{\sqrt{\mathbf{C} \mathbf{C}^\dagger + m^2} + m} \mathbf{C} v_K \\ -v_K \end{pmatrix}$$
(2.15)
$$N_{KK'}^+ = \frac{1}{2} < v_K \left[ \frac{\mathbf{C}^\dagger \mathbf{C}}{(\sqrt{\mathbf{C}^\dagger \mathbf{C} + m^2} + m)^2} + 1 \right] v_{K'} >$$

For $\mathbf{H}^-$ the single particle wave functions of interest are

$$\psi_K^{L-} = \frac{1}{\sqrt{2N_{KK}^-}} \begin{pmatrix} u_K \\ \mathbf{C}^\dagger \frac{1}{\lambda_K^{L-} - m} u_K \end{pmatrix} = \frac{1}{\sqrt{2N_{KK}^-}} \begin{pmatrix} u_K \\ \mathbf{C}^\dagger \frac{1}{-\sqrt{\mathbf{C} \mathbf{C}^\dagger + m^2} - m} u_K \end{pmatrix}$$
$$= \frac{1}{\sqrt{2N_{KK}^-}} \begin{pmatrix} u_K \\ -\frac{1}{\sqrt{\mathbf{C}^\dagger \mathbf{C} + m^2} + m} \mathbf{C}^\dagger u_K \end{pmatrix}$$
(2.16)
$$N_{KK'}^- = \frac{1}{2} < u_K \left[ \frac{\mathbf{C} \mathbf{C}^\dagger}{(\sqrt{\mathbf{C} \mathbf{C}^\dagger + m^2} + m)^2} + 1 \right] u_{K'} >$$

Note that the matrices $N^+$ and $N^-$ are diagonal.

The many body states $|L\pm>_A$ are Slater determinants of the single particle wave functions $\psi_K^{L\pm}$ and the overlap

$$_A<L-|L+>_A = \det_{KK'}(<\psi_K^{L-}|\psi_{K'}^{L+}>)$$
$$= \frac{1}{\sqrt{\det N^- \det N^+}} \det_{KK'} <u_K|\frac{1}{\sqrt{\mathbf{CC}^\dagger + m^2} + |m|}\mathbf{C}|v_{K'}> \quad (2.17)$$

The above equation is not yet very useful because the states $v_K$ and $u_K$ also depend on the operator $\mathbf{C}$. However, the overlap will not change if we perform unitary transformations (but with $\det_L \det_R = 1$ rather than a phase; this restriction is meaningless because the single particle wave functions were defined only up to phase – more about this later) independently in the spaces $V_L$ and $V_R$. This means that the bases of eigenvectors, $\{v_K\}$ and $\{u_K\}$, can be replaced by arbitrary orthonormal bases of $V_L$ and $V_R$. All the dependence on $\mathbf{C}$ becomes now explicit and we have formally established the connection

$$\det \mathbf{C}(A) \Leftrightarrow {_A}<L-|L+>_A \quad (2.18)$$

By an identical line of reasoning we can also arrive at the following relation connecting $\det \mathbf{C}^\dagger$ and the overlap of the many body states $|R\pm>_A$ obtained by filling all the $\lambda_K^{R\pm}$ states.

$$_A<R-|R+>_A = \det_{KK'}(<\psi_K^{R-}|\psi_{K'}^{R+}>)$$
$$= \frac{1}{\sqrt{\det N^- \det N^+}} \det_{KK'} <v_K|\frac{1}{\sqrt{\mathbf{C}^\dagger\mathbf{C} + m^2} + |m|}\mathbf{C}^\dagger|u_{K'}> \quad (2.19)$$

Note that $|R\pm>$ represent the Dirac sea for $-\mathbf{H}^\pm$. We have the other formal connection

$$\det \mathbf{C}^\dagger(A) \Leftrightarrow {_A}<R-|R+>_A \; . \quad (2.20)$$

Comparing (2.17) and (2.19) it is clear that indeed (2.18) and (2.20) are conjugates of each other.

The overlap is an adequate replacement for $\det \mathbf{C}$ since, for states $v_i$ for which the norm of $\mathbf{C}v_i$ is small relative to $m$, all the additional factors that appear in the overlap are constants. In a good choice of gauge fixing the smallness of the norm of $\mathbf{C}v_i$ means, for small gauge fields, that the typical momentum modes contained in the state $v_i$ are small relative to $m$. We anyhow will have to regularize the expressions to make them well defined and any regularization induces mutilations of high momentum modes. Therefore, the fact that the overlap has not come out exactly $\det \mathbf{C}$ even at the formal level is of no concern.

Formally, the expression is also gauge invariant. Let $\mathcal{G}_L(G)$ denote the unitary representation of a gauge group element $G$ in $V_L$ and $\mathcal{G}_R(G)$ denote the representation of the same gauge group element $G$ in $V_R$. The structure of $\mathbf{C}$ is such that one assumes $(\mathcal{G}_R(G))^\dagger \mathbf{C}(A_\mu)\mathcal{G}_L(G) = \mathbf{C}(A_\mu^G)$; then the overlap is gauge invariant. The phases of $\det_{ij}<u_R^{(i)}(\mathcal{G}_R(G))^\dagger u_R^{(j)}>$ and of $\det_{ij}<$

$v_L^{(i)} \mathcal{G}_L(G) v_L^{(j)} >$ cancel in the expression below:

$$\det_{ij} \left( < u_R^{(i)} \frac{1}{\sqrt{\mathbf{C}(A_\mu^G)(\mathbf{C}(A_\mu^G))^\dagger + m^2} + |m|} \mathbf{C}(A_\mu^G) v_L^{(j)} > \right) =$$

$$\det_{ij} \left( < u_R^{(i)} \frac{1}{\sqrt{\mathbf{C}(A_\mu)(\mathbf{C}(A_\mu))^\dagger + m^2} + |m|} \mathbf{C}(A_\mu) v_L^{(j)} > \right) \quad (2.21)$$

$$\det_{ij} \left( < u_R^{(i)} (\mathcal{G}_R(G))^\dagger u_R^{(j)} > \right) \det_{ij} \left( < v_L^{(i)} \mathcal{G}_L(G) v_L^{(j)} > \right)$$

The above equation holds because the sets $\{v_L^{(i)}\}$ and $\{u_R^{(i)}\}$ of the spaces $V_L$ and $V_R$ respectively are complete.

The overlap is still only a formal expression representing $\det \mathbf{C}$. We know that for small and slowly varying gauge potentials (with respect to the mass $m$) we would, in perturbation theory, get to leading order in $\frac{1}{m}$ expressions identical to the possibly divergent, ordinary, unregulated Feynman diagrams. Thus we have not lost anything by replacing the previous formal expression for the chiral determinant by the overlap. We claim that we have made even some progress: It will be easier to turn the overlap into a well defined object rather than $\det \mathbf{C}$. Moreover, the formulation is such that we will not be tempted to consider an eigenvalue problem for $\mathbf{C}$ and as a result break Lorentz invariance.

We can easily find gauge invariant ways to regulate the Hamiltonian problems. This may come as surprise because a gauge invariant regularization of a chiral anomalous theory must be pathological. When $\mathbf{H}^\pm$ gets regularized the simple structure it has in terms of the split $V = V_R \oplus V_L$ will be mangled. More precisely, with respect to the split of $V$ we had:

$$\mathbf{H}^\pm = \begin{pmatrix} \mathbf{B}^\pm & \mathbf{C} \\ \mathbf{C}^\dagger & -\mathbf{B}^\pm \end{pmatrix} \quad [\mathbf{B}^\pm, \mathbf{C}] = 0 \quad (2.22)$$

Upon regularization, the commutativity made explicit above will be lost. We cannot therefore use the explicit construction of the overlap in terms of the two independent bases given earlier. As a result, in particular, the "proof" of gauge invariance of the overlap may break down. We still can easily define the second quantized system and say that we need the overlap between the two ground states. This defines the *absolute value* of the overlap but not its phase. The Hamiltonians can be regularized in a gauge invariant manner but the overlap is not guaranteed to come out gauge invariant. Physically it is clear that the control of "high momentum modes", the usual objective of ultraviolet regularization, is attained by regularizing the Hamiltonians. The possibility of anomalies is kept open by the existence of an additional ambiguity in the phase choices of the two second quantized ground states. The fact that usually one associates anomalies with the regularization is misleading. If the single ambiguity available to the theory is the fact that it has an uncontrolled ultraviolet behavior then anomalies are forced to use that opening to creep out. The overlap formulation is making it possible to deliberately leave an ambiguity in the regularized theory so that the anomalies can be reflected by it. In this way the problem of anomalies is separated from the problem of regularization and the truly ultraviolet infinities can be controlled by standard means. In short, the overlap expression provides a route to particular regularizations that break gauge invariance in a sort of minimal way. The obstacles to making these regularizations non–perturbative (i.e. having them defined for any vector potentials $A$, without any restrictions, like, for example, on the topological sector) disappear.

## 3. Zeros in nontrivial backgrounds

If the gauge background cannot be deformed to zero, it has non–zero topological charge and the chiral determinant vanishes. Thus, for most physics questions, we can do without a definition of the phase of the two second quantized ground states of the Hamiltonians for gauge fields not deformable to zero. Before regularization, at the formal level, it is easy to see that the overlap will vanish when the operators $\mathbf{C}$ or $\mathbf{C}^\dagger$ have zero eigenstates. Consider a gauge field of instanton number one. $\mathbf{C}$ has a zero mode ($\mathbf{C} v_0 = 0$) but $\mathbf{C}^\dagger$ has no zero mode. For all $|\lambda_K^{L\pm}| > m$, the solutions $v_K$ and $u_K$ in (2.14) are paired. But there is one solution $v_0$ with $\lambda_K^{L\pm} = -m$ that is not paired. This implies that there is one more $\psi_K^{L+}$ in (2.15) than there are $\psi_K^{L-}$ in (2.16) and the overlap matrix $< \psi_K^{L-} \psi_{K'}^{L+} >$ is rectangular. The determinant of a rectangular matrix vanishes identically (the determinant is a function of the rows (columns) of a matrix which is linear in each and totally antisymmetric; if the matrix is rectangular there aren't enough columns (rows) for a non–vanishing totally antisymmetric function to exist). In a second quantized formalism the vanishing is attributed to the fact that one of the vacua has less filled states than the other. After regularization this argument will be more difficult to see because of the loss of the simple structure of $\mathbf{H}^\pm$. Nevertheless, the mechanism for such a vanishing to occur is sufficiently general that it can survive: All that has to happen is that the number of negative energy levels for one sign of the mass term be different from the number of negative energy levels for the other sign of the mass term. Even if the detailed simple connection between $\mathbf{C}$ and $\mathbf{H}^\pm$ is lost upon regularization, the departure from the unregularized structure is small for slowly varying states and the imbalance in the number of filled states should be preserved. To alter the imbalance some low momentum state would have to travel a large distance when the regularization is turned on because for any sign of the mass there is a large gap ($\geq 2m$) in the spectrum of $\mathbf{H}^\pm$.

## 4. The phase of the Chiral determinant

Upon regularization, it is necessary to define the phase of the chiral determinant in the zero topological sector. This definition is arrived at by completely defining the two ground states making up the overlap. These states are parametrically dependent on the background vector potential. The phase of the overlap is the phase of a reasonably generic complex valued function defined over the space of vector potentials. There are points in that space where the overlap will vanish (these are "accidental" vanishing points – we are in the trivial topological sector). Generically, the phase will non-trivially wind around the vortex in the space of gauge potentials where the overlap stays zero. This is important because the inevitability of anomalies in the continuum in the class of acceptable regularizations can be established by showing that there are discs in the space of vector potentials whose boundaries live completely on a single gauge orbit, but, nevertheless, the phase of the chiral determinant must non-trivially wind when one goes around the boundary, since the chiral determinant has a zero of generic type in the interior of the disc. It would therefore be wrong to replace the definition of the chiral determinant with a definition that treats the absolute value and the phase completely independently. Upon regularization, such a definition might lead to an expression where the phases wind around points where the absolute value does not vanish, and points where the absolute value vanishes but the phase does not wind around. Therefore, the overlap, if properly defined as a complex valued function over the space of vector potentials, has the potential to preserve the topological understanding of anomalies in the continuum [10].

We wish to abstract from the formal overlap some properties about its phase that can survive regularization and the associated distortion of the structure of $\mathbf{H}^{\pm}$ relative to the split $V = V_R \oplus V_L$. Assuming no accidental degeneracies the rays to which the states $|L\pm>_A$ and $|R\pm>_A$ correspond are uniquely defined. To pick a state along the ray we need one additional condition. One way to single out a state is to fix the phase of its overlap with some fixed reference states $|\Omega_L\pm>$ and $|\Omega_R\pm>$. The state $|\Omega_L\pm>$ needs to be "transverse" to the ray $e^{i\phi}|L\pm>_A$, i.e. $<\Omega_L\pm|L\pm>_A \neq 0$ and then we can impose $<\Omega_L\pm|L\pm>_A > 0$.* Similarly, we can impose $<\Omega_R\pm|R\pm>_A > 0$. It is important that the choice preserves the following properties of the formal determinant:

1. The chiral determinants of opposite chiralities are conjugates of each other, i.e.,

$$\det \mathbf{C}(A) = \left[\det \mathbf{C}^{\dagger}(A)\right]^{*}.$$

2. For a gauge potential $A_{\mu}(x)$ and its Lorentz transformed partner $A'_{\mu}(x) = \Lambda_{\mu\nu} A_{\nu}(\Lambda^{-1}x)$, the chiral determinants are formally related by

$$\det(\mathbf{C}(A)) = \det(\mathbf{C}(A')).$$

3. For a gauge potential $A_{\mu}(x)$ and the partner under parity

$$A'_{\mu}(x) = \{-\vec{A}(-\vec{x}, x_4), A_4(-\vec{x}, x_4)\},$$

(in two dimensions $x_4$ should be replaced by $x_2$ above) the chiral determinants are formally related by

$$[\det(\mathbf{C}(A'))]^{*} = \det(\mathbf{C}(A)).$$

4. For a gauge potential $A_{\mu}(x)$ and its partner under charge conjugation $A'_{\mu}(x) = -A^{*}_{\mu}(x)$, the chiral determinants are formally related in 2-D by

$$\det(\mathbf{C}(A')) = \det(\mathbf{C}(A)),$$

and in 4-D by**

$$[\det(\mathbf{C}(A'))]^{*} = \det(\mathbf{C}(A)).$$

---

* It is possible that the overlap with the reference state vanishes accidentally and our definition for the phase breaks down. We believe that the gauge potentials where these accidents happen can be safely ignored because the reference states are not that special.

** This property has an important consequence in 4-D. The chiral determinant will be real when $A$ is restricted to a real or pseudoreal representation (in which case there exists a matrix in the internal space $R$ such that $A' = RAR^{-1}$).

5. For a gauge potential $A_\mu(x)$ and its partner under a global gauge transformation $A'_\mu(x) = g^\dagger A_\mu(x) g$ where $g \in U(N)$, the chiral determinants are formally related by

$$\det(\mathbf{C}(A')) = \det(\mathbf{C}(A)).$$

We show below in a sequence of Lemmas that the Wigner-Brillouin phase choice for the many body states preserves the above properties of the chiral determinant when written as an overlap of these two many body states. The Wigner-Brillouin phase choice is simply demanding that the overlaps ${}_0<R\pm|R\pm>_A$ and ${}_0<L\pm|L\pm>_A$ are real and positive. The important point for us is that only quite general properties of $\mathbf{H}^\pm$ are used in establishing 1–5 for the overlap. Even if regularization spoils the simple structure of $\mathbf{H}^\pm$ (which has to happen as we explained before), the properties of the overlap under the symmetries can be preserved because the Hamiltonians will still obey the general properties used in proving the Lemmas. Of course, for a lattice regularization one must give up on Lorentz invariance, but one still has the usual invariance under the appropriate discrete crystallographic group.

It is worthwhile to insist on the discrete symmetries: For example, in four dimensions, the combined requirements of translational, Lorentz and global gauge covariance together with the above behavior under parity and charge conjugation imply that the phase of the overlap, when expanded in powers of the vector potential $A_\mu$ starts at third power. This is a desirable property for a regularization to have.

Now we proceed to show that 1–5 hold for the overlap formula with the Wigner-Brillouin phase choice for the many body states. Essentially, 1–5 are consequences of the fact that $|R\pm>_0$ and $|L\pm>_0$ are invariant under the symmetries under consideration.

In what follows the following Lemma about a general unitary matrix will be useful.

**Lemma 4.0** *Let*

$$U = \begin{pmatrix} A & B \\ C & D \end{pmatrix}$$

*be an $N \times N$ unitary matrix with $A$ being a $K \times K$ square matrix ($0 < K < N$). Then*

$$\det U = \frac{\det A}{\det D^\dagger}$$

*Proof:*
We write
$$\det U = \det A \det(D - CA^{-1}B).$$
For a unitary matrix, one has the relation
$$D - CA^{-1}B = {D^\dagger}^{-1}.$$

Inserting this in the previous equation proves the Lemma ∎

The single particle eigenfunctions for the gauge free case are chosen to satisfy the following two properties:
1. Since
$$-[\gamma_4]_{\alpha\gamma}\mathbf{H}^\pm((-\vec{x},x_4)\gamma i,(-\vec{y},y_4)\delta j;0)[\gamma_4]_{\delta\beta} = \mathbf{H}^\pm(x\alpha i, y\beta j;0) \tag{4.1}$$

we set
$$\psi_K^{L\pm}(x\alpha i; 0) = [\gamma_4]_{\alpha\beta} \psi_K^{R\pm}((-\vec{x}, x_4)\beta i; 0) \tag{4.2}$$

$\gamma_4$ is replaced by $\gamma_2$ in 2-D.

2. Given the single particle states for the $A = 0$ problem let $\mathcal{U}(+m; 0| -m; 0)$ be the unitary matrix connecting the basis at $+m$ to the basis at $-m$. We assume that the single particle states have been chosen so that ${}_0\!<R-|R+>_0$ is real and positive. Then from (4.2), ${}_0\!<L-|L+>_0$ is also real and positive. From Lemma 4.0, we then have

$$\det \mathcal{U}(-m; 0| + m; 0) = 1 \tag{4.3}$$

**Lemma 4.1** If ${}_0\!<R\pm|R\pm>_A$ and ${}_0\!<L\pm|L\pm>_A$ are real and positive then

$$_A\!<R-|R+>_A = {}_A\!<L+|L->_A$$

*Proof:*
From Lemma 4.0, we have

$$\det \mathcal{U}(\pm m; 0| \pm m; A) = \frac{{}_0\!<R\pm|R\pm>_A}{{}_A\!<L\pm|L\pm>_0}. \tag{4.4}$$

Since ${}_0\!<R\pm|R\pm>_A$ and ${}_0\!<L\pm|L\pm>_A$ are real and positive (4.4) implies that

$$\det \mathcal{U}(\pm m; 0| \pm m; A) = 1 \tag{4.5}$$

Since
$$\mathcal{U}(-m; A| + m; A) = \mathcal{U}(-m; A| - m; 0)\mathcal{U}(-m; 0| + m; 0)\mathcal{U}(+m; 0| + m; A), \tag{4.6}$$

(4.3) and (4.5) implies that
$$\det \mathcal{U}(-m; A| + m; A) = 1 \tag{4.7}$$

By Lemma 4.0, we have

$$\det \mathcal{U}(-m; A| + m; A) = \frac{{}_A\!<R-|R+>_A}{{}_A\!<L+|L->_A} \tag{4.8}$$

and this along with (4.7) proves the Lemma ∎

**Lemma 4.2** *For a gauge potential $A_\mu(x)$ let its Lorentz transformed partner be*

$$A'_\mu(x) = \Lambda_{\mu\nu} A_\nu(\Lambda^{-1} x).$$

*If ${}_0\!<R\pm|R\pm>_A$ and ${}_0\!<R\pm|R\pm>_{A'}$ are real and positive then*

$$_{A'}\!<R-|R+>_{A'} = {}_A\!<R-|R+>_A .$$

*Proof:*
Since,
$$\mathbf{H}^{\pm}[x\alpha i, y\beta j; A'] = \left[e^{i\omega(\Lambda)}\right]_{\alpha\gamma} \mathbf{H}^{\pm}[(\Lambda^{-1}x)\gamma i, (\Lambda^{-1}y)\delta j; A]\left[e^{-i\omega(\Lambda)}\right]_{\delta\beta} \quad (4.9)$$

where $\gamma_\mu \Lambda_{\mu\nu} = e^{i\omega(\Lambda)}\gamma_\nu e^{-i\omega(\Lambda)}$, we have the following relation between the eigenvectors of the two Hamiltonians (assuming no degeneracies):

$$\psi_K^{R\pm}(x\alpha i; A') = e^{i\phi_K^{R\pm}} \left(e^{i\omega(\Lambda)}\right)_{\alpha\beta} \psi_K^{R\pm}(\Lambda^{-1}x\beta i; A) \quad (4.10)$$

In the above equation $\phi_K^{R\pm}$ are undetermined phases. For the case when $A = 0$, (4.9) implies degeneracies and therefore

$$\tilde{\psi}_K^{R\pm}(x\alpha i; 0) = \left(e^{i\omega(\Lambda)}\right)_{\alpha\beta} \psi_K^{R\pm}(\Lambda^{-1}x\beta i; 0) \quad (4.11)$$

defines another set of eigenvectors for the free Hamiltonians. The transformation matrix relating the vectors $\psi_K^{R\pm}$ and $\tilde{\psi}_K^{R\pm}$ will be block diagonal and will only mix vectors in a degenerate subspace. Further, the individual blocks will all have unit determinant. Since the state $|R\pm>$ is obtained by filling all states in a given degenerate subspace, using $\psi_K^{R\pm}$ or $\tilde{\psi}_K^{R\pm}$ will result in the same state $|R\pm>$. From (4.10) and (4.11) we have

$$_0<R\pm|R\pm>_{A'} = {_0<R\pm|R\pm>_A}\, e^{i\sum_K \phi_K^{R\pm}} \quad (4.12)$$

We are given that $_0<R\pm|R\pm>_A$ and $_0<R\pm|R\pm>_{A'}$ are real and positive and therefore we have

$$\sum_K \phi_K^{R\pm} = 0 \quad (4.13)$$

From (4.10) we have

$$_{A'}<R-|R+>_{A'} = {_A<R-|R+>_A}\, e^{i\sum_K (\phi_K^{R+} - \phi_K^{R-})} \quad (4.14)$$

and this along with (4.13) proves the Lemma ∎

**Lemma 4.3** *For a gauge potential $A_\mu(x)$ and the partner under parity*

$$A'_\mu(x) = \{-\vec{A}(-\vec{x}, x_4), A_4(-\vec{x}, x_4)\}$$

*(in two dimensions $x_4$ should be replaced by $x_2$ above), if $_0<R\pm|R\pm>_A$, $_0<L\pm|L\pm>_A$, $_0<R\pm|R\pm>_{A'}$ and $_0<L\pm|L\pm>_{A'}$ are real and positive then*

$$_{A'}<R-|R+>_{A'} = {_A<L-|L+>_A}\,.$$

*Proof:*
For the gauge potential $A_\mu(x)$ and its partner under parity, $A'_\mu(x)$ we have the relation

$$\mathbf{H}^{\pm}[x\alpha i, y\beta j; A'] = -[\gamma_4]_{\alpha\gamma} \mathbf{H}^{\pm}[(-\vec{x}, x_4)\gamma i, (-\vec{x}, x_4)\delta j; A][\gamma_4]_{\delta\beta} \quad (4.15)$$

In the above equation, $\gamma_4$ should be replaced by $\gamma_2$ in 2-D and the proof that follows trivially extends to 2-D. The eigenstates for $A'$ are related to the eigenstates for $A$ by

$$\psi_K^{L\pm}(x\alpha i; A') = e^{i\phi_K^{L\pm}} [\gamma_4]_{\alpha\beta} \psi_K^{R\pm}(-\vec{x}x_4\beta i; A)$$
$$\psi_K^{R\pm}(x\alpha i; A') = e^{i\phi_K^{R\pm}} [\gamma_4]_{\alpha\beta} \psi_K^{L\pm}(-\vec{x}x_4\beta i; A) \quad (4.16)$$

In the above equations, $\phi_K^{L\pm}$ and $\phi_K^{R\pm}$ are undetermined phases. Using (4.2) and (4.16) we get

$$_0<L\pm|L\pm>_{A'} = {}_0<R\pm|R\pm>_A \, e^{i\sum_K \phi_K^{L\pm}}$$
$$_0<R\pm|R\pm>_{A'} = {}_0<L\pm|L\pm>_A \, e^{i\sum_K \phi_K^{R\pm}} \quad (4.17)$$

Since $_0<R\pm|R\pm>_A$, $_0<L\pm|L\pm>_A$, $_0<R\pm|R\pm>_{A'}$ and $_0<L\pm|L\pm>_{A'}$ are real and positive, it follows that

$$\sum_K \phi_K^{L\pm} = \sum_K \phi_K^{R\pm} = 0 \quad (4.18)$$

From (4.16) we have

$$_{A'}<R-|R+>_{A'} = {}_A<L-|L+>_A \, e^{\sum_K (\phi_K^{R+} - \phi_K^{R-})} \quad (4.19)$$

and this along with (4.18) proves the Lemma ∎

**Lemma 4.4** *For a gauge potential $A_\mu(x)$ and its partner under charge conjugation $A'_\mu(x) = -A_\mu^*(x)$, if $_0<R\pm|R\pm>_A$, $_0<L\pm|L\pm>_A$, $_0<R\pm|R\pm>_{A'}$ and $_0<L\pm|L\pm>_{A'}$ are real and positive then*

$$_{A'}<R-|R+>_{A'} = {}_A<R-|R+>_A$$

*in 2-D and*

$$_{A'}<R-|R+>_{A'} = {}_A<L-|L+>_A$$

*in 4-D.*

*Proof:*
For the gauge potential $A_\mu(x)$ and its partner under charge conjugation, $A'_\mu(x)$, we have the following relation in 2-D

$$\mathbf{H}^\pm[x\alpha i, y\beta j; A'] = -\left[[\gamma_1]_{\alpha\gamma} \mathbf{H}^\pm[x\gamma i, y\delta j; A][\gamma_1]_{\delta\beta}\right]^* \quad (4.20)$$

and the following relation in 4-D

$$\mathbf{H}^\pm[x\alpha i, y\beta j; A'] = \left[[\Sigma_2]_{\alpha\gamma} \mathbf{H}^\pm[x\gamma i, y\delta j; A][\Sigma_2]_{\delta\beta}\right]^* \quad (4.21)$$

where

$$\Sigma_2 = \begin{pmatrix} \sigma_2 & 0 \\ 0 & -\sigma_2 \end{pmatrix} \quad (4.22)$$

In 2-D, (4.20) gives (assuming no degeneracies)

$$\psi_K^{L\pm}(x\alpha i; A') = e^{i\phi_K^{L\pm}} [\gamma_1]_{\alpha\beta} \left[\psi_K^{R\pm}\right]^*(x\beta i; A)$$
$$\psi_K^{R\pm}(x\alpha i; A') = e^{i\phi_K^{R\pm}} [\gamma_1]_{\alpha\beta} \left[\psi_K^{L\pm}\right]^*(x\beta i; A) \quad (4.23)$$

In the above equations, $\phi_K^{L\pm}$ and $\phi_K^{R\pm}$ are undetermined phases. For the case when $A = 0$, the relations corresponding to (4.23) are

$$\psi_K^{L\pm}(x\alpha i; 0) = e^{i\chi_K^\pm}[\gamma_1]_{\alpha\beta}[\psi_K^{R\pm}]^*(x\beta i; 0)$$
$$\psi_K^{R\pm}(x\alpha i; 0) = e^{i\chi_K^\pm}[\gamma_1]_{\alpha\beta}[\psi_K^{L\pm}]^*(x\beta i; 0)$$
(4.24)

In the above equations, $\chi_K^\pm$ are undetermined phases. From (4.24) we have

$$_0<R-|R+>_0 = [_0<L-|L+>_0]^* e^{i\sum_K(\chi_K^+ - \chi_K^-)}$$
(4.25)

The second property of the many body states in the free case, namely $_0<R-|R+>_0$ and $_0<L-|L+>_0$ are real and positive, implies that

$$\sum_K(\chi_K^+ - \chi_K^-) = 0$$
(4.26)

From (4.23) and (4.24) we have

$$_0<R\pm|R\pm>_{A'} = [_0<L\pm|L\pm>_A]^* e^{i\sum_K(\phi_K^{R\pm} - \chi_K^\pm)}$$
$$_0<L\pm|L\pm>_{A'} = [_0<R\pm|R\pm>_A]^* e^{i\sum_K(\phi_K^{L\pm} - \chi_K^\pm)}$$
(4.27)

Since $_0<R\pm|R\pm>_{A'}$, $_0<L\pm|L\pm>_{A'}$, $_0<R\pm|R\pm>_A$ and $_0<L\pm|L\pm>_A$ are real and positive, it follows from (4.27) that

$$\sum_K \phi_K^{R\pm} = \sum_K \phi_K^{L\pm} = \sum_K \chi_K^\pm$$
(4.28)

From (4.23) we have

$$_{A'}<R-|R+>_{A'} = [_A<L-|L+>_A]^* e^{\sum_K(\phi_K^{R+} - \phi_K^{R-})}$$
$$= {}_A<R-|R+>_A \, e^{\sum_K(\phi_K^{R+} - \phi_K^{R-})}$$
$$= {}_A<R-|R+>_A \, e^{\sum_K(\chi_K^+ - \chi_K^-)}$$
$$= {}_A<R-|R+>_A$$
(4.29)

which proves the Lemma in 2-D. In the above equation we have used Lemma 4.0 in the second equality, (4.28) in the third and (4.26) in the last equality.

In 4-D (4.21) gives

$$\psi_K^{R\pm}(x\alpha i; A') = e^{i\phi_K^{R\pm}}[\Sigma_2]_{\alpha\beta}[\psi_K^{R\pm}]^*(x\beta i; A)$$
(4.30)

where $\phi_K^{R\pm}$ are arbitrary phases. For the case when $A = 0$, we have the situation where

$$\tilde{\psi}_K^{R\pm}(x\alpha i; 0) = e^{i\chi_K^{R\pm}}[\Sigma_2]_{\alpha\beta}[\psi_K^{R\pm}]^*(x\beta i; 0)$$
(4.31)

are another set of eigenvectors. $\chi_K^{R\pm}$ are phases chosen in such a way that the many body states $|R\pm>$ formed out of $\psi_K^{R\pm}$ and $\tilde{\psi}_K^{R\pm}$ are the same. From (4.31) we have the relation

$$_0<R-|R+>_0 = [_0<R-|R+>_0]^* e^{i\sum_K(\chi_K^{R+} - \chi_K^{R-})}$$
(4.32)

We have chosen the states $|R\pm>_0$ so that $_0<R-|R+>_0$ is real (c.f. second property of the free states) and therefore it follows that

$$\sum_K \chi_K^{R+} = \sum_K \chi_K^{R-} \tag{4.33}$$

From (4.30) and (4.31) we have

$$_0<R\pm|R\pm>_{A'} = {_0<R\pm|R\pm>_A}^* e^{i\sum_K(\phi_K^{R\pm} - \chi_K^{R\pm})} \tag{4.34}$$

and since $_0<R\pm|R\pm>_{A'}$ and $_0<R\pm|R\pm>_A$ are real, we get

$$\sum_K \phi_K^{R\pm} = \sum_K \chi_K^{R\pm} \tag{4.35}$$

From (4.30) we also have

$$\begin{aligned}
_{A'}<R-|R+>_{A'} &= [_A<R-|R+>_A]^* e^{\sum_K(\phi_K^{R+} - \phi_K^{R-})} \\
&= {_A<L-|L+>_A}\, e^{\sum_K(\phi_K^{R+} - \phi_K^{R-})} \\
&= {_A<L-|L+>_A}\, e^{\sum_K(\chi_K^{R+} - \chi_K^{R-})} \\
&= {_A<L-|L+>_A}
\end{aligned} \tag{4.36}$$

where we have used Lemma 4.0 in the second equality, (4.35) in the third and (4.33) in the last equality. This proves the Lemma in 4-D ∎

**Lemma 4.5** *For a gauge potential $A_\mu(x)$ and its partner under a global gauge transformation $A'_\mu(x) = g^\dagger A_\mu(x) g$ where $g \in U(N)$, if $_0<R\pm|R\pm>_A$ and $_0<R\pm|R\pm>_{A'}$ are real and positive then*

$$_{A'}<R-|R+>_{A'} = {_A<R-|R+>_A}$$

*Proof:*

For the gauge potential $A_\mu(x)$ and its partner under a global gauge transformation, $A'_\mu(x)$ we have

$$\mathbf{H}^\pm(x\alpha i, y\beta j; A') = [g^\dagger]_{ik} \mathbf{H}^\pm(x\alpha k, y\beta l; A)[g]_{lj} \tag{4.37}$$

As a consequence of this we have the following relation

$$\psi_K^{R\pm}(x\alpha i; A') = e^{i\phi_K^{R\pm}}[g^\dagger]_{ij}\psi_K^{R\pm}(x\alpha j; A) \tag{4.38}$$

For $A = 0$ we have degeneracies and therefore

$$\tilde{\psi}_K^{R\pm}(x\alpha i; 0) = [g^\dagger]_{ij}\psi_K^{R\pm}(x\alpha j; 0) \tag{4.39}$$

is another set of eigenvectors for the free Hamiltonians. The transformation matrix relating $\psi_K^{R\pm}$ and $\tilde{\psi}_K^{R\pm}$ will be block diagonal and will mix only states in a degenerate subspace. The states $|R\pm>_0$ are obtained by filling all the states in a subspace. From (4.38) and (4.39) we have

$$_0<R\pm|R\pm>_{A'} = {_0<R\pm|R\pm>_A}\, e^{i\sum_K(\phi_K^{R\pm} + \alpha)} \tag{4.40}$$

where $e^{i\alpha} = [\det g]^{\frac{1}{N}}$ Since ${}_0< R \pm |R\pm >_A$ and ${}_0< R \pm |R\pm >_{A'}$ are real and positive it follows from (4.40) that

$$\sum_K \phi_K^{R\pm} = \sum_K \alpha \tag{4.41}$$

From (4.38) and (4.41) we have

$$\begin{aligned}{}_{A'}< R - |R+ >_{A'} &= {}_A< R - |R+ >_A \, e^{\sum_K (\phi_K^{R+} - \phi_K^{R-})} \\ &= {}_A< R - |R+ >_A\end{aligned} \tag{4.42}$$

and this proves the Lemma ∎

## 5. Expectation values of fermionic operators

In section 2, we obtained a formula for the chiral determinant as an overlap of two many body states. To the chiral Dirac operator $\mathbf{C}(A)$ in the presence of an arbitrary hermitian matrix valued vector potential $A$, two single particle Hamiltonians $\mathbf{H}^\pm(A)$ given by

$$\mathbf{H}^\pm(A) = \begin{pmatrix} \pm m & \mathbf{C}(A) \\ \mathbf{C}^\dagger(A) & \mp m \end{pmatrix} \tag{5.1}$$

were associated.

$$\mathcal{H}^\pm(A) = \sum_{x\alpha i, y\beta j} a^\dagger_{x\alpha i} \mathbf{H}^\pm(x\alpha i, y\beta j; A) a_{y\beta j} \tag{5.2}$$

are two many body Hamiltonians for non-interacting fermions with the single particle Hamiltonians given by (5.1). $a_{x\alpha i}$ and $a^\dagger_{x\beta i}$ are single particle destruction and creation operators obeying canonical commutation relations

$$\{a_{x\alpha i}, a_{y\beta j}\} = 0; \quad \{a^\dagger_{x\alpha i}, a^\dagger_{y\beta j}\} = 0; \quad \{a_{x\alpha i}, a^\dagger_{y\beta j}\} = \delta_{xy}\delta_{\alpha\beta}\delta_{ij}. \tag{5.3}$$

It was shown that

$$\begin{aligned}\det \mathbf{C}(A) &\Leftrightarrow {}_A< L - |L+ >_A \\ \det \mathbf{C}^\dagger(A) &\Leftrightarrow {}_A< R - |R+ >_A\end{aligned} \tag{5.4}$$

were proper definitions of the chiral determinants. The $|L\pm >_A$ were ground states of $\mathcal{H}^\pm(A)$ and the $|R\pm >_A$ were ground states of $-\mathcal{H}^\pm(A)$. Explicitly, referring to (2.11) which defines the single particle eigenfunctions of $\mathbf{H}^\pm(A)$, if we define new creation operators

$$[b_K^{R\pm}]^\dagger = \sum_{x\alpha i} a^\dagger_{x\alpha i} \psi_K^{R\pm}(x\alpha i; A); \quad [b_K^{L\pm}]^\dagger = \sum_{x\alpha i} a^\dagger_{x\alpha i} \psi_K^{L\pm}(x\alpha i; A), \tag{5.5}$$

then

$$|R\pm> = \prod_K [b_K^{R\pm}]^\dagger |0>; \quad |L\pm> = \prod_K [b_K^{L\pm}]^\dagger |0> \tag{5.6}$$

where $|0>$ is the bare vacuum annihilated by all the destruction operators $a_{x\alpha i}$. If we define

$$O^{RR}_{KK'}(A) = \sum_{x\alpha i} [\psi^{R-}_K(x\alpha i; A)]^* \psi^{R+}_{K'}(x\alpha i; A)$$
$$O^{RL}_{KK'}(A) = \sum_{x\alpha i} [\psi^{R-}_K(x\alpha i; A)]^* \psi^{L+}_{K'}(x\alpha i; A)$$
$$O^{LR}_{KK'}(A) = \sum_{x\alpha i} [\psi^{L-}_K(x\alpha i; A)]^* \psi^{R+}_{K'}(x\alpha i; A) \qquad (5.7)$$
$$O^{LL}_{KK'}(A) = \sum_{x\alpha i} [\psi^{L-}_K(x\alpha i; A)]^* \psi^{L+}_{K'}(x\alpha i; A)$$

then the overlaps are given by

$$_A<R-|R+>_A = \det O^{RR}(A); \quad _A<L-|L+>_A = \det O^{LL}(A). \qquad (5.8)$$

We wish to construct generating functionals for the expectation values of physical fermionic operators in an arbitrary gauge background. If the gauge background is trivial topologically the generating functional will factorize into a product of generating functionals, one factor for each gauge group multiplet. Then one can construct each of these factors independently by replacing the formal path integrals

$$\int [d\Psi^R][d\bar{\Psi}^R] \exp\Big[\bar{\Psi}^R \mathbf{D} \frac{1+\gamma_5}{2} \Psi^R + \sum_{x\alpha i}(\bar{\eta}^R_{x\alpha i}\Psi^R_{x\alpha i} + \eta^R_{x\alpha i}\bar{\Psi}^R_{x\alpha i})\Big]$$

$$\int [d\Psi^L][d\bar{\Psi}^L] \exp\Big[\bar{\Psi}^L \mathbf{D} \frac{1-\gamma_5}{2} \Psi^L + \sum_{x\alpha i}(\bar{\eta}^L_{x\alpha i}\Psi^L_{x\alpha i} + \eta^L_{x\alpha i}\bar{\Psi}^L_{x\alpha i})\Big]$$

by well defined objects, $Z^R(\bar{\eta}^R, \eta^R; A)$ and $Z^L(\bar{\eta}^L, \eta^L; A)$ respectively. $\Psi^R$ and $\Psi^L$ are four component Dirac spinors that represent the physical two component right handed and left handed fermions and the terms bilinear in them are the two chiral components of the Dirac operator as defined in (2.3). The bilinear terms contain chiral projectors ($\frac{1\pm\gamma_5}{2}$) and are therefore not invertible. This implies that in the first (second) of the two equations above only half the modes are propagating and represent a physical single right (left) handed fermion. $\eta^R$, $\eta^L$, $\bar{\eta}^R$ and $\bar{\eta}^L$ are Grassmann variables and are introduced in the usual manner to facilitate the definition of fermionic expectation values. The expressions for $Z^L$ and $Z^R$ are

$$Z^R(\bar{\eta}^R, \eta^R; A) = {}_A<R-|\exp\Big[\sum_{x\alpha i}(\bar{\eta}^R_{x\alpha i}a_{x\alpha i} + \eta^R_{x\alpha i}a^\dagger_{x\alpha i})\Big]|R+>_A$$
$$Z^L(\bar{\eta}^L, \eta^L; A) = {}_A<L-|\exp\Big[\sum_{x\alpha i}(\bar{\eta}^L_{x\alpha i}a_{x\alpha i} + \eta^L_{x\alpha i}a^\dagger_{x\alpha i})\Big]|L+>_A \qquad (5.9)$$

Completely antisymmetrized $n$-point functions of fermions are obtained by differentiation with respect to the associated Grassmann variables. Note that the dependence on $\bar{\eta}^R \frac{1-\gamma_5}{2}$, $\bar{\eta}^L \frac{1+\gamma_5}{2}$, $\eta^R \frac{1+\gamma_5}{2}$ and $\eta^L \frac{1-\gamma_5}{2}$ is no longer as trivial as it was in the formal path integral expressions. Nevertheless, we have not added any extra physical modes: Consider an $n$-point function and order the operators inside so that all the annihilation operators appear to the left and the creation operators appear to the right. In doing this there will be many terms with either $n$ or less operator insertions. Express the fermion creation and annihilation operators in the $\left[b_K^{R\pm}\right]^\dagger; \left[b_K^{L\pm}\right]^\dagger; b_K^{R\pm}; b_K^{L\pm}$ basis. Then, from (5.5) and (5.9) it becomes clear that only half of the modes will propagate. Although we are representing a single right (left) handed fermion by a four component Dirac spinor we do have the correct number of physical propagating modes. The explicit formulae in subsection (5.1) will make this clear.

If the gauge background is not trivial topologically the fermionic functional integral of the source term no longer factorizes and we must look for a definition that includes all the physical fermionic fields in the theory together. Eq. (5.9) generalizes easily, multiplying the expressions for each gauge multiplet. The product of the vacua matrix elements can be written as the matrix element of a complete source term between the two full vacuum states, each given by a tensor product over the multiplet vacua:

$$Z^{\text{total}}(\bar{\eta}, \eta; A) =$$
$$_A<0-|\exp\Big[\sum_{x\alpha i I_R}(\bar{\eta}_{x\alpha i}^{R,I_R} a_{x\alpha i}^{R,I_R} + \eta_{x\alpha i}^{R,I_R}(a^{R,I_R})_{x\alpha i}^\dagger) + \sum_{x\alpha i I_L}(\bar{\eta}_{x\alpha i}^{L,I_L} a_{x\alpha i}^{L,I_L} + \eta_{x\alpha i}^{L,I_L}(a^{L,I_L})_{x\alpha i}^\dagger)\Big]|0+>_A$$
$$|0\pm>_A = \bigotimes_{I_R}|R, I_R; \pm>_A \bigotimes_{I_L}|L, I_L; \pm>_A$$
(5.10)

The remaining task is to define the phases of the two complete vacua $|0\pm>_A$. This will be done in section (5.2).

*5.1. Zero topological sector*

In the zero topological sector one can write $Z^R$ and $Z^L$ in (5.9) as

$$Z^R(\bar{\eta}^R, \eta^R; A) = \exp[W^R(\bar{\eta}^R, \eta^R; A)]_A<R-|R+>_A$$
$$Z^L(\bar{\eta}^L, \eta^L; A) = \exp[W^L(\bar{\eta}^L, \eta^L; A)]_A<L-|L+>_A$$
(5.11)

The states $|R\pm>_A$ and $|L\pm>_A$ are well defined since their phases have been fixed by the Wigner-Brillouin choice (that is demanding that their overlap with the corresponding free states be real and positive). In particular, the two point functions are fully defined and are

$$<\bar{\Psi}_{x\alpha i}^R \Psi_{y\beta j}^R>_A \Leftrightarrow \frac{1}{2}\frac{_A<R-|(a_{x\alpha i}^\dagger a_{y\beta j} - a_{y\beta j} a_{x\alpha i}^\dagger)|R+>_A}{_A<R-|R+>_A}$$
$$<\bar{\Psi}_{x\alpha i}^L \Psi_{y\beta j}^L>_A \Leftrightarrow \frac{1}{2}\frac{_A<L-|(a_{x\alpha i}^\dagger a_{y\beta j} - a_{y\beta j} a_{x\alpha i}^\dagger)|L+>_A}{_A<L-|L+>_A}$$
(5.12)

By referring to the formal path integrals we see that these are the inverses of the chiral operators $\mathbf{C}^\dagger$ and $\mathbf{C}$ respectively. Therefore, we would expect to have

$$<\bar{\Psi}_{x\alpha i}^R \Psi_{y\beta j}^R>_A = -\Big[<\bar{\Psi}_{y\beta j}^L \Psi_{x\alpha i}^L>_A\Big]^*$$
(5.13)

We proceed to show that the precisely defined right hand sides of (5.12) also satisfy (5.13). We use

(5.5) to write $a_{x\alpha i}$ and $a^\dagger_{x\alpha i}$ as

$$a_{x\alpha i} = \sum_K \psi_K^{R\pm}(x\alpha i;A) b_K^{R\pm} + \sum_K \psi_K^{L\pm}(x\alpha i;A) b_K^{L\pm}$$
$$a^\dagger_{x\alpha i} = \sum_K \left[\psi_K^{R\pm}(x\alpha i;A)\right]^* [b_K^{R\pm}]^\dagger + \sum_K \left[\psi_K^{L\pm}(x\alpha i;A)\right]^* [b_K^{L\pm}]^\dagger \quad (5.14)$$

From (5.6) it follows that

$$[b_K^{R+}]^\dagger |R+>_A = 0; \quad b_K^{L+}|R+>_A = 0; \quad {}_A<R-|b_K^{R-} = 0; \quad {}_A<R-|[b_K^{L-}]^\dagger = 0; \quad (5.15)$$

Then

$$\frac{1}{2}{}_A<R-|R+>_A \delta_{xy}\delta_{\alpha\beta}\delta_{ij} - \frac{1}{2}{}_A<R-|(a^\dagger_{x\alpha i}a_{y\beta j} - a_{y\beta j}a^\dagger_{x\alpha i})|R+>_A$$
$$= {}_A<R-|a_{y\beta j}a^\dagger_{x\alpha i}|R+>_A$$
$$= \sum_{K_1,K_1'} \psi_{K_1'}^{L-}(y\beta j)\left[\psi_{K_1}^{L+}(x\alpha i)\right]^* <0|\prod_{K'} b_{K'}^{R-} b_{K_1'}^{L-} [b_{K_1}^{L+}]^\dagger \prod_K [b_K^{R+}]^\dagger |0>$$
$$= \left[\sum_{K_1,K_1'} \left[\psi_{K_1'}^{L-}(y\beta j)\right]^* \psi_{K_1}^{L+}(x\alpha i) <0|\prod_{K'} b_{K'}^{L-} [b_{K_1'}^{L-}]^\dagger b_{K_1}^{L+} \prod_K [b_K^{L+}]^\dagger |0>\right]^* \quad (5.16)$$
$$= \left[{}_A<L-|a^\dagger_{y\beta j}a_{x\alpha i}|L+>_A\right]^*$$
$$= \left[\frac{1}{2}{}_A<L-|(a^\dagger_{y\beta j}a_{x\alpha i} - a_{x\alpha i}a^\dagger_{y\beta j})|L+>_A + \frac{1}{2}{}_A<L-|L+>_A \delta_{xy}\delta_{\alpha\beta}\delta_{ij}\right]^*$$

We have used (5.3) in the first and last equality and (5.14) and (5.15) in the second and fourth equality. The matrix elements appearing after the second equality are overlaps bewteen two states, each obtained by filling all the positive energy states and one negative energy state for both Hamiltonians respectively. The matrix elements appearing after the third equality are overlaps between two states, each obtained by filling all the negative energy states but one for each of the two Hamiltonians. There is a one to one correspondence between the two overlaps given by a particle hole transformation. More specifically, the two overlaps are related by Lemma (4.0):

$$\det \mathcal{U}(-m;A|+m;A) = \frac{<0|\prod_{K'} b_{K'}^{R-} b_{K_1'}^{L-} [b_{K_1}^{L+}]^\dagger \prod_K [b_K^{R+}]^\dagger |0>}{\left[<0|\prod_{K'} b_{K'}^{L-} [b_{K_1'}^{L-}]^\dagger b_{K_1}^{L+} \prod_K [b_K^{L+}]^\dagger |0>\right]^*}. \quad (5.17)$$

By (4.7) the left hand side above is unity and this establishes the third equality in (5.16). Dividing (5.16) by ${}_A<R-|R+>_A = \left({}_A<L-|L+>_A\right)^*$ we obtain the analogue of (5.13).

An explicit expression for the two-point function can be obtained by writing the determinants representing the matrix elements in (5.16) in terms of the matrices $O$ defined in (5.7). We employ the identity $\det \begin{pmatrix} X & Y \\ Z & V \end{pmatrix} = \det X \det(V - ZX^{-1}Y)$ where now $Y$ is a column vector and $Z$ is a row vector and $V$ is a number in the second equality in (5.16) to obtain:

$$\frac{{}_A\!< R - |a_{y\beta j} a^\dagger_{x\alpha i}|R+>_A}{{}_A\!< R - |R+>_A} = \sum_{K_1, K_1'} \psi^{L-}_{K_1'}(y\beta j)\left[\psi^{L+}_{K_1}(x\alpha i)\right]^* \left[O^{LL}_{K_1' K_1} - \sum_{K, K'} O^{LR}_{K_1' K}\left[O^{RR}\right]^{-1}_{KK'} O^{RL}_{K'K_1}\right] \tag{5.18}$$

Using the explicit expressions in (5.7) and the completeness relation for the eigenvectors of $\mathbf{H}^\pm(A)$, (5.18) reduces to

$$\frac{{}_A\!< R - |a^\dagger_{x\alpha i} a_{y\beta j}|R+>_A}{{}_A\!< R - |R+>_A} = \sum_{KK'} \psi^{R+}_{K'}(y\beta j; A)[O^{RR}]^{-1}_{K'K}\left[\psi^{R-}_K(x\alpha i; A)\right]^* \tag{5.19}$$

The right side of the above formula has a simple interpretation. Since we are computing the propagator of a right handed fermion we expect to get $[O^{RR}]^{-1}$. The two wavefunctions carry out a change of bases from the two bases of the negative energy subspaces of $\mathbf{H}^\pm$ to a common coordinate basis. It is evident that the matrix $O^{RR}$ is treated as a representative of a mapping between two distinct spaces. The right hand side viewed as a matrix element implies that the associated matrix has a rank equal to the dimension of the space spanned by the index $K$ which is nothing but the basis for a single right handed fermion.

(5.19) can be recast in a form that makes the relation (5.13) transparent: Since

$$\mathcal{U}(-m; A| + m; A) = \begin{pmatrix} O^{RR} & O^{RL} \\ O^{LR} & O^{LL} \end{pmatrix}$$

is a unitary matrix it follows that

$$\left[O^{RR}\right]^{-1} = \left[O^{RR}\right]^\dagger - \left[O^{LR}\right]^\dagger \left(\left[O^{LL}\right]^\dagger\right)^{-1} \left[O^{RL}\right]^\dagger \tag{5.20}$$

Employing again completeness we obtain the following identity:

$$\sum_{KK'} \left[\psi^{R+}_{K'}(y\beta j; A)[O^{RR}]^{-1}_{K'K}\left[\psi^{R-}_K(x\alpha i; A)\right]^* + \psi^{L-}_K(y\beta j; A)\left[(O^{LL})^\dagger\right]^{-1}_{KK'}\left[\psi^{L+}_{K'}(x\alpha i; A)\right]^*\right]$$
$$= \delta_{xy}\delta_{\alpha\beta}\delta_{ij} \tag{5.21}$$

Using (5.21) the expression for the two-point function in (5.19) can be rewritten as

$$\frac{1}{2}\frac{{}_A\!< R - |(a^\dagger_{x\alpha i} a_{y\beta j} - a_{y\beta j} a^\dagger_{x\alpha i})|R+>_A}{{}_A\!< R - |R+>_A}$$
$$= \frac{1}{2}\sum_{KK'} \left[\psi^{R+}_{K'}(y\beta j; A)[O^{RR}]^{-1}_{K'K}\left[\psi^{R-}_K(x\alpha i; A)\right]^* - \psi^{L-}_K(y\beta j; A)\left[(O^{LL})^\dagger\right]^{-1}_{KK'}\left[\psi^{L+}_{K'}(x\alpha i; A)\right]^*\right] \tag{5.22}$$

To obtain the two-point function for the left-handed fermion, all the $R$'s have to be replaced by $L$'s in the above equation and the relation (5.13) is made transparent.

We will presently show that $W^R(\bar\eta^R, \eta^R; A)$ is bilinear and has only terms connecting $\bar\eta^R$ and $\eta^R$. The same is true for the left handed fermions. This has the following consequences: The $n$-point functions should have an equal number of fermion creation and annihilation operators for them to be non-zero. The $n$-point functions are sums of products of the two-point function in (5.12). The proof is straightforward but slightly complicated because the expectation value is between two different states $|R\pm>$. From (5.5), we use

$$[b_K^{R+}]^\dagger = \sum_{x\alpha i} a^\dagger_{x\alpha i} \psi_K^{R+}(x\alpha i; A); \quad [b_K^{L-}]^\dagger = \sum_{x\alpha i} a^\dagger_{x\alpha i} \psi_K^{L-}(x\alpha i; A), \quad (5.23)$$

to express all the $a^\dagger_{x\alpha i}$ in terms of $[b_K^{R+}]^\dagger$ and $[b_K^{L-}]^\dagger$ and we use

$$b_K^{R-} = \sum_{x\alpha i} a_{x\alpha i} \left[\psi_K^{R-}(x\alpha i; A)\right]^*; \quad b_K^{L+} = \sum_{x\alpha i} a_{x\alpha i} \left[\psi_K^{L+}(x\alpha i; A)\right]^*, \quad (5.24)$$

to express all the $a_{x\alpha i}$ in terms of $b_K^{R-}$ and $b_K^{L-}$. The above transformations are non-singular (and therefore valid) if the overlap $_A<R-|R+>_A$ is non-zero. One way to see this is as follows. In (5.23) we know that all the $\psi_K^{R+}(x\alpha i; A)$ are linearly independent and so are all the $\psi_K^{L-}(x\alpha i; A)$. All we need to show is that we cannot write any $\psi_K^{R+}(x\alpha i; A)$ as a linear combination of $\psi_K^{L-}(x\alpha i; A)$. If this were possible then $b_K^{R-}|R->_A = 0$ and we would have $_A<R-|R+>_A = 0$. Similar arguments can be used for (5.24) also. Therefore we can write

$$a_{x\alpha i} = \sum_K \phi_K^{R-}(x\alpha i; A) b_K^{R-} + \sum_K \phi_K^{L+}(x\alpha i; A) b_K^{L+}$$
$$a^\dagger_{x\alpha i} = \sum_K \left[\phi_K^{R+}(x\alpha i; A)\right]^* [b_K^{R+}]^\dagger + \sum_K \left[\phi_K^{L-}(x\alpha i; A)\right]^* [b_K^{L-}]^\dagger \quad (5.25)$$

where $\phi_K^{R\pm}(x\alpha i; A)$ and $\phi_K^{L\pm}(x\alpha i; A)$ are obtained by inverting the relations (5.23) and (5.24). That is

$$\sum_K \phi_K^{R-}(x\alpha i; A)\left[\psi_K^{R-}(y\beta j; A)\right]^* + \sum_K \phi_K^{L+}(x\alpha i; A)\left[\psi_K^{L+}(y\beta j; A)\right]^* = \delta_{xy}\delta_{\alpha\beta}\delta_{ij}$$
$$\sum_K \left[\phi_K^{R+}(x\alpha i; A)\right]^* \psi_K^{R+}(y\beta j; A) + \sum_K \left[\phi_K^{L-}(x\alpha i; A)\right]^* \psi_K^{L-}(y\beta j; A) = \delta_{xy}\delta_{\alpha\beta}\delta_{ij} \quad (5.26)$$

Let

$$\mathcal{Q}_- = \sum_{x\alpha i; K} \left[\bar\eta^R_{x\alpha i}\phi_K^{R-}(x\alpha i; A) b_K^{R-} + \eta^R_{x\alpha i}\left[\phi_K^{L-}(x\alpha i; A)\right]^* \left[b_K^{L-}\right]^\dagger\right]$$
$$\mathcal{Q}_+ = \sum_{x\alpha i; K} \left[\bar\eta^R_{x\alpha i}\phi_K^{L+}(x\alpha i; A) b_K^{L+} + \eta^R_{x\alpha i}\left[\phi_K^{R+}(x\alpha i; A)\right]^* \left[b_K^{R+}\right]^\dagger\right] \quad (5.27)$$

From (5.3), (5.23), (5.24) and (5.26) it follows that the commutator

$$[\mathcal{Q}_+, \mathcal{Q}_-] = \sum_{x\alpha i; y\beta j} \eta^R_{x\alpha i}\bar\eta^R_{y\beta j} \sum_K \left[\phi_K^{R-}(x\alpha i)\left[\psi_K^{R-}(y\beta j)\right]^* + \psi_K^{R+}(x\alpha i)\left[\phi_K^{R+}(y\beta j)\right]^* - \delta_{xy}\delta_{\alpha\beta}\delta_{ij}\right]$$
$$= -\sum_{x\alpha i; y\beta j} \eta^R_{x\alpha i}\bar\eta^R_{y\beta j} \sum_K \left[\phi_K^{L+}(x\alpha i)\left[\psi_K^{L+}(y\beta j)\right]^* + \psi_K^{L-}(x\alpha i)\left[\phi_K^{L-}(y\beta j)\right]^* - \delta_{xy}\delta_{\alpha\beta}\delta_{ij}\right]$$
$$(5.28)$$

is a $c$-number. There are only terms that mix $\eta^R$ and $\bar\eta^R$ in the commutator. All other terms are zero because $a^2_{x\alpha i}=0$ and $[a^\dagger_{x\alpha i}]^2=0$. Upon inserting (5.25) in (5.9) and using (5.27) we see that

$$\exp[W^R(\bar\eta^R,\eta^R;A)]_A\!<\!R-|R+>_A\; =\; _A\!<\!R-|\exp[\mathcal{Q}_-+\mathcal{Q}_+]|R+>_A$$
$$=\exp\Bigl(\frac{1}{2}[\mathcal{Q}_+,\mathcal{Q}_-]\Bigr)_A\!<\!R-|\exp[\mathcal{Q}_-]\exp[\mathcal{Q}_+]|R+>_A$$
$$=\exp\Bigl(\frac{1}{2}[\mathcal{Q}_+,\mathcal{Q}_-]\Bigr)_A\!<\!R-|R+>_A \tag{5.29}$$

We have split the exponential in the second equality and used the fact that the commutator in (5.28) is a $c$-number. In the third equality we have used (5.15). Therefore,

$$W^R(\bar\eta^R,\eta^R;A)=\frac{1}{2}[\mathcal{Q}_+,\mathcal{Q}_-] \tag{5.30}$$

where the expression for the right hand side is in (5.28) and is a bilinear in $\eta^R$ and $\bar\eta^R$.

*5.2. Non-zero topological sectors*

Our construction for nonzero topological charge will apply only to situations where a local fermionic operator $\mathcal{F}$ exists with the following properties: It is a Lorentz scalar; it is translational and gauge invariant; it can connect a state $|0+>_A$ to a state $|0->_A$ when $A$ carries topological charge. We shall take $\mathcal{F}$ to be the most local and least dimension operator satisfying these requirements. Such an operator will be referred to as a 't Hooft [11] vertex since it would exhibit the (minimal) violation of global symmetries due to instanton–like gauge fields. If such an operator does not exist we have no physical reason to try to define fermionic expectation values in nontrivial gauge backgrounds. Typically, there will be one or few such operators. They make up a fundamental set in the sense that any nontrivial gauge background that would give a nonvanishing expectation value to some fermionic operator will also give a nonvanishing expectation value to a monomial made out of the fundamental 't Hooft vertices $\mathcal{F}$ and their hermitian conjugates. Henceforth we shall assume, mainly for notational simplicity, that there is only one such operator $\mathcal{F}$.

We shall define the phases of the states $|0\pm>_A$ in (5.9) by replacing the physical fields in $\mathcal{F}$ by the matching creation and annihilation operators as in (5.9), and denoting the resulting operator by $\hat{\mathcal{F}}$ demanding $_A\!<\!0\pm|\hat{\mathcal{F}}^n|0\pm>_A>0$ or $_A\!<\!0\pm|[\hat{\mathcal{F}}^\dagger]^n|0\pm>_A>0$ for some $n\geq 0$. One may unify our phase definitions to one equation that is satisfactory for all topological sectors:

$$_A\!<\!0\pm|\Bigl[e^{\hat{\mathcal{F}}}+e^{\hat{\mathcal{F}}^\dagger}\Bigr]|0\pm>_0 \quad\text{is real and positive.} \tag{5.31}$$

In a topological sector with topological charge $q$, $Z$ in (5.9) can be written as

$$Z(\bar\eta,\eta;A)=\exp[W_q(\bar\eta,\eta;A)]z(\bar\eta,\eta;A). \tag{5.32}$$

$W_q$ is bilinear and only terms connecting an $\eta$ to an $\bar\eta$ appear. $z$ is a monomial in a subset of the Grassmann sources $\bar\eta$ and $\eta$ of a power increasing with $|q|$. $z$ reflects the insertions needed to "soak up" the zero modes in path integral formulations. This structure can be derived by following the steps described in the previous sub-section and therefore we omit the details.

# 6. The chiral determinant in perturbation theory

For a gauge background that can be smoothly deformed to zero we can use perturbation theory. Since it is generally easier to carry out perturbative computations using Feynman diagram techniques [3], rather than Schrödinger time independent perturbation theory (which could have been directly applied to the overlap [4]), we take a detour and first derive an Euclidean path integral expression for the overlap formula. Once we have a path integral expression, the Feynman rules can be easily written down.

To write the overlap $_A< L - |L+ >_A$ as a path integral we proceed as follows. For generic perturbative gauge fields, we expect $_0< L \pm |L\pm >_A$ to be non-zero. Since $|L\pm >_A$ are ground states of $\mathcal{H}^\pm(A)$ given by (5.2), we can write them as

$$|L\pm >_A = \lim_{T_\pm \to \infty} \frac{e^{-T_\pm \mathcal{H}^\pm(A)}|L\pm >_0}{\sqrt{_0< L \pm |e^{-2T_\pm \mathcal{H}^\pm(A)}|L\pm >_0}} \tag{6.1}$$

$|L\pm >_A$ so constructed satisfies the Wigner-Brillouin phase choice. From (6.1), the overlap can be written as

$$_A< L - |L+ >_A = \lim_{T_\pm \to \infty} \frac{_0< L - |e^{-T_- \mathcal{H}^-(A)} e^{-T_+ \mathcal{H}^+(A)}|L+ >_0}{\sqrt{_0< L - |e^{-2T_- \mathcal{H}^-(A)}|L- >_0}\sqrt{_0< L + |e^{-2T_+ \mathcal{H}^+(A)}|L+ >_0}} \tag{6.2}$$

To convert the above expression to a path integral, we first write down the one particle Schrödinger wave equation with the Hamiltonian $\mathbf{H}^\pm$ (c.f. (2.10)):

$$i\partial_t \psi = \gamma_5[\pm m + \gamma_\mu(\partial_\mu + iA_\mu(x))]\psi \tag{6.3}$$

We view this as a five dimensional Dirac equation and use the five-dimensional Minkowski metric $(+,-,-,-,-)$. The five dimensional $\gamma$ matrices are taken as $\Gamma_0 = \gamma_5$, $\Gamma_\mu = -i\gamma_\mu$, $\mu = 1,2,3,4$. We now want to recast the above equation in the form of the five dimensional Dirac equation. To this end we multiply by $\Gamma_0$ and obtain:

$$(i\Gamma^a \partial_a - \Gamma^a A_a \mp m)\psi = 0 \tag{6.4}$$

Here $a = 0,1,2,3,4$, $A_0 = 0$ and $A_\mu(x)$ is independent of $t$. The Minkowski path integral corresponding to the second quantized system equivalent to the above Dirac equation and incorporating Fermi statistics is:

$$\int [d\bar\psi][d\psi] e^{-i \int \bar\psi[\Gamma^a(i\partial_a - A_a) \mp m]\psi} \tag{6.5}$$

We now go to five dimensional Euclidean space, replacing $t$ by $-is$ to obtain the Euclidean path integral

$$\int [d\bar\psi][d\psi] e^{-\int \bar\psi[\gamma_5 \partial_s + \gamma_\mu(\partial_\mu + iA_\mu) \pm m]\psi} \tag{6.6}$$

Using (6.6) the three factors in (6.2) can be written as a path integral and the result is

$$_A< L - |L+ >_A = \lim_{T_\pm \to \infty} \frac{\int [d\psi][d\bar\psi] e^{-\int_{s=-T_-}^{s=T_+} ds L_E(\bar\psi,\psi;A,m,s)}}{\sqrt{\int [d\psi][d\bar\psi] e^{-\int_{s=-2T_-}^{s=0} ds L_E(\bar\psi,\psi;A,m,s)}} \sqrt{\int [d\psi][d\bar\psi] e^{-\int_{s=0}^{s=2T_+} ds L_E(\bar\psi,\psi;A,m,s)}}} \tag{6.7}$$

where the $s-$ dependent Euclidean Lagrangian is

$$L_E(\bar\psi,\psi;A,m,s) = \int [dx]\bar\psi(x,s)\Big[\gamma_5\partial_s + \gamma_\mu(\partial_\mu + iA_\mu(x)) + m\epsilon(s)\Big]\psi(x,s). \qquad (6.8)$$

Note that the Lagrangians in the denominator are in fact independent of $s$. This is because in one factor $s$ is always negative and in the other factor $s$ is always positive. This reflects the fact that one Hamitonian is $\mathcal{H}^-$ and the other is $\mathcal{H}^+$. On the other hand in the numerator the Lagrangian changes as $s$ goes from the negative side to positive side and this accounts for the two Hamiltonians $\mathcal{H}^\pm$ present in the numerator.

(6.7) is still unregularized. For finite $T_\pm$ gauge invariance is broken by the boundaries, no matter what boundary conditions we pick. When the $T_\pm$ are taken to infinity this breaking disappears from the *absolute* magnitude of the expression; at the same time the free propagator is specified by requiring decay at infinite $s$. The phase of the expression is determined and will typically come out gauge dependent. Even before considering regularization we need to understand how in perturbation theory the infinite $s$ ranges do not induce additional infinities.*

In perturbation theory one computes the logarithm of the overlap. At a fixed order one has three diagrams to consider: one from the numerator and two from the two factors in the denominator. Each of these diagrams has $n$ vertices and each vertex has, among other labels, an $s$ label, $s_1, s_2, .....s_n$. The integrals over the $s_i$ variables are to be done in the intervals prescribed and the limits on $T_\pm$ of the three diagrams are taken in a correlated way. To obtain the rules directly in the large $T_\pm$ limit, change variables in each of the diagrams from $s_1, s_2, ....s_n$ to a center of mass coordinate $\bar s$ and $n-1$ orthogonal linear combinations; the complicated bounds of integration for the latter simplify in the limit $T_\pm \to \infty$ and there the ranges of the relative variables are independent and run over the whole real axis. The integrals over the relative coordinates are to be performed first; as long as the center of mass is kept finite no divergences will appear from these integrals. These integrals can be done independently for each of the three diagrams. One is now left with three $s$ integrals: one over $\bar s$ coming from the numerators and two over $\bar s_\pm$ coming from the denominators. The square roots in the denominator attach a factor $\frac{1}{2}$ to the $\bar s_\pm$ integrals. However, the diagrams coming from the denominator, once the integrals over the relative coordinates have been carried out over their infinite ranges have no $\bar s_\pm$ dependence left as the systems corresponding to the denominator are homogeneous. (This will be only approximately true if we smooth out the behavior of the mass term but this has no effect on the finiteness.) The factors of one half simply cancel against the factors of 2 in the ranges and the constant terms are subtracted, respectively, from the positive $s$ and negative $s$ portions of the $\bar s$ integrands rendering the corresponding integrals finite. Since the denominators are real the subtractions only affect the real part of the diagrams. The imaginary part of the diagrams generated by the numerator can be summed over $s$ without encountering divergences, i.e. are well defined.

The above defines the unregulated Feynman diagrams of the theory. One may ask whether the path integral representations, or, equivalently, the limiting expressions leading to them can be taken over to apply also for $A$ fields that cannot be deformed to zero. The answer is emphatically no: The path integrals, the limiting expressions and the related Feynman rules are to be used only for perturbative $A$'s. For example, if the background is an instanton, the limiting expressions would be saturated by an excited state on one of the sides rather than vanish as they should. So, only the overlap formula is to be taken outside perturbation theory.

---

\* For a detailed discussion see recent work by S. Aoki and R. Levien [12].

## 7. Regularizing the Overlap

We turn now to the problem of regularizing the overlap formulae (2.18) and (2.20) on the lattice. We choose the lattice because on it the dynamics of the gauge fields can be defined non–perturbatively and we wish to extend this definition to chiral fermions.

More precisely, our task is the following: Assume, for definiteness, that Euclidean spacetime is a torus. On it we define a gauge connection that may be in a nontrivial topological sector. We embed a finite hypercubic lattice in this torus and define link variables $U_\mu(x)$ as the parallel transporters along the link connecting $x$ to $x + \hat{\mu}$. We wish to define two Hamiltonian matrices $\mathbf{H}^\pm$ with indices as in (5.2) (except $x$ and $y$ now label sites and have a finite range) such that, following through with the construction of the overlap will produce a functional of the link variables which will vanish if the continuum connection was in a nontrivial sector and will agree with continuum perturbation theory in the topologically trivial sector. By this we mean that if the external connection is small and slowly varying on the lattice scale (i.e. the lattice is fine enough), a double expansion in it and its derivatives has the form one might have obtained by regularizing the determinant of the chiral Dirac operator by some acceptable continuum method. If the continuum connection is in a topologically nontrivial sector, the lattice overlap should vanish due to unequal fillings of the two Dirac seas, as explained in section 3. In addition we require that if we follow through with the construction of the generating functional of fermion field expectation values (eq. (5.9)) for the lattice Hamiltonians in an any topological sector, the correct continuum behavior is obtained when the external momenta of the fermi fields are small relative to the inverse lattice spacing. We also require to preserve the symmetries mentioned in section 4, more specifically, eq. (4.9) for the hypercubic space symmetry group, eq. (4.15), eq. (4.20) in 2-D, eq. (4.21) in 4-D and eq. (4.37). The task is to find local matrices $\mathbf{H}^\pm$ (in the sense that matrix elements connecting sites that are too far away vanish exponentially) satisfying the above requirements. These matrices are parametrically dependent on the link variables with no restrictions on them and so is the rest of the construction. Thus we obtain a non–perturbative definition.

The free Hamiltonian matrices must be traceless by (4.15) and so must be the Hamiltonians in the presence of parity invariant gauge potentials. It causes no complications to make the Hamiltonian matrices traceless for all gauge fields. Then the structure is as in eq. (2.22),

$$\mathbf{H}^\pm = \begin{pmatrix} \mathbf{B}^\pm & \mathbf{C} \\ \mathbf{C}^\dagger & -\mathbf{B}^\pm \end{pmatrix}, \qquad (7.1)$$

however we shall see that we need to allow for $[\mathbf{B}^\pm, \mathbf{C}] \neq 0$.

On the lattice $\mathbf{C}$ becomes a finite matrix of fixed dimension and $\mathbf{C}^\dagger$ its adjoint under the ordinary inner product. Thus $\mathbf{C}$ and $\mathbf{C}^\dagger$ have the same rank. Under a local gauge transformation $\mathbf{C}$ will transform by conjugation and therefore $\det \mathbf{C}$ would be gauge invariant. These properties show that $\det \mathbf{C}$ would be an incorrect replacement for the chiral determinant since there would be no anomalies and the behavior of $\mathbf{C}$ in topologically nontrivial backgrounds would be wrong.

The mechanism by which lattice regularizations avoid contradiction with the known continuum properties can be seen already for the free field case and goes under the name of "doubling" [7]. Indeed, in four dimensions, relations (4.9), (4.20) and (4.21) require $\mathbf{C}$ to be of the form

$$\mathbf{C}(x\alpha j, y\beta k) = \sum_\mu \sigma_\mu(\alpha, \beta)[f(U_\mu, U_\mu^\dagger; \mu)](xj, yk) \tag{7.2}$$

with

$$\{[f(U_\mu, U_\mu^\dagger; \mu)](xj, yk)\}^* = [f(U_\mu^*, U_\mu^T; \mu)](xj, yk) \tag{7.3}$$

For $U_\mu \equiv 1$ the simplest choice for $f$ is

$$[f(1, 1; \mu)](xj, yk) = \frac{1}{2}\delta_{jk}[\delta_{y,x+\hat\mu} - \delta_{x,y+\hat\mu}] \tag{7.4}$$

Upon Fourier transformation of (7.4), $\mathbf{C}(p)$ becomes proportional to $\sum_\mu \sigma_\mu f(p; \mu)$ with $f(p; \mu) = \sin p_\mu \equiv \bar p_\mu$. $\mathbf{C}(p)$ has 16 zeros rather than only one at $p = 0$ as we would have liked. Linearizing around the zeros we get 16 copies of the continuum $\sigma \cdot p$ but with some of the signs of the $\sigma_\mu$'s switched. 8 of the $\sigma_\mu$ sets thus obtained transform as the original set and the other 8 transform as the set $\sigma_\mu^\dagger$. We see that we have in reality not one Weyl fermion of a given handedness but 8 right handed ones and 8 left handed ones. It is impossible to couple the link variables in a local and gauge invariant way to only some of these particles since they are represented by the Fourier modes of the same local field. Therefore, upon gauging one ends up with a vector theory and there is no contradiction with the know continuum properties. The structure of the regular lattice, as reflected in the toroidal topology of the momentum space (in the infinite volume limit) ensures that no other choice for the functions $f(p; \mu)$, even giving up some of the symmetries we required, will eliminate the doubling (probably a better term would be mirroring) of the Weyl fermions. In summary, $\mathbf{C}$ is not an acceptable discretization of the corresponding continuum operator because it describes too many fermion species avoiding the reproduction of anomalies and because it cannot have an analytical index.

The above considerations mean that one cannot let the matrices $\mathbf{B}^\pm$ stay constants like in the formal treatment. If we did, the derivations presented there hold here too, and the overlap is simply related to the determinant of $\mathbf{C}$, but the latter does not describe the theory we want. Let us first see how to rescue the overlap in the free case, but without changing $\mathbf{C}$. We need to break the tight relationship between the overlap and $\det \mathbf{C}$ for momenta away from the origin. If $\mathbf{B}^+$ is a negative constant and $\mathbf{C} \approx 0$, the vacuum of $\mathbf{H}^+$ contains filled single particle states distinguished by having a vanishing bottom half; similarly if $\mathbf{B}^-$ is a positive constant and $\mathbf{C} \approx 0$, the vacuum of $\mathbf{H}^-$ contains filled single particle states distinguished by having a vanishing top half. Thus every single particle state filled in one of the vacua is orthogonal to every single particle state filled in the other vacuum and the zero of $\mathbf{C}$ propagates to a zero in the overlap. We want this to happen only for the zeros at $p = 0$. Therefore, we make the matrix $\mathbf{B}^\pm$ space dependent, introducing a derivative term, such that in Fourier space $\mathbf{B}^-(p)$ be positive everywhere, but $\mathbf{B}^+(p)$ be negative only near $p = 0$ and positive near all the other zeros of $\mathbf{C}$:

$$\mathbf{B}^\pm(x, y) = \frac{1}{2}\sum_\mu [2\delta_{x,y} - \delta_{y,x+\hat\mu} - \delta_{x,y+\hat\mu}] \mp m \qquad 0 < m < 1 \tag{7.5}$$

This shows that the well known Wilson mass term eliminates unwanted fermionic excitations for the overlap also. Note that we have not put in any dependence on the spin indices; this helps preserving the symmetry properties we discussed above. We have also ignored the group indices since they are immaterial in the free case.

We still need to show that upon gauging indeed the continuum behavior in perturbation theory and in nontrivial topological backgrounds is reproduced. This is the main goal of the entire paper. Here we just wish to explain that the reasons given before for **C** being an incorrect discretization no longer apply. In order to preserve gauge invariance when the gauge fields are turned on the derivatives appearing in $\mathbf{B}^\pm$ and **C** must be covariantized; therefore $\mathbf{B}^\pm$ and **C** no longer commute. It becomes now impossible to write down an explicit formula for the overlap in terms of smooth functions of the matrices $\mathbf{B}^\pm$ and **C**: the derivation given for the continuum, formal problem, breaks down. The overlap is still *implicitly* defined, but it is no longer guaranteed that one can make it smooth in the link variables expanded around $U_\mu \equiv 1$ and gauge invariant at the same time. It is also clear that the possible gauge breaking is restricted to the phase. With the Wigner Brillouin choice the overlap will still obey the set of invariances discussed in section 4 and this provides useful restrictions as mentioned there. Thus anomalies can be reproduced.

Also, the possibility of correctly reproducing the effects of nontrivial topological sectors is opened up. For this to work one must see that unlike the free Hamiltonians, the ones with gauge fields turned on do not have impenetrable (gauge field independent) gaps in their spectrum. Indeed, on a lattice the space of the link variables is connected (the topological difference between the continuum $A_\mu$ representing an instanton say and $A_\mu \equiv 0$ is lost upon passage to the link variables on the embedded lattice) so it must be possible for an eigenvalue to go through zero as the gauge field is changed. This has to be true of at least one of the Hamiltonians (we shall see later that indeed it is true only of $\mathbf{H}^+$ but not of $\mathbf{H}^-$ and this is just enough). Unequal filling levels of the two vacua are able to ensure the robust vanishing of the chiral determinant reflecting the known related continuum property.

In summary the two Hamiltonians are given by (7.1) with

$$\mathbf{C}(x\alpha j, y\beta k) = \sum_\mu \sigma_\mu(\alpha, \beta) \frac{1}{2}[\delta_{y,x+\hat{\mu}} U_\mu^{ij}(x) - \delta_{x,y+\hat{\mu}}(U_\mu^\dagger(y))^{ij}] \qquad (7.6)$$

$$\mathbf{B}^\pm(x\alpha i, y\beta j) = \frac{1}{2}\delta_{\alpha\beta} \sum_\mu [2\delta_{x,y}\delta_{ij} - \delta_{y,x+\hat{\mu}} U_\mu^{ij}(x) - \delta_{x,y+\hat{\mu}}(U_\mu^\dagger(y))^{ij}] \mp m \equiv B \mp m \qquad (7.7)$$

and define the two many body ground states with phases fixed by the Wigner Brillouin choice relative to the appropriate reference states (depending on topology). The two ground states can then be used to compute the overlap or fermionic expectation values as appropriate. Regularized versions of the symmetries discussed in section 4 hold. The two Hamiltonians are local and typically have a gap. Each one of the ground states should be local in the sense that sampling it with various observables should generate local functionals of the gauge fields. Nonlocal terms in the action can appear once the overlap of the two different ground states is constructed.

The action is known to include nonlocal terms of a special kind if the theory is anomalous. It is therefore important to see in more detail how anomalies appear in the overlap. We are considering here perturbative anomalies and therefore the background field is assumed to have zero topological charge.* Let $|R\pm>_U$ be many body states obtained by filling all the positive energy states of $\mathbf{H}^\pm(U)$ respectively, with arbitrary overall phases. The Wigner–Brillouin phase choice is implemented by the replacement

$$|R\pm>_U \to |R\pm>_U \frac{{}_U<R\pm|R\pm>_1}{|{}_U<R\pm|R\pm>_1|} \tag{7.8}$$

---

\* To be sure, the anomalies have to appear also when the background is topologically non-trivial.

The chiral determinant for a single righthanded fermion is the overlap of the two $\pm$ states on the righthand side of (10.15). Eliminating a gauge field independent constant, the effective action induced by a single righthanded fermion becomes

$$e^{\Gamma^R(U)} = \frac{{}_1<R-|R->_U}{|{}_1<R-|R->_U|} \frac{{}_U<R-|R+>_U}{{}_1<R-|R+>_1} \frac{{}_U<R+|R+>_1}{|{}_U<R+|R+>_1|} \tag{7.9}$$

Under a gauge transformation,

$$U_\mu^g(x) = g^\dagger(x) U_\mu(x) g(x+\hat{\mu}), \tag{7.10}$$

the Hamiltonians transform as

$$\mathbf{H}^\pm(x\alpha i, y\beta j; U^g) = G^\dagger(x\alpha i; x'\alpha' i') \mathbf{H}^\pm(x'\alpha' i', y'\beta' j'; U) G(y'\beta' j', y\beta j) \tag{7.11}$$

where

$$G(x\alpha i; y\beta j) = g^{ij}(x) \delta_{xy} \delta_{\alpha\beta}. \tag{7.12}$$

Therefore the single particle eigenfunctions at $U$ and $U^g$ are related by

$$\psi(U^g) = G^\dagger \psi(U) e^{i\phi(U,g)} \tag{7.13}$$

and the induced relations on the many body states are

$$|R\pm>_{U^g} = \mathcal{G}^\dagger |R\pm>_U e^{i\Phi_\pm(U,g)} \tag{7.14}$$

$\mathcal{G}$ is defined by $G$ and is the same for both $\pm$ states. When (7.14) is used in (7.9) the unknown phases $\Phi_\pm(U,g)$ drop out. In the continuum the anomalies completely determine the dependence of the action on coordinates along gauge orbits. This dependence is given by the gauged Wess–Zumino Lagrangian [13]. If our regularization is correct it must also provide a lattice version of this continuum Lagrangian. The lattice Wess–Zumino action is a functional of the link fields $U_\mu(x)$ and the group valued variables $g(x)$. (When $g$ is infinitesimal the dependence on $U_\mu$ and $g$ is restricted by Zumino's consistency conditions.) Define

$$\mathcal{W}_\pm(U,g) \equiv \frac{{}_U<R\pm|\mathcal{G}|R\pm>_1}{{}_U<R\pm|R\pm>_1} \tag{7.15}$$

Then the lattice form of the exponent of the gauged Wess–Zumino model is:

$$e^{\Gamma^R(U^g) - \Gamma^R(U)} = \frac{\mathcal{W}_-^*(U,g) \mathcal{W}_+(U,g)}{|\mathcal{W}_-^*(U,g) \mathcal{W}_+(U,g)|} \tag{7.16}$$

To obtain the anomaly one expands the logarithm of the above pure phase around $g \equiv 1$ to linear order. There are two additive contributions, one from the negative side and the other from the positive side. As we shall see later, in the continuum limit only the contribution from the positive side survives.

The locality of the $\mathbf{H}^\pm$ comes at some price: If one sets up perturbation theory from a path integral representation of the overlap one has to deal with a continuous time parameter and the action has a theta function discontinuity. This induces some technical complications that are likely soluble*. In our previous work we avoided the ambiguities related to the theta function discontinuity by discretizing the fifth coordinate also (we use language appropriate to 4-D, but everything holds in 2-D also). In this framework one deals with two transfer matrices roughly equivalent to the exponents of our Hamiltonians. The structure of the transfer matrices is quite similar to the structure of the Hamiltonians, but strict locality is lost. To be sure, the expressions are still local in the sense that the couplings between far away sites are suppressed exponentially.

For completeness we recall here the relevant formulae for the transfer matrices obtained in [4]: The many body problems are still written as in (5.2) but the new $\mathbf{H}^\pm$'s are given implicitly, in terms of their exponents. There is no loss in doing this because the single particle problems are solved by diagonalizing the exponents now, the single difference being that the boundary between filled and empty states in the vacuum is at eigenvalue equal to one. Using our previous expressions for $\mathbf{C}$ and $\mathbf{B}^\pm$ we now have:

$$T^\pm \equiv \exp(\mathbf{H}^\pm) = \begin{pmatrix} \frac{1}{1+\mathbf{B}^\pm} & \frac{1}{1+\mathbf{B}^\pm}\mathbf{C} \\ \mathbf{C}^\dagger \frac{1}{1+\mathbf{B}^\pm} & \mathbf{C}^\dagger \frac{1}{1+\mathbf{B}^\pm}\mathbf{C} + 1 + \mathbf{B}^\pm \end{pmatrix} \tag{7.17}$$

By (7.7) is is clear that the matrices $1 + \mathbf{B}^\pm$ are positive definite for any $0 < m < 1$. Therefore the inverses can be expanded around unity and one sees that the expressions, while not local in the strict sense are satisfactorily local, as mentioned above.** Numerically, this promises fast convergence of the conjugate gradient algorithm used to work out the inverses, thus the penalty for not having strict locality isn't that harsh as one might fear at first sight. Still, it is easier to work with (7.1).

## 8. Topological charge on the lattice.

The topological charge to be associated with a set of link variables is defined so that the fermionic reaction is the same as in the continuum: It is given by the difference in the number of negative energy levels of $\mathbf{H}^+$ less the number of negative energy levels of $\mathbf{H}^-$. Clearly this is a gauge invariant definition. A direct consequence of the connectedness of the space of $U_\mu$'s is the existence of backgrounds for which at least one of the Hamiltonians has a zero eigenvalue. In the context of "lattice topology" such gauge field configurations are usually referred to as "singular".

If the number of sites is $V$ the dimension of $\mathbf{H}^\pm$ is $2^{\frac{d}{2}}V\mathrm{N} \times 2^{\frac{d}{2}}V\mathrm{N}$ and, for $U_\mu \equiv \mathbf{1}_\mathrm{N}$ there are exactly $2^{\frac{d}{2}}V\mathrm{N}/2$ negative (positive) eigenvalues for each sign. As a matter of fact, $\mathbf{H}^-$ has $2^{\frac{d}{2}}V\mathrm{N}/2$ negative (positive) eigenvalues for *any* gauge background. To prove this, we note that it is true for the free case and next show that $\mathbf{H}^-$ can never have a zero eigenvalue: Let us assume that such an eigenvalue existed and that the corresponding eigenvector is $\begin{pmatrix} u \\ v \end{pmatrix}$:

$$\begin{aligned} \mathbf{B}^- u + \mathbf{C} v &= 0 \\ \mathbf{C}^\dagger u - \mathbf{B}^- v &= 0 \end{aligned} \tag{8.1}$$

---

\* B. Blok, private communication [14]
\*\* In our previous work [4–6] we used the notation $\mathbf{B}^\pm$ for the quantities $1 + \mathbf{B}^\pm$ in (7.8).

Multiply the first equation by $u^\dagger$ from the left and the second by $v^\dagger$ from the left. Complex conjugate the second equation, use the fact that $\mathbf{B}^-$ is hermitian and subtract the two equations. You get $v^\dagger \mathbf{B}^- v + u^\dagger \mathbf{B}^- u = 0$ which contradicts the rather obvious positivity of $\mathbf{B}^-$. Similarly one can show that for any gauge field $U_\mu$ all eigenvalues $E$ of $\mathbf{H}^-$ satisfy $u^\dagger(\mathbf{B}^- - E)u + v^\dagger(\mathbf{B}^- + E)v = 0$, implying $|E| \geq m$, so $\mathbf{H}^-$ always has a gap around zero. Clearly the proof breaks down for $\mathbf{H}^+$ because $\mathbf{B}^+$ isn't necessarily positive.

Consider now $\mathbf{B}^-$ say but let $m = \mu$ be a varying parameter with the gauge background fixed. Denote the corresponding $\mathbf{H}^-$ by $\mathbf{H}(\mu)$. As long as $\mu$ is positive no level has crossed zero; when $\mu$ becomes negative the gap can close. When $\mu = -m$ $\mathbf{B}^-$ becomes $\mathbf{B}^+$ and $\mathbf{H}(\mu)$ becomes $\mathbf{H}^+$. The imbalance of negative energy eigenvalues of $\mathbf{H}^+$ is equal to the number of odd eigenvalue crossings occuring for $\mathbf{H}(\mu)$ from positive to negative less the number of odd eigenvalue crossings from negative to positive when $\mu$ is monotonically varied from $m$ to $-m$. A crossing is "odd" if the first nonvanishing derivative with respect to $\mu$ of the respective eigenvalue at the point where the eigenvalue vanishes is of odd order.

Generically, crossings are odd with the first derivative nonzero. So, in practice, in order to determine what the topological charge of a given lattice gauge configuration all we need to do is to follow the eigenvalue flow of

$$\mathbf{H}(\mu) = \begin{pmatrix} B + \mu & \mathbf{C} \\ \mathbf{C}^\dagger & -B - \mu \end{pmatrix} \tag{8.2}$$

through zero. By looking at the lowest eigenstate of $\mathbf{H}^2(\mu)$ and checking the expectation value of $\mathbf{H}\mu)$ in the respective eigenvector a relatively efficient numerical method can be devised to measure the topological charge of a given gauge field configuration.

Linear crossings of zero for eigenvalues of $\mathbf{H}(\mu)$ occur at values $\mu$ where

$$\mathbf{H}(\mu)\psi = 0; \qquad \psi^\dagger \gamma_5 \psi \neq 0 \tag{8.3}$$

Rather than look at the hermitian eigenvalue problem (8.3) one can make the $\mu$ dependence explicit by considering the equivalent eigenvalue problem

$$\gamma_5 \mathbf{H}(\mu)\psi = 0 \quad \Rightarrow \quad \begin{pmatrix} B & \mathbf{C} \\ -\mathbf{C}^\dagger & B \end{pmatrix} \psi \equiv \mathbf{D}\psi = -\mu\psi \qquad \psi^\dagger \gamma_5 \psi \neq 0 \tag{8.4}$$

We shall assume that all crossings are generic which means that the first derivative of the crossing eigenvalue is not zero at the crossing point. Then the last equation means that the topological charge can be obtained by looking at the usual lattice Dirac operator with a Wilson term (its coefficient, usually denoted by $r$, is unity – a preferred value for other reasons) and adding or subtracting unity for each *real* eigenvalue in the interval $(-m, m)$ depending on the sign of $\psi^\dagger \gamma_5 \psi$. The lattice Dirac operator $\mathbf{D}$ is hermitian under the indefinite inner product $(\psi, \phi) = \psi^\dagger \gamma_5 \phi$. Therefore, $\mathbf{D}$ can have non-real eigenvalues $\lambda$ only if $(\psi_\lambda, \psi_\lambda) = 0$. Thus, in the generic case, the original definition in terms of the imbalance in the number of positive versus negative eigenvalues of $\mathbf{H}^+$, is equivalent to:

$$n_{\text{top}} = \lim_{\epsilon \to 0} \sum_{\{i;\ |\Re \lambda_i| < m\}} \frac{\psi_i^\dagger \gamma_5 \psi_i}{|\psi_i^\dagger \gamma_5 \psi_i| + \epsilon}; \qquad \mathbf{D} \psi_i = \lambda_i \psi_i \qquad (8.5)$$

$\mathbf{D}$ can be thought of as a lattice version of an approximately massless and antihermitian continuum Dirac operator. The above equations can then be viewed as a regularized version of the following continuum relation:*

$$n_{\text{top}} = \lim_{\epsilon \to 0} \sum_i \frac{\int f_i^\dagger \gamma_5 f_i}{|\int f_i^\dagger \gamma_5 f_i| + \epsilon}; \qquad \gamma \cdot (\partial + iA) f_i = \lambda_i f_i \qquad (8.6)$$

This is another indication that the lattice definition of the topological charge obtained from the overlap is reasonable.**

Clearly the definition contains some arbitrariness associated with the choice of $m$. In the continuum, this arbitrariness should have no effect if one measures the topological susceptibility for example. Most crossings reflecting real continuum instanton–like fields should take place for small values of the parameter $\mu$. It may appear that we have an additional example of an ambiguity in our definition because we could use the transfer matrices instead of the Hamiltonians. However, a closer look reveals that the two definitions are identical if the parameter $m$ here is identified with the parameter $m$ there. To see this we need to go into some detail:

First we show that, just like $\mathbf{H}^-$, $T^-$ has an impenetrable gap around unity. Let us represent the eigenvector of $T^-$ corresponding to the eigenvalue $\lambda$ by $\begin{pmatrix} u \\ -v \end{pmatrix}$. The eigenvalue equations

$$\frac{1}{1+\mathbf{B}^-} u - \frac{1}{1+\mathbf{B}^-} \mathbf{C} v = \lambda u$$
$$\mathbf{C}^\dagger \frac{1}{1+\mathbf{B}^-} u - \mathbf{C}^\dagger \frac{1}{1+\mathbf{B}^-} \mathbf{C} v - (1+\mathbf{B}^-) v = -\lambda v \qquad (8.7)$$

are equivalent to:

$$\mathbf{C}^\dagger u = -v + \frac{1+\mathbf{B}^-}{\lambda} v$$
$$\mathbf{C} v = u - \lambda (1+\mathbf{B}^-) u \qquad (8.8)$$

---

\* In (8.3) (8.4) and (8.5) the notation $\psi^\dagger \gamma_5 \psi$ implies also a sum over the sites $x$, while in (8.6), following continuum conventions, the integral over $x$ is made explicit.

\*\* Another fermionic definition of the topological charge has been given in [15]. These authors elected to focus on the continuum formula $n_{\text{top}} = m \int tr[\gamma_5 \frac{1}{\gamma \cdot (\partial + iA) + m}]$ replacing it by $n_{\text{top}} = \kappa_p m \int \sum_i \frac{\bar{f}_i \gamma_5 f_i}{m + \lambda_i - M_c}$ where $\mathbf{D} f_i = \lambda_i f_i$, $\bar{f}_i \mathbf{D} = \lambda_i \bar{f}_i$ and $\int \bar{f}_i f_i = 1$. For the full definition of the parameters $\kappa_p$ and $M_c$ we refer to the source. For smooth backgrounds one can take $\kappa_p = 1$ and $M_c = -min_i(\Re \lambda_i)$. The parameter $m$ is to be extrapolated to zero. For such smooth backgrounds we expect good agreement with our definition.

Multiply the first equation by $v^\dagger$ from the left, the second equation by $u^\dagger$ from the left and complex conjugate it. Subtracting the so obtained equations one obtains:

$$v^\dagger \left[ \frac{1+\mathbf{B}^-}{\lambda} - 1 \right] v + u^\dagger \left[ \lambda(1+\mathbf{B}^-) - 1 \right] u = 0 \tag{8.9}$$

The above is impossible if $\lambda$ is such that $\mathbf{B}^- + 1 > max(\lambda, \frac{1}{\lambda})$. Since $\mathbf{B}^- > m$ there is a gap for

$$\frac{1}{1+m} < \lambda < 1+m \tag{8.10}$$

.

Thus, just like in the Hamiltonian case, the topological charge in the transfer matrix formulation will again be given by the number of (assumed generic) crossings of unity of eigenvalues of $T(\mu)$ as we vary $\mu$ from $-m$ to $m$. $T(\mu)$ is defined as $T^-$ with the parameter $m$ replaced by $\mu$, so $\mu$ interpolates between $T^-$ and $T^+$. We shall show now that generic crossings occur for $\mathbf{H}(\mu)$ and $T(\mu)$ at the same time and with the same sign, and thus obtain the equivalence of the two definitions for the topological charge for generic gauge configurations.

The unit eigenvalue condition for $T(\mu)$ where the eigenvector is taken as $\psi = \begin{pmatrix} u \\ -v \end{pmatrix}$ is

$$\begin{aligned} \mathbf{C}^\dagger u - \mathbf{B}^- v &= 0 \\ \mathbf{C} v + \mathbf{B}^- u &= 0 \end{aligned} \tag{8.11}$$

This equation is identical to (8.1). To see the relation between the signs we need to compute the $\mu$ derivative of the crossing eigenvalue at the crossing point. Using the eigenvalue equation it is easy to see that

$$<\psi|\frac{\partial T}{\partial \mu}|\psi> = \begin{pmatrix} u \\ -v \end{pmatrix}^\dagger \begin{pmatrix} -\frac{1}{[1+\mathbf{B}(\mu)]^2} & -\frac{1}{[1+\mathbf{B}(\mu)]^2}\mathbf{C} \\ -\mathbf{C}^\dagger \frac{1}{[1+\mathbf{B}(\mu)]^2} & 1 - \mathbf{C}^\dagger \frac{1}{[1+\mathbf{B}(\mu)]^2}\mathbf{C} \end{pmatrix} \begin{pmatrix} u \\ -v \end{pmatrix} = v^\dagger v - u^\dagger u \tag{8.12}$$

which is the desired relation. This concludes the derivation of the equivalence of the two definitions of the topological charge. The continuum interpretation of the hamiltonian definition thus holds for the transfer matrix definition also.

For $\mathbf{H}(\mu)$ it was useful to consider $\mathbf{H}^2(\mu)$ because for it the states which will potentially cross are located near the lower end of the spectrum. When working with the transfer matrix version of the overlap the role of $\mathbf{H}^2(\mu)$ is played by $T(\mu) + T^{-1}(\mu)$. It is easy to find the inverse of the transfer matrices in (7.8):

$$(T^\pm)^{-1} = \exp(-\mathbf{H}^\pm) = \begin{pmatrix} \mathbf{C}\frac{1}{1+\mathbf{B}^\pm}\mathbf{C}^\dagger + 1 + \mathbf{B}^\pm & -\mathbf{C}\frac{1}{1+\mathbf{B}^\pm} \\ -\frac{1}{1+\mathbf{B}^\pm}\mathbf{C}^\dagger & \frac{1}{1+\mathbf{B}^\pm} \end{pmatrix} \tag{8.13}$$

We see that the inverses are just as local as the transfer matrices were; this comes as no surprise since the inverses generate propagation in the opposite direction in the auxiliary "time" space.

# 9. Construction of anomaly free theories

The ground work in the previous sections makes the procedure for constructing vector and chiral theories quite straightforward: Sections 2 and 4 have established the formalism for treating a single right (left) handed fermion coupled to a gauge field on the lattice. The method to compute fermionic expectation values was described in section 5. The lattice regularized theory was described in section 7. This forms the basis for constructing vector theories and chiral theories. In this section we construct such theories and show that several fundamental properties of these theories are reproduced.

We start by constructing a single flavor massless vector theory. We decompose the Dirac fermions in the single gauge multiplet into their left handed and right handed components. To each chiral part ($R$ or $L$) we associate a set of creation and annihilation operators $[a^R_{x\alpha i}]^\dagger$; $[a^L_{x\alpha i}]^\dagger$; $a^R_{x\alpha i}$; $a^L_{x\alpha i}$. Both sets separately obey the canonical commutation relations given in (5.3) and the two sets anticommute with each other. From each set of creation and annihilation operators we construct a many body Hamiltonian as given in (5.2). The single particle Hamiltonians appearing in both of these many body Hamiltonians are the same and are given by (7.1), (7.6) and (7.7). The single particle Hamiltonians are isomorphic because both the right and left handed fermions are in the same representation of the gauge group, carry the same charge and are coupled to the same gauge field. The two ground states that constitute the overlap are denoted* by $|V\pm>_U$. $|V\pm>_U$ are direct products of $|R\pm>_U$ and $|L\pm>_U$ where $|R\pm>_U$ ($|L\pm>_U$) are the specific many body states associated with the Hamiltonian of the right (left) handed fermion as described in the beginning of section 5. The overlap is a product of the two overlaps, one for the right handed and the other for the left handed fermion and are given by (5.4).

To compute expectation values of fermions we first decompose them into left and right components. In the massless case the left and right contribution factorize and the individual expectation values are then computed by following the procedure described in section 5. The generating functional is given by

$$Z^V(\bar\eta^R, \eta^R, \bar\eta^L, \eta^L; U) = {}_U\!<V-|\exp\left[\bar\eta^R a^R + \eta^R [a^R]^\dagger + \bar\eta^L a^L + \eta^L [a^L]^\dagger\right]|V+>_U \qquad (9.1)$$

In the above equation, we have suppressed the sum over $x\alpha i$ explicitly shown in (5.10). In the zero topological sector the result is of the form

$$Z^V(\bar\eta^R, \eta^R, \bar\eta^L, \eta^L; U) = {}_U\!<V-|V+>_U \exp\left[-\bar\eta^R G^R(U)\eta^R - \bar\eta^L G^L(U)\eta^L\right] \qquad (9.2)$$

where $G^L(U) = -\left[G^R(U)\right]^\dagger$ and the explicit expression for $G^R(U)$ is given in (5.22). It should be kept in mind here that the sources are for Dirac fermions and that only half the modes of $G^R(U)$ propagate and these correspond to the physical right handed chiral fermion.

The massive vector theory is thought of as a perturbation of the massless case. The mass term that couples the right handed fermion to the left handed fermion is treated in a manner similar to

---

* The gauge fields are denoted by the link variables $U$ to show that the definition is on the lattice.

the fermionic sources. The generating functional for a massive theory is written as

$$Z^V(\bar{\eta}^R, \eta^R, \bar{\eta}^L, \eta^L; U, m_f) = {}_U\!< V-|\exp\left[m_f[a^R]^\dagger a^L\right]\exp\left[\bar{\eta}^R a^R + \eta^R[a^R]^\dagger \right.$$
$$\left. + \bar{\eta}^L a^L + \eta^L[a^L]^\dagger\right]\exp\left[m_f[a^L]^\dagger a^R\right]|V+>_U \quad (9.3)$$

where $m_f$ is the fermion mass (here assumed to be positive for simplicity). The mass term can be rewritten in terms of fermionic source terms:

$$Z^V(\bar{\eta}^R, \eta^R, \bar{\eta}^L, \eta^L; U, m_f) = \int [d\bar{\xi}^R][d\xi^R][d\bar{\xi}^L][d\xi^L] {}_U\!< V-|\exp\left[\sqrt{m_f}(\xi^R[a^R]^\dagger + \bar{\xi}^L a^L)\right]$$
$$\exp\left[\bar{\eta}^R a^R + \eta^R[a^R]^\dagger + \bar{\eta}^L a^L + \eta^L[a^L]^\dagger\right] \quad (9.4)$$
$$\exp\left[\sqrt{m_f}(\xi^L[a^L]^\dagger + \bar{\xi}^R a^R)\right]|V+>_U \exp[\xi^R\bar{\xi}^L + \xi^L\bar{\xi}^R]$$

In the zero topological sector (9.2) can be used in (9.4) and the integration in (9.4) can be carried out. This results in the following expression for the generating functional:

$$Z^V(\bar{\eta}^R, \eta^R, \bar{\eta}^L, \eta^L; U, m_f) = {}_U\!< V-|V+>_U \det K(U, m_f) \exp\left[-\bar{\eta}^V G^V(U, m_f)\eta^V\right] \quad (9.5)$$

where

$$\eta^V = \begin{pmatrix} \eta^R \\ \eta^L \end{pmatrix}; \quad \bar{\eta}^V = (\,\bar{\eta}^L \quad \bar{\eta}^R\,) \quad (9.6)$$

The matrix $K(U, m_f)$ in (9.5) is given by

$$K(U, m_f) = \begin{pmatrix} -1 & m_f \tilde{G}^L(U) \\ -m_f \tilde{G}^R(U) & -1 \end{pmatrix} \quad (9.7)$$

where

$$\tilde{G}^L(U) = \frac{1}{2} - G^L(U); \quad \tilde{G}^R(U) = \frac{1}{2} + G^R(U); \quad \left[\tilde{G}^L(U)\right]^\dagger = \tilde{G}^R(U) \quad (9.8)$$

From (9.5), (9.7) and (9.8), the expression for the vector determinant is

$${}_U\!< V-|V+>_U \det K(U, m_f) = {}_U\!< R-|R+>_U {}_U\!< L-|L+>_U$$
$$\det\left[1 + m_f^2 \tilde{G}^L(U)\left[\tilde{G}^L(U)\right]^\dagger\right] \quad (9.9)$$

Since ${}_U\!< R-|R+>_U = {}_U\!< L+|L->_U$ (c.f. Lemma 4.1) it follows that the above formula for the vector determinant is real and positive for all gauge fields. This is true at the level of a single flavor and holds whether the fermion is massless or massive. Since the determinant is real it is also gauge invariant because the only gauge breaking in the right and left overlaps were in their phases. (9.9) should be compared with the formal expression for the vector determinant

$$\det \mathbf{D} = \det \mathbf{C} \det \mathbf{C}^\dagger \det\left[1 + m_f^2 \mathbf{C}^{-1}[\mathbf{C}^\dagger]^{-1}\right] \quad (9.10)$$

The matrix $G^V(U, m_f)$ in (9.5) is given by

$$G^V(U, m_f) = \begin{pmatrix} 0 & G^L(U) \\ -[G^L(U)]^\dagger & 0 \end{pmatrix}$$
$$+ \begin{pmatrix} m_f \frac{\tilde{G}^L(U)[\tilde{G}^L(U)]^\dagger}{1+m_f^2 \tilde{G}^L(U)[\tilde{G}^L(U)]^\dagger} & m_f^2 \frac{\tilde{G}^L(U)[\tilde{G}^L(U)]^\dagger}{1+m_f^2 \tilde{G}^L(U)[\tilde{G}^L(U)]^\dagger} \tilde{G}^L(U) \\ -m_f^2 [\tilde{G}^L(U)]^\dagger \frac{\tilde{G}^L(U)[\tilde{G}^L(U)]^\dagger}{1+m_f^2 \tilde{G}^L(U)[\tilde{G}^L(U)]^\dagger} & m_f \frac{[\tilde{G}^L(U)]^\dagger \tilde{G}^L(U)}{1+m_f^2 [\tilde{G}^L(U)]^\dagger \tilde{G}^L(U)} \end{pmatrix} \quad (9.11)$$

The first term is the propagator for massless fermions and the second term includes all the effects due to the mass. In a manner similar to the relation between (9.9) and (9.10), the Dirac propagator (9.11) also has a formal analogue expression. The diagonal terms in (9.11) lead to a version of the well known [16] spectral criterion for $\bar{\Psi}\Psi$ to have non–zero expectation value when $m_f \to 0$: Let the eigenvalue density of $\tilde{G}^L(U)[\tilde{G}^L(U)]^\dagger$, after averaging over gauge fields (including the fermion determinant), be $\rho(\lambda)$ where the eigenvalues $\lambda$ are non–negative. $\bar{\Psi}\Psi$ maintains a nonzero expectation value as $m_f \to 0$ if for large $\lambda$ we have $\rho(\lambda) \sim \frac{1}{\lambda^{5/2}}$.

Turning now to an arbitrary number of flavors, $N_f$, amounts to a trivial extension. We can think about the physical Dirac fermions as basic objects whose propagator is $G^V(U, m_f)$.* The extension of relation (9.8) (equivalently (5.13)) to the Dirac propagator is

$$\left[\gamma_5 G^V(U, m_f) \gamma_5\right]^\dagger = G^V(U, m_f) \tag{9.12}$$

The above equation is satisfied by (9.11) because the diagonal terms are hermitian and the off diagonal terms are related by one being the negative hermitian conjugate of the other. Having this equation hold in a regularized framework independently of whether the theory is massless or not is useful when more flavors are considered because positivity of the regularized determinant (c.f. (9.9)) and relation (9.12) are the only ingredients needed to prove the well known mass inequalities between mesons and baryons. The original proofs in the path integral formalism go through directly [17]. In the lattice case, the improvement relative to Weingarten's work is that a restriction on the number of flavors is no longer needed.**

The massless multiflavor case has at the formal path integral level a $U(N_f) \times U(N_f)$ symmetry, the two factors acting on the left and right components independently. It is important to see how these symmetries are realized after regularization by our method. This can be easily done because the symmetry is there for each gauge field, even before carrying out the full functional integral. The symmetry can be seen in the formulae for the fermion generating functional (5.9) when properly generalized to many flavors. For each one of the symmetry transformations an associated canonical transformation on the creation/annihilation operators $a, a^\dagger$ exists and is independent of the background gauge fields. If the total vacua $|V\pm>_U$ are both annihilated by the generator of the above canonical transformation the generating functional will exhibit the symmetry by its functional dependence on the sources $\bar\eta, \eta$. For the $SU(N_f) \times SU(N_f)$ part of the symmetry group it is easy to ascertain that the symmetry indeed holds. The unimodularity of the matrices in flavor space plays a role in deriving the invariance of the two vacua. Similarly, one can easily convince oneself that the vectorial $U(1)_V$ is also preserved, although the two vacua now rotate by a phase. Indeed, the phase coming from $|V->_U$ exactly cancels the phase coming from $|V+>_U$ and overall invariance holds. This is true even if there are individual imbalances in the level of filling due to nontrivial gauge backgrounds. The effects of the imbalances in the righthanded and lefthanded sectors cancel each other. For an axial $U(1)_A$ transformation the two imbalances add to each other and therefore $U(1)_A$ is not a symmetry if the background has topological charge. This is just what we would have liked to see happen. Let us take a background carrying topological charge equal to unity. The axial transformation formally known as $\Psi \to e^{i\alpha\gamma_5}\Psi$ translates into $a^{L,I} \to e^{-i\alpha}a^{L,I}$ and $a^{R,I} \to e^{i\alpha}a^{R,I}$ for $I = 1, 2, \cdots N_f$. The vacua transform as $|V+>_U \to e^{2i\alpha N_f}|V+>_U$ and $|V->_U \to |V->_U$. For a fermion expectation value to be nonzero it has to transform in the opposite way. That is what a 't Hooft vertex does.

---

\* One may easily replace the common parameter $m_f$ by a more general mass matrix connecting each physical righthanded gauge multiplet to some identical multiplet of lefthanded nature. This matrix can be brought into diagonal form with the help of some canonical transformations on the creation/annihilation operators.

\*\* Traditional lattice regularizations of vector theories involve a parameter $\kappa$ that has to be fine tuned to a specific value $\kappa_c$ (dependent on the lattice gauge coupling) in order to have finite or zero mass quarks in the continuum: While the numerical values of $\kappa_c$ are not known exactly, we know that when the continuum limit is approached and the lattice gauge coupling goes to zero, $\kappa_c$ is such that the positivity of the fermionic determinant per individual flavor is lost and one needs an even number of flavors to preserve the needed positivity of the gauge integration measure.

If $N_f = 1$ the 't Hooft vertex is, in terms of physical fields, given by $\Psi_R^\dagger \Psi_L$, which together with its hermitian conjugate becomes the mass operator. If there are $N_f$ flavors, the 't Hooft operator is made up of $2N_f$ fermi fields. When the 't Hooft operator aquires an expectation valus this indicates a breaking of the axial $U(1)_A$, down to a discrete subgroup. The breaking occurs even when the volume is finite and is clearly distinct from a spontaneous breaking. The global symmetry $U_R(N_f) \times U_L(N_f)$ formally present in the massless case gets explicitly broken down to $SU_R(N_f) \times SU_L(N_f) \times U_V(1) \times Z_2$.

When a mass term is turned on terms with the quantum numbers needed to give a nonzero matrix element between the two vacua for any gauge background will be generated by expanding in the mass. Thus, the vectorial determinant will not vanish for any topological charge, as expected.

We have shown that our regularization preserves the topological effects one is familiar with from the continuum as far as quantum numbers go and we shall see more of this below. However, we ought to remember that the continuum reasoning is credible because one can imagine setups where one can carry out semiclassical computations with confidence and indeed show that quantities that could be non–zero indeed are. (For example, we may enclose the system in a small Euclidean volume, so that the effective couplings never get strong.) The magnitude of some of the above topological effects can then be computed using instanton techniques. If our regularization is indeed right we have to convince ourselves that one could carry out the instanton computations directly on the lattice and that the results would be the ones we know from the continuum. The main point to check is whether the overlap indeed behaves in a way similar to the continuum determinant when the background is a "dilute gas of instantons and anti–instantons" [18]. The dilute gas approximation was important in at least two respects: It showed explicitly how clustering is restored after integration over all topological sectors and by this reassured us that indeed a unitary local field theory can incorporate the somewhat unfamiliar situation of having to carry out a functional integral over a disconnected set of fields of the same type. The dilute gas approximation also indicated how a theory with no massless particles, like QCD for example, while having no sensitivity to the boundary conditions still manages to exhibit effects reflecting topological properties of the gauge field space. The dilute gas approximation achieved the above by reintroducing extensivity in the thermodynamical sense via a factorization of the chiral determinant into approximately local factors when the gauge background consists of widely separated instantons and anti–instantons. Similar issues arise when one considers below the insertion of 't Hooft vertices and their exponentiation. In the following sections we shall try to verify that indeed the overlap behaves as a determinant in the sense that it factorizes in dilute backgrounds. We shall also try to see how effects depending on the explicit form of the continuum fermionic zero modes get reproduced by the overlap.

The definition of the phases of the states given in section 5.2 for all topological sectors also applies to the vector theory with $N_f$ flavors. $\mathcal{F}$ used to define the phase is the 't Hooft operator that breaks the axial $U_A(1)$ symmetry. In the definition $\mathcal{F}$ can be replaced by $e^{i\theta}\mathcal{F}$ with $\theta$ being some real parameter. This just replaces the reality condition of the Wigner-Brillouin in non-zero sectors with a condition where the phase is fixed at $n\theta$ in the sector with charge $n$. The only difference would be that different topological sectors will now be weighted by $e^{in\theta}$ which is equivalent, in four dimensions, to the addition of a $\frac{i\theta}{8\pi^2}\int tr(\tilde{F}F)$ term to the $\frac{1}{4g^2}\int tr(FF)$ term of the gauge action. This is the usual $\theta$ term that one would like to introduce in a vector theory. The phase factor in front of $\mathcal{F}$ can be absorbed by redefining the fermion operators by $a^{R,I} \to e^{-i\frac{x_I \theta}{2N_f}} a^{R,I}$ and $a^{L,I} \to e^{i\frac{x_I \theta}{2N_f}} a^{L,I}$ with $\sum_I x_I = N_f$. The many body Hamiltonians are invariant under this redefinition and only the mass terms get affected. The ability to move the $\theta$–parameter dependence from the $\int \tilde{F}F$ term to the mass matrix is familiar from the continuum and we see that it holds exactly in our regularization. In particular, by appropriately choosing the constants $x_I$ we see that the vacuum energy will be $\theta$ independent if one of the Dirac fermions is massless.

The construction of a chiral theory follows in the same manner. Each multiplet of a given handedness is represented by one of the building blocks we have constructed. Since the right and left handed fermions are not paired in a chiral theory the resulting overlap need not be real whether the theory under consideration is anomaly free or not. As will be verified later on, each block carries the correct amount of "consistent" anomaly. The difference between an anomaly free theory and an anomalous one is that the leading operators in the imaginary gauge breaking terms cancel in the former (i.e. the coefficient of the induced gauge Wess-Zumino action vanishes in the anomaly free case). However, even in the anomaly free case exact gauge invariance does not get restored for finite lattice spacings: Higher order terms that break the gauge invariance could still be present on the lattice, but their effects are expected to disappear in the continuum limit. This point will be discussed again in a later section.

The standard model can be constructed by following the same procedure: The new ingredient are the Yukawa interactions and they can be added in a manner that is similar to the addition of the mass term in the vector case. There is one major difference beyond the obvious space time dependence and gauge group representation: The mass term in the vectorial case would eliminate all the zeros of the fermionic determinant in backgrounds carrying non–trivial topological charge. However, in the standard model the Yukawa term couples fermions transforming as doublets under $SU(2)$ to fermions transforming as singlets under $SU(2)$, so only one of them sees the topological charge of the $SU(2)$ gauge field. Therefore, no term in the expansion in the Yukawa interaction can undo the vanishing of the overlap. One still needs to insert a 't Hooft vertex to get a nonzero matrix element. This feature is important in understanding baryon decay in the standard model. There have been several recent papers proving the resilience of the topology induced zeros when there are Yukawa interactions present [19].

If we consider $SU(N)$ gauge theory interacting with a single adjoint multiplet of lefthanded Weyl fermions (gluinos) the continuum limit should come out to be supersymmteric. Unlike in previous approaches no fine tuning is needed because the fermions are strictly massless (see [20]). However, we have not yet detected any remainders of the continuum supersymmetry on the lattice. Still, we would hope that this observation is an indication that progress on the problem of constructing lattice models that have some remnants of supersymmtery could be made using our methods as an inspiration.

## 10. Abelian Gauge models in Two Dimensions

Massless fermions in two dimensions differ in several related respects from massless fermions in four dimensions: Anomalies can be cancelled only between left and right movers. Charge conjugation does not turn a left mover into a right mover. Anomalous gauge theories are renormalizable. Anomaly free theories can be regularized to all orders in perturbation theory gauge invariantly by adding a finite number of Pauli–Villars Dirac fields of wrong statistics.* Abelian models are soluble since the determinants are exactly known, and moreover, turn out to be quadratic in the gauge fields rendering all the models of massless fermions interacting with abelian gauge fields gaussian in the trivial topological sector. Like in four dimensions, some information on the slightly more complicated higher sectors can be obtained from the zero topological sector by assuming clustering. Even truly chiral abelian gauge models (with nontrivial anomaly cancellations – not of the vector type) have no parity violating terms in the chiral determinant, i.e., in Euclidean space the action has no imaginary part.

In spite of the above we believe that tests of the overlap regularization in two dimensional chiral abelian models are of significance. The objectives of the tests fall into two classes: One is to establish that the lattice overlap correctly reproduces the chiral determinant (both its imaginary part and its real part) and the salient features of the fermion two–point function in backgrounds that carry zero total topological charge. This will be the subject of subsections 10.2 and 10.4. The second is to show that the lattice overlap correctly reproduces the continuum features of fermions in topologically non–trivial backgrounds. This has to be done at the "classical" level, i.e. showing that the correct effects of fermionic zero modes with the right quantum numbers are reproduced (in subsection 10.3), and also at the "quantum" level. By "quantum" we mean the needed features that enabled 't Hooft to replace the fermionic effects caused by instantons by an effective local vertex. To very good accuracy, in four dimensional spontaneously broken gauge theories, if 't Hooft's vertex is included instantons can be ignored, and the theory is perfectly "normal", obeying clustering. The replacement works not only because one has the right number of zero modes, but also because of the form of their asymptotic decay at large Euclidean distances from the instanton core. Thus, it is important to show, at least in some simple cases, that some general properties of the shape of the continuum zero modes are preserved by our lattice regularization. This will be done in subsections 10.4 – 10.6. In subsection 10.7 we shall show that non–trivial global topology, and the associated zero modes, do not spoil the perturbative anomaly equations.

---

\* We thank S. Yankielowicz for pointing this out to us [21].

The simple tests that we have carried out for specific backgrounds strongly indicate that the complete dynamical regularized quantum model will have the correct continuum limit. We know of no other regularization method that has gone as far. We believe that any credible method for latticizing chiral fermions with gauge interactions is required to have an explicit mechanism showing how continuum instanton effects are reproduced. The paragraph at the end of section 6. shows by example how even seemingly perfect perturbative behavior is still an insufficient test. At the level of quantum numbers this requirement has been emphasized in [22].

The tests we shall present in detail below will all be for the Hamiltonian, local version of the lattice overlap. Similar tests have been carried out in some cases in the transfer matrix formalism and were equally successful. In particular, results of tests of the anomaly and eigenvalue flows dealt with in subsections 10.2 and 10.3 below have been previously reported in [5] and [6]. We shall always work on tori and use antiperiodic boundary conditions for the fermions. These conditions will be implemented by adding the appropriate $U(1)$ gauge factors to the link variables. These phases will never be shown explicitly in our equations; whenever a link configuration is constructed, the fields displayed in the equations differ from the ones in our programs by the above phase factors.

*10.1 Review of Chiral Schwinger Model [23], [21]*

The partition function of the Chiral Schwinger Model is formally given by:

$$\mathcal{Z} = \int dA_\mu d\bar{\psi} d\psi e^{-\int d^2 x \left[ \frac{1}{4g^2} F_{\mu\nu}^2 - \sum_{r=1}^{R} \bar{\psi}_r^R \sigma_\mu^* (\partial_\mu + iq_r^R A_\mu) \psi_r^R - \sum_{l=1}^{L} \bar{\psi}_l^L \sigma_\mu (\partial_\mu + iq_l^L A_\mu) \psi_l^L \right]}$$
$$\equiv \int dA_\mu d\bar{\psi} d\psi e^{-S[\bar{\psi},\psi,A_\mu]} \tag{10.1}$$

$\sigma_1 = 1$ and $\sigma_2 = i$ as defined in (2.6). We have included several single component gauge invariant fermions of both chiralities. We are tacitly assuming that we are working on a torus and that the infinite volume is to be taken whenever possible. The parameter $g$ has dimensions of mass and so does the vector potential $A_\mu$. The field $F_{\mu\nu} = \partial_\mu A_\nu - \partial_\nu A_\mu \equiv \epsilon_{\mu\nu} E$ has dimensions of mass squared and the topological invariant $\frac{1}{2\pi} \int d^2 x E$ takes integral values. The fermion charges are integers also. The path integral over $A_\mu$ includes a discrete sum over all the sectors. In the following we shall focus on expectation values of operators to which only the zero topological sector contributes. This simplifies the evaluation of the chiral determinant.

Gauge fields of zero topological charge can be assumed to have a field strength that vanishes at infinity. The gauge potential can be decomposed into two real dimensionless scalar fields that also vanish at infinity and are defined by:

$$A_\mu = \partial_\mu \chi + \epsilon_{\mu\nu} \partial_\nu \phi \tag{10.2}$$

The integral over $A_\mu$ is replaced by an integral over $\chi$ and $\phi$. $\chi$ should be thought of as an angular variables and if the theory were gauge invariant, the integral over $\chi$ could be dropped for gauge invariant observables. There is no field dependence in the Jacobian associated with the $A \to \chi, \phi$ change of variables. The following change of variables is done for the fermions:

$$\psi_r^R = e^{-iq_r^R \chi} e^{q_r^R \phi} \psi_r^{R'} \quad \bar{\psi}_r^R = e^{iq_r^R \chi} e^{-q_r^R \phi} \bar{\psi}_r^{R'} \quad \psi_l^L = e^{-iq_l^L \chi} e^{-q_l^L \phi} \psi_l^{L'} \quad \bar{\psi}_r^L = e^{iq_l^L \chi} e^{q_l^L \phi} \bar{\psi}_l^{L'} \tag{10.3}$$

Under these changes of variables the action changes to:

$$S = \int d^2x \left[ \frac{1}{2g^2}(\partial^2\phi)^2 - \sum_{r=1}^{R} \bar{\psi}_r^{R'} \sigma^* \cdot \partial \psi_r^{R'} - \sum_{l=1}^{L} \bar{\psi}_l^{L'} \sigma \cdot \partial \psi_l^{L'} \right] \tag{10.4}$$

To complete the change of variables one needs the Jacobian of the transformation on the fermions. Naïvely it would be unity, but Lorentz invariant regularizations will spoil this.

The Jacobian comes from the ratios of the appropriate chiral determinants and their free counterparts. Since we are in zero topological sector we can evaluate the determinants in perturbation theory. Note that the right–handed determinants depend only on $\sigma^* \cdot A \equiv A_1 - iA_2$ while the left handed ones only on $\sigma \cdot A \equiv A_1 + iA_2$. Any one loop fermion diagram with three external gauge fields or more is convergent and therefore (we assume that a regulator has been introduced and subsequently removed) unchanged by regularization. In particular it is gauge invariant. But one cannot make a gauge invariant quantity out of $A_1 - iA_2$ alone, without using also $A_1 + iA_2$. So all orders higher than the second vanish. We cannot demand the regulator to preserve gauge invariance and make the two loop diagrams finite for arbitrary charges. But we can find regulators that ensure the gauge invariance of the absolute value of the chiral determinants. Also, a regulator can be chosen to make the absolute value of the determinant parity even and the phase parity odd as described in section 4. It goes without saying that the regulator is chosen to preserve Lorentz invariance. These requirements when combined with dimensional considerations limit the form of the logarithm of the chiral determinant to: $\Gamma_{r,l}^{R,L}[A] = \int d^2x[c_1 E \frac{1}{\partial^2} E + ic_2 E \frac{1}{\partial^2} \partial \cdot A]$ where $c_{1,2}$ are real constants. Their values are determined by the anomaly equations:

$$\partial_\mu \frac{\delta \Gamma_r^R}{\delta A_\mu(x)} = \frac{iq_r^{R^2}}{4\pi} E(x) \qquad \partial_\mu \frac{\delta \Gamma_l^L}{\delta A_\mu(x)} = -\frac{iq_l^{L^2}}{4\pi} E(x)$$
$$\epsilon_{\mu\nu} \partial_\mu \frac{\delta(\Gamma_r^R + \Gamma_r^{R*})}{\delta A_\nu(x)} = \frac{q_r^{R^2}}{2\pi} E(x) \qquad \epsilon_{\mu\nu} \partial_\mu \frac{(\delta\Gamma_l^L + \delta\Gamma_l^{L*})}{\delta A_\nu(x)} = \frac{q_l^{L^2}}{2\pi} E(x) \tag{10.5}$$

Note that the regulator induced dependences on both $A_1 \pm iA_2$ in the chiral determinants: it is easy to see that a Pauli Villars regulator, for example, will do this.

Therefore the Jacobian is given by:

$$J = \exp\left[ \frac{\sum_r q_r^{R^2} + \sum_l q_l^{L^2}}{4\pi} \int d^2x \phi \partial^2 \phi + i \frac{\sum_r q_r^{R^2} - \sum_l q_l^{L^2}}{4\pi} \int d^2x \phi \partial^2 \chi \right] \tag{10.6}$$

Our decision to choose a regulator that puts all the gauge breaking into the parity odd part of the chiral determinant (i.e. in its phase) fixes the parameter $a$ of Jackiw and Rajaraman [23] to unity for unit charge. The anomaly free case where $\sum_r q_r^{R^2} = \sum_l q_l^{L^2} \equiv Q^2$ is a gauge invariant theory and has several additional $U(1)$ symmetries, depending on the number of fermions. One of these extra global $U(1)$'s is always anomalous.

The most famous case is the original Schwinger model where $R = L = 1$ and $q^R = q^L = 1$. Here the anomalous $U(1)$ is axial charge. The symmetry is explicitly violated in nonzero topological backgrounds which give a non–zero expectation value to the axially charged, gauge invariant, Lorentz scalar operators $V \equiv \bar{\psi}^L \psi^R$ and $\bar{V} \equiv \bar{\psi}^R \psi^L$. This can be seen by computing the expectation value $< V(x)\bar{V}(0) >$ which gets contributions only from the zero topological charge. Using the transformation to free fields one establishes that when $x^2 \to \infty$ the correlator $<V(x)\bar{V}(0)>$ does not vanish, implying via clustering that $<V> \neq 0$. Since in the zero topological sector $<V> = 0$, clustering requires the inclusion of all topological sectors in the path integral. The calculation in more detail is:

$$<V(x)V(0)> = <\bar{\psi}^{L'}(x)\psi^{L'}(0)\psi^{R'}(x)\bar{\psi}^{R'}(0)><e^{2(\phi(x)-\phi(0))}> = \frac{1}{(2\pi)^2}\frac{1}{|\sigma \cdot x|^2}e^{4G(\frac{g^2x^2}{\pi})} \quad (10.7)$$

where

$$G(\frac{g^2x^2}{\pi}) = \int \frac{d^2p}{(2\pi)^2}\frac{1-e^{ip \cdot x}}{\frac{p^2}{\pi}+\frac{p^4}{g^2}} =$$
$$\frac{1}{4}\int_0^\infty \frac{dt}{t}(1-e^{-\frac{1}{t}})(1-e^{-\frac{tg^2x^2}{4\pi}}) = \quad (10.8)$$
$$\frac{1}{4}\left[\int_0^\infty \frac{dt}{t}e^{-\frac{1}{t}-\frac{tg^2x^2}{4\pi}} + 2\gamma + \log\frac{g^2x^2}{4\pi}\right]$$

The integral in the last line decreases exponentially as $x^2 \to \infty$ and the constant $\gamma$ is .577215.... From this one gets, assuming clustering, the well-known result:

$$|<V>| = \frac{1}{2\pi}\left(\frac{ge^\gamma}{2\sqrt{\pi}}\right) \quad (10.9)$$

The simplest example that contains three fermion fields will have two globally conserved $U(1)$'s and another global $U(1)$ will be anomalous. The anomalous $U(1)$ can be chosen to be the fermion number. To preserve local gauge invariance one needs a Pythagorean relation between the charges. The standard example is the 3-4-5 model: $R = 2, L = 1, q_1^R = 3, q_2^R = 4, q^L = 5, Q = 5$. An analogue of the operator $V$ of the Schwinger model is $V \equiv V_1^R V_2^R V^L$:

$$V_1^R = \psi_1^R[(\sigma \cdot \partial)\psi_1^R][(\sigma \cdot \partial)^2\psi_1^R] \quad V_2^R = \psi_2^R[(\sigma \cdot \partial)\psi_2^R][(\sigma \cdot \partial)^2\psi_2^R][(\sigma \cdot \partial)^3\psi_2^R]$$
$$V^L = \bar{\psi}^L[(\sigma^* \cdot \partial)\bar{\psi}^L][(\sigma^* \cdot \partial)^2\bar{\psi}^L][(\sigma^* \cdot \partial)^3\bar{\psi}^L][(\sigma^* \cdot \partial)^4\bar{\psi}^L] \quad (10.10)$$

These operators are gauge invariant in spite of employing ordinary rather than covariant derivatives, because the extra terms vanish by antisymmetry. Under an Euclidean "boost" $\psi^R \to e^{i\alpha/2}\psi^R$, $\bar{\psi}^R \to e^{i\alpha/2}\bar{\psi}^R$, $\psi^L \to e^{-i\alpha/2}\psi^L$, $\bar{\psi}^L \to e^{-i\alpha/2}\bar{\psi}^L$ and $\sigma \cdot \partial \to e^{i\alpha}\sigma \cdot \partial$, $\sigma^* \cdot \partial \to e^{-i\alpha}\sigma^* \cdot \partial$. Therefore $V$ is a Lorentz scalar. $\bar{V}$ is obtained from $V$ by interchanging every $\psi$ with the corresponding $\bar{\psi}$ and every $\sigma \cdot \partial$ with $\sigma^* \cdot \partial$. $V$ is gauge invariant and also chargeless under the global non-anomalous $U(1)$ which acts by

$$\psi_1^R \to e^{i\theta q_2^R}\psi_1^R, \quad \bar{\psi}_1^R \to e^{-i\theta q_2^R}\bar{\psi}_1^R$$
$$\psi_2^R \to e^{-i\theta q_1^R}\psi_2^R, \quad \bar{\psi}_2^R \to e^{i\theta q_1^R}\bar{\psi}_2^R \quad (10.11)$$
$$\psi^L \to \psi^L, \quad \bar{\psi}^L \to \bar{\psi}^L$$

When substituting the fermion fields in $V$ and $\bar{V}$ with the free fields and the $\phi$ exponentials the derivatives have to be taken as acting on the free fermion fields, to avoid vanishing by Fermi statistics. Denoting by $V'$ and by $\bar{V}'$ the expressions one obtains from $V$ and $\bar{V}$ when one substitutes the $\psi'$'s for the corresponding $\psi$'s one finds:

$$<V(x)\bar{V}(0)> = <V'(x)\bar{V}'(0)>e^{4Q^2G(\frac{g^2Q^2x^2}{\pi})} \quad (10.12)$$

Free field theory leads to:

$$< V'(x)\bar{V}'(0) > = \frac{f(|q_1^R|)f(|q_2^R|)f(|q^L|)}{(2\pi)^{|q_1^R|+|q_2^R|+|q^L|}(x^2)^{Q^2}}$$

$$f(|q|) = \det_{0 \leq i,j \leq |q|-1}[(i+j)!] = [\prod_{i=0}^{|q|-1}(i!)]^2 \tag{10.13}$$

Exactly like in the Schwinger model one determines from clustering and the large $x^2$ asymptotics of $G$ that $< V > \neq 0$. Since $V$ carries fermion number 2 this means that the model will allow for reactions in which the total fermion number changes by two units. These reactions will take place in backgrounds that carry topological charge 1.

$$|<V>| = \sqrt{\frac{f(|q_1^R|)f(|q_2^R|)f(|q^L|)}{(2\pi)^{|q_1^R|+|q_2^R|+|q^L|}}}\left(\frac{e^{\gamma}gQ}{2\sqrt{\pi}}\right)^{Q^2} \tag{10.14}$$

For completeness, let us record the effect of finite temperatures on the expectations $|<V>|$. The derivations are essentially identical and one ends up having to replace the factors $\frac{gQe^{\gamma}}{2\sqrt{\pi}}$ at zero temperature by $\frac{gQe^{\gamma}}{2\sqrt{\pi}}\exp[\int_0^{\infty}d\tau(1-e^{\frac{gQe^{\gamma}}{2T\sqrt{\pi}}\cosh\tau})^{-1}]$ at finite temperature $T$. For very high temperatures the new factor simplifies to $2\pi Te^{-\frac{\pi\sqrt{\pi}T}{gQ}}$. The finite temperature results are important in two respects: Simulations of the full dynamical system at finite temperatures would be easier, since the number of lattice points needed might be quite small. In addition, the behavior of the chiral condensate of the ordinary Schwinger model, $< V + \bar{V} >$, with temperature emphasizes the fact that the axial symmetry is broken explicitly by the anomaly and not spontaneously (at finite temperature the system is effectively one–dimensional and any spontaneous breaking of a symmetry that occurs at zero temperature would have had to completely disappear).

Both in the Schwinger model and in the 3-4-5 model and both at zero and non–zero temperatures the $U(1)$ violating processes can be understood as a consequence of the zero modes of the various chiral Dirac operators in a gauge background of nontrivial topological charge. Since the integral over the scalar field is Gaussian, for each observable there is one particular background gauge configuration that contains all the dependence on the observable. The expectation value of the observable is determined by one particular gauge field configuration and the fermionic determinant in it. If the chiral fermion regularization can be shown to preserve continuum features like the approximate zero mode structure in some fixed backgrounds made out of superpositions of instanton and anti-instanton like configurations, it is quite likely that the full model, including the fluctuations of the gauge field, has been regularized in a way that reproduces all the continuum features. We already pointed out that due to the quadratic nature of the induced gauge field action, once anomalies cancel, even in a chiral model, the chiral determinant is real, i.e. no parity violating processes between the gauge bosons are induced. Thus in the two dimensional abelian models, genuinely chiral models and massless vector models do not differ as much as they do in four dimensions. For this reason, any regularization that preserves the factorization of the chiral determinant into contributions from each fermion of each charge and chirality and has correct parity behavior would most likely reproduce the anomaly–free chiral models correctly if it reproduces the vector model correctly. In the following subsections we will show that the overlap regularization of the abelian models reproduces all the necessary perturbative and non-perturbative features in fixed gauge backgrounds.

*10.2 Perturbative gauge backgrounds*

We intend to check that the lattice regularized overlap reproduces correctly the perturbative features. We shall pick a continuum smooth background on a torus given by a plane wave with a small number of nodes and a small amplitude. This background will be discretized using sets of lattices of decreasing coarseness. We wish to show that in the limit the overlap will be equal, as a complex number, to the continuum formulae reviewed in the previous subsection. Also, we shall show that when we vary the background by a gauge transformation the variation of the imaginary part is correctly reproduced, again in the continuum limit. (The real part is strictly gauge invariant and therefore unaffected by the variation.) It is sufficient to verify this for one chirality and one charge. We pick right handed fermions with unit charge.

We embed an $L \times L$ lattice in the continuum torus whose physical size is $l \times l$. The lattice spacing $a$ is given by $aL = l$. To compute the effective action in perturbative backgrounds, we choose

$$A_\mu(x) = A_\mu^0 \cos \frac{2\pi k \cdot x}{l} \tag{10.15}$$

as our gauge field on the continuum torus. $A_\mu^0$ has dimensions of mass and is held fixed. The $k_\mu$ are fixed integers and determine the momentum of the gauge field. The electric field, $E = \partial_1 A_2 - \partial_2 A_1$, associated with this gauge field is given by

$$E(x) = -\frac{2\pi}{l}(k_1 A_2^0 - k_2 A_1^0) \sin \frac{2\pi k \cdot x}{l} \tag{10.16}$$

Referring to (10.6) we see that, ignoring field independent constants, the effective action for this particular gauge field is given by

$$\begin{aligned} \text{Real}[\Gamma^R(A)] &= -\frac{(k_1 A_2^0 l - k_2 A_1^0 l)^2}{8\pi(k_1^2 + k_2^2)} \\ \text{Imag}[\Gamma^R(A)] &= \frac{(k_1 A_1^0 l + k_2 A_2^0 l)(k_1 A_2^0 l - k_2 A_1^0 l)}{8\pi(k_1^2 + k_2^2)} \end{aligned} \tag{10.17}$$

and the gauge variation of the effective action is given by

$$\frac{\partial \Gamma^R(A)}{\partial \chi(x)} = \frac{i}{2l}(k_1 A_2^0 - k_2 A_1^0) \sin \frac{2\pi k \cdot x}{l} \tag{10.18}$$

If we choose

$$\chi(x) = \chi^0 \sin \frac{2\pi k \cdot x}{l} \tag{10.19}$$

then (10.18) gives

$$-\frac{i}{\pi} \frac{\partial \Gamma^R(A)}{\partial \chi^0} = \frac{1}{4\pi}(k_1 A_2^0 l - k_2 A_1^0 l) \tag{10.20}$$

On the lattice, the link element corresponding to the gauge field (10.15) is

$$U_\mu(n) = \exp\left[i \int_0^1 A_\mu(x + ta\hat{\mu}) dt\right] = \exp\left[i \frac{l A_\mu^0}{L} \frac{\sin \frac{p_\mu}{2}}{\frac{p_\mu}{2}} \cos(p \cdot n + \frac{p_\mu}{2})\right] \sim \exp\left[i \frac{l A_\mu^0}{L} \cos(p \cdot n + \frac{p_\mu}{2})\right] \tag{10.21}$$

with lattice momentum

$$p_\mu = \frac{2\pi k_\mu}{L}; \quad 0 \le k_\mu << L. \tag{10.22}$$

$n = x/a$ labels the discrete sites on the lattice. Note that the $lA_\mu^0$ are dimensionless numbers. The gauge transformation (10.19) is represented on the lattice by

$$g(n) = e^{i\chi^0 \sin(p \cdot n)} \tag{10.23}$$

We compute the overlap on several lattices with increasing $L$'s and extrapolate to the continuum, $L \to \infty$, with the intention to verify (10.17). An exhaustive verification for all $lA_\mu^0$ and $k_\mu$ would establish that the lattice regularized overlap reproduces all the exact continuum results for perturbative gauge fields.

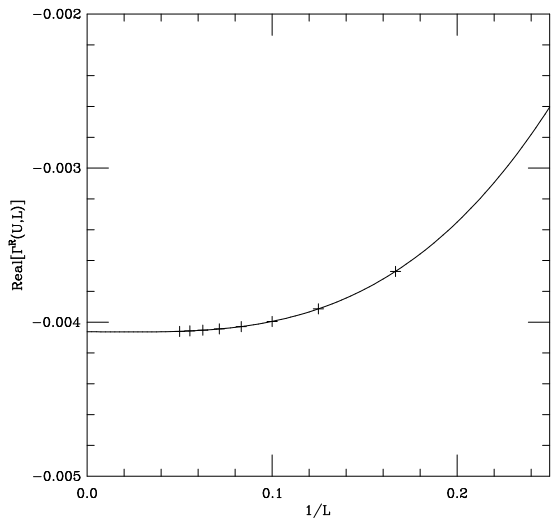

**Figure 10.1** The real part of the induced action as a function of number of lattice points for a particular smooth configuration.

As an example we choose $lA_1^0 = lA_2^1 = 0.32$ and $k_1 = 1$ and $k_2 = 0$. We set $m = 0.9$ in the Hamiltonians $\mathbf{H}^\pm$. Recall that the parameter $m$ is an additional ultraviolet cutoff in our regularization procedure. This parameter does not have to be fine tuned, except being kept large relative to physical scales. The continuum limit is independent of the precise value of $m$. The effective action is computed on a finite lattice by using (7.9). The finite dimensional Hamiltonians are diagonalized for the gauge field background given by (10.21) and the corresponding single particle eigenfunctions are obtained. From this the many body states $|R\pm>_U$ are constructed. The real part is computed for various $L$'s and plotted as a function of $\frac{1}{L}$ in Figure 10.1. Extraction of the continuum limit yields a value of $-0.00406(2)$ which is in good agreement with the righthand side of the first line in (10.17), namely $-\frac{(0.32)^2}{8\pi} = -0.0040744$. The imaginary part is computed for various $L$'s and plotted as a function of $\frac{1}{L}$ in Figure 10.2. Extraction of the continuum limit yields a value of $0.00409(2)$ which is in good agreement with the righthand side of the second line in (10.17), namely $\frac{(0.32)^2}{8\pi} = 0.0040744$.

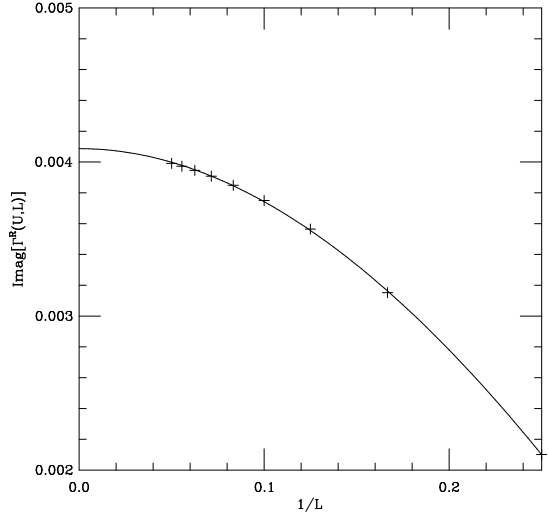

**Figure 10.2** The imaginary part of the induced action as a function of number of lattice points for a particular smooth configuration.

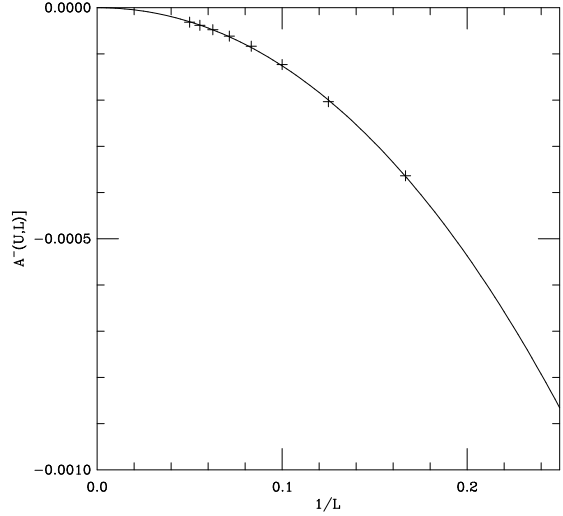

**Figure 10.3** The contribution to the anomaly coming from the "inactive" side, $\mathcal{A}^-$ as a function of number of lattice points for a particular smooth configuration.

The gauge variation of the effective action on an $L \times L$ lattice, following (7.16), is expressed as

$$-\frac{i}{\pi}\frac{\partial \Gamma^R(U^g, L)}{\partial \chi^0}\bigg|_{\chi^0=0} \equiv \mathcal{A}(U, L) = \mathcal{A}^-(U, L) + \mathcal{A}^+(U, L) \qquad (10.24)$$

with

$$\mathcal{A}^{\pm}(U, L) = \mp \frac{i}{\pi}\frac{\partial}{\partial \chi^0} \log\left[\frac{{}_U\!< R + |\mathcal{G}|R+ >_1}{|{}_U\!< R + |\mathcal{G}|R+ >_1|}\right] \qquad (10.25)$$

$\mathcal{A}^{\pm}(U, L)$ are the contributions to the anomalies from $|R\pm>$ respectively. $\mathcal{G}$ is as defined in (7.10)–

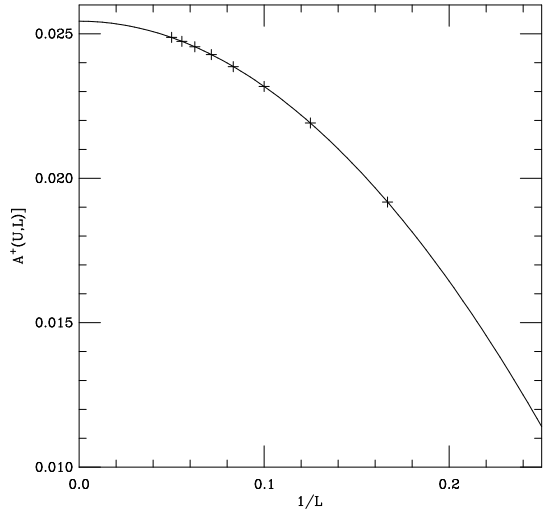

**Figure 10.4** The contribution to the anomaly coming from the "active" side, $\mathcal{A}^+$ as a function of number of lattice points for a particular smooth configuration.

(7.14) with the $g$ appearing in (7.10) given by (10.23). It has been observed before that $\mathbf{H}^+$ is the "active" side. For instance, $\mathbf{B}^+$ changes sign as a function of momentum in order to remove the doublers on the lattice (see section 7) while $\mathbf{B}^-$ always stays positive. Also, only $\mathbf{H}^+$ can become gapless and therefore give rise to mismatches in the number of filled states (see section 8). Now we shall see that only $\mathcal{A}^+(U)$ contributes to the anomaly in the continuum limit: In Figure 10.3 we plot $\mathcal{A}^-(U)$ as a function of $\frac{1}{L}$ and see that $\lim_{L \to \infty} \mathcal{A}^-(U) = 0$ indicating that the state $|R-\rangle_U$ does not contribute to the anomaly. In Fig (10.4), we plot $\mathcal{A}^+(U)$ as a function of $\frac{1}{L}$. Extraction of the continuum limit from $\mathcal{A}(U)$ yields a value of $0.0254(1)$ which is in agreement with the righthand side of (10.18), namely $\frac{0.08}{\pi} = 0.0254648$. Of course, since the anomaly is evaluated as a derivative of a functional, it comes in the "consistent" form which for the present abelian computation only means that the numerical coefficients in the first row of (10.5) are $\frac{1}{4\pi}$ rather than $\frac{1}{2\pi}$.

It is rather obvious from the above that if the anomalies cancel between different fermion species in the continuum, (as for example they do in the 3-4-5 model – see subsection 10.1) we expect the imaginary part on the lattice to be very small for gauge fields typical to small bare gauge coupling. We tested this on a $4 \times 4$ lattice, for the 3-4-5 model, with a set of background link variables whose phases were randomly drawn from the interval $(-\frac{\pi}{10}, \frac{\pi}{10})$ and found that the residual imaginary part was smaller by a factor of about one hundred than the individual contributions from each species.

*10.3 Topological charge*

We explained in section 3 in general terms and in section 8 in detail how the overlap encodes the topological properties of continuum gauge field space. The purpose of this subsection is to show how this works in simple examples. We shall also show numerically that the lattice definition of topological charge is robust under the addition of reasonable amounts of noise to a given smooth configuration. This indicates that the classical concept of continuum topological charge when latticized by our method maintains its relevance at the quantum level. While this is always assumed, it is not a trivial feature because dominating configurations in a path integral are very different from smooth continuous fields.

We start by defining a continuum gauge field on the torus which carries unit topological charge, but, unlike an instanton, has uniform field strength. Later, in subsection 10.6, we shall see that when the configuration has the topological charge density more localized the needed mismatches also occur. Our tests will show that our regularization reproduces the Atiyah–Singer index theorem, i.e. the number of effective zero modes depends only on the topology. Even if the field strength is uniform the vector potential cannot be a smooth function on the torus because the related $U(1)$ bundle is nontrivial. To define the gauge field it is enough to cut the torus along one of its main circles, and make the vector potential change discontinuously across the cut by a gauge transformation. In the direction that has been left uncut the gauge potential is periodic and hence smooth. The gauge transformation at the cut is nontrivial having unit winding when one goes around the cut circle. If we now consider the chiral Dirac operator as a mapping from the set of fermion fields with a discontinuity across the cut given by the above gauge transformation we discover that the images are in the same set. Thus, the mapping itself is entirely smooth in the sense that a small deformation in the source field is mapped into a small deformation of the image field. When we put this on the lattice the link fields must reflect both the above discontinuity and the the fermion boundary conditions.* The following definitions contain a parameter $q$. Integer values of $q$ correspond to topological charge $q$ and smooth Dirac operators, while the intermediate values of $q$ correspond to Dirac operators that have real discontinuities on the torus and are thus forbidden in the continuum. Their existence on the lattice is the reason that strictly speaking there is no topology on the lattice and any configuration of gauge fields can be continuously deformed into any other.

A uniform electric field $E = \frac{2\pi q}{L^2}$ on a torus has a topological charge equal to $q$. Such an electric field is realized on a finite lattice by the following configuration:

$$U_1(n_1, n_2) = \begin{cases} 1 & \text{if } n_1 \neq L-1 \\ \exp\left[-i\frac{2\pi n_2 q}{L}\right] & \text{if } n_1 = L-1 \end{cases} \; ; \quad U_2(n_1, n_2) = \exp\left[i\frac{2\pi n_1 q}{L^2}\right] \qquad (10.26)$$

The lattice sites are $(n_1, n_2)$ with $0 \leq n_{1,2} \leq L-1$. If $q$ is an integer then the above configuration produces a uniform electric field. The configuration is periodic in the direction 2. The $U$'s living on links in the 2 direction implement the linearly raising vector potential needed to generate a constant electric flux. When $n_1$ goes from $L-1$ to zero we note a discontinuity in $U_2$ by a gauge transformation. The torus is cut be severing the direction 1 links $(L-1, n_2)$–$(0, n_2)$. The $U_1$'s living on these links implement the boundary gauge transformation on the fermion field. It is easy to check that, if $q$ is an integer, the parallel transporters around all elementary plaquettes are equal to each other and to $e^{i\frac{2\pi q}{L^2}}$. Thus, physically, the configuration is smooth. This is not true when $q$ is not an integer (consider the plaquette whose corners are $(L-1, L-1)(L-1, 0)(0, 0)(0, L-1)$); however the configuration is still perfectly acceptable to our construction. The field configuration is well defined for any value of $q$ and we would expect the lattice topological charge to change discontinuously as one varies $q$.

---

\* We remind the reader that we have implicit underlying antiperiodic boundary conditions – what is being discussed here is on top of those boundary conditions.

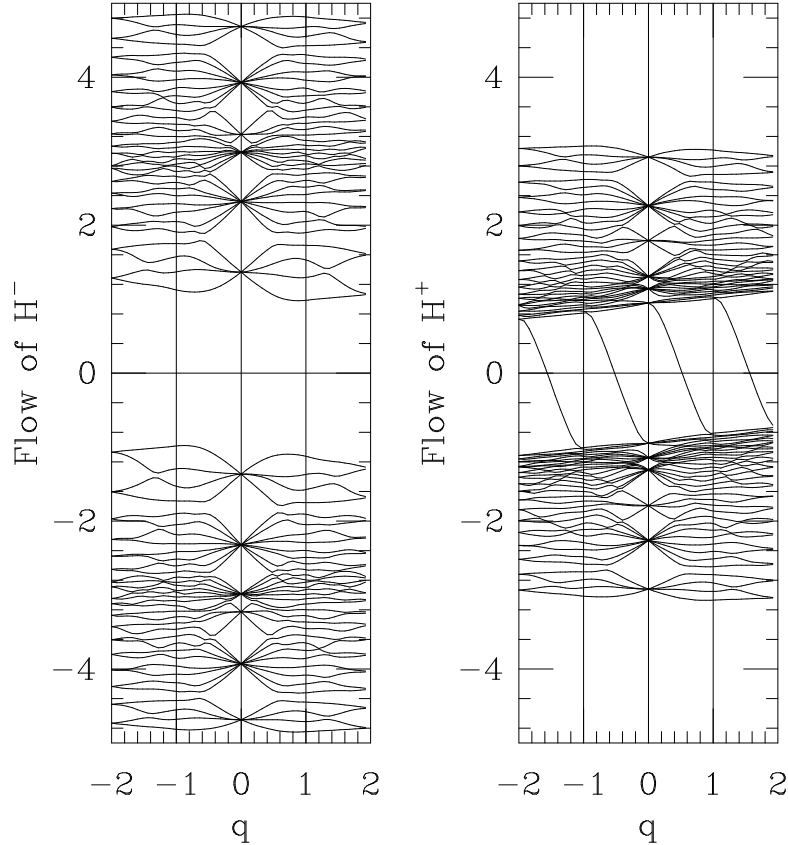

**Figure 10.5** Spectra of $\mathbf{H}^{\pm}$ as function of $q$. The integer values of $q$ correspond to smooth gauge field configurations of topological charge $q$.

In what follows we consider a $6 \times 6$ lattice and $m = 0.9$ in $\mathbf{H}^{\pm}$. In Figure 10.5, we plot the spectral flow of $\mathbf{H}^{-}$ as a function of $q$. We see a gap in the spectrum for all values of $q$ and there are equal number of eigenvalues, namely 36, on either side of the gap. This is in agreement with the proof given in section 8 that $\mathbf{H}^{-}$ has a gap for all gauge fields. Actually, the bounds on the gap ($|E| \leq m$) derived there are seen to be very tight.** At $q = 0$ the gauge configuration is trivial and there are four–fold and eight–fold degeneracies coming from lattice symmetries and spin. These degeneracies are lifted when $q$ is turned on. Levels repel when $q$ is varied and the overall range stays pretty constant.

In Figure 10.5, we plot the spectral flow of $\mathbf{H}^{+}$ as a function of $q$. Here we see the gap closing as one moves away from $q = 0$. At $q = 0$ there are an equal number of positive and negative eigenvalues. Between $0.52 < q < 1.55$ we have 35 positive eigenvalues and 37 negative eigenvalues. In this region both the overlaps $_U\!\!<R-|R+>_U$ and $_U\!\!<L-|L+>_U$ vanish. The insertion of one creation (annihilation) operator in the right (left) handed overlap renders it non-zero showing that these gauge configurations indeed have a lattice topological charge of one unit. The plot is therefore separated into sectors defined by $-2 < q < -1.55$, $-1.55 < q < -0.52$, $-0.52 < q < 0.52$, $0.52 < q < 1.55$, and $1.55 < q < 2$ with lattice topological charges equal to $-2$, $-1$, $0$, $1$ and $2$ respectively.

---

** To be precise one must add finite volume effects that are dependent on the implicit antiperiodic boundary conditions.

We also note that, except for the crossing eigenvalue, the gap is similar in magnitude to the one of $\mathbf{H}^-$ and both gaps are indeed of the order $2m$ as expected from the continuum, formal, overlap. The range of eigenvalues of $\mathbf{H}^+$ is restricted relative to the one of $\mathbf{H}^-$ at the top and bottom. This is easily understood as a result of the change of sign in front of $m$: indeed the difference between the $\mathbf{H}^\pm$ is of order $2m$. This squeezes the levels for $\mathbf{H}^+$ into smaller ranges. Note the discernible sloping of the gap in Figure 10.5: Extrapolating, it shows that we cannot make $q$ too large and maintain the connection to continuum topological charge. This is just a cutoff effect, and we know such must exist since there is no nontrivial topology on the lattice. Another example where cutoff effects are seen to affect the lattice–topology continuum–topology connection is a configuration that has a small concentrated amount of electric flux easily missed by a too coarse embedded lattice.

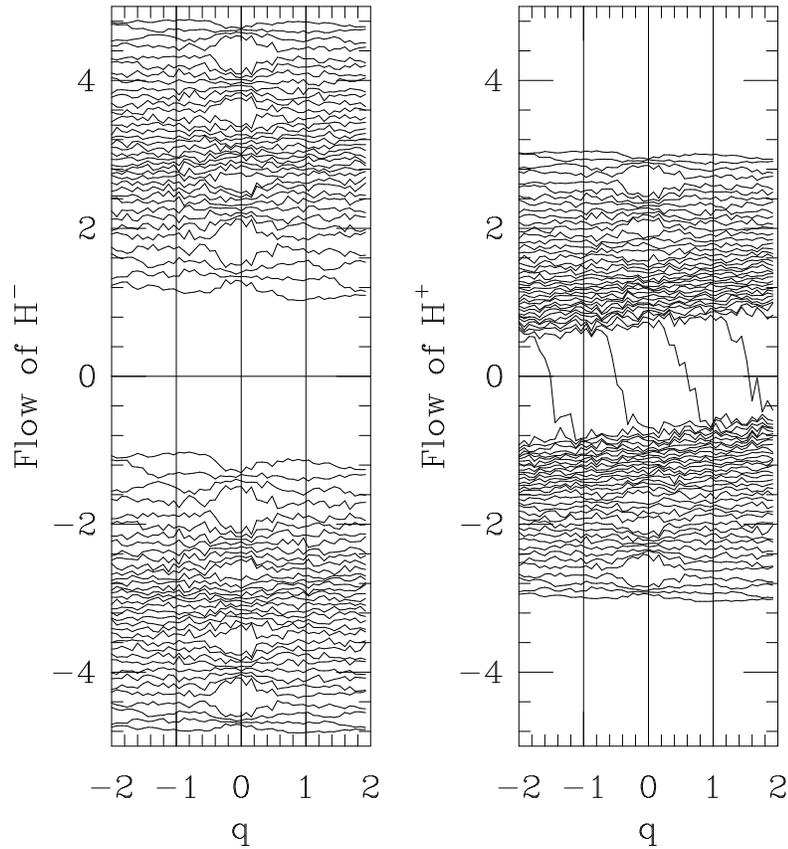

**Figure 10.6** Same as Figure 10.4, but now at each new value of $q$ the corresponding link variables have been multiplied by independent random phase factors.

In a Monte–Carlo simulation the topological charge is a fluctuating quantum variable. Therefore topology changing events have to take place during the evolution. The frequency of these events determines the relative normalizations of the contributions coming from the different sectors. These normalizations are expected to come out in such a way that the equilibrium configurations satisfy clustering. Imagine a sequence of configurations labeled by $q$ as in figure 10.5. It is very unlikely that such a smooth evolution will ever occur. A slightly more realistic scenario is one in which one takes each configuration labeled by a $q$ (not necessarily an integer) and adds to it noise. We took the lattice configurations (10.26), and, for each $q$ multiplied all links by factors $e^{i\phi}$ where $\phi$ was randomly drawn from the interval $(-\frac{\pi}{6},\frac{\pi}{6})$. This is a sizable amount of noise but not an overwhelming one. The eigenvalue flows obtained from the new set of configurations labeled by $q$ are plotted in figure 10.6. The main message is that all the gross features of the noise free flows of figure 10.5 survive. In particular, we get all the crossings we had there, thus providing evidence for the robustness of our definition of lattice topological charge.

*10.4 Instanton–anti-instanton background*

We continue to look at large but smooth gauge abelian backgrounds in this subsection. We now want to see what happens if the total topological charge is zero, but there are lumps of local concentration of electrical flux of sufficient size.* As a specific example we consider a single right handed fermion in a gauge background made up of an instanton "ring" separated from an anti-instanton "ring" along a torus whose other direction is a small circle representing a finite physical temperature. We project this configuration onto an $N_T \times L$ lattice where $1/N_T$ is the temperature in dimensionless units. The usage of finite temperature is mainly for technical reasons (it reduces the size of the matrices we have to diagonalize), but is quite relevant physically also. The link variables that define the field configuration are

$$U_1(n_1,n_2) = 1; \quad U_2(n_1,n_2) = \begin{cases} \exp\left[\frac{2\pi i}{N_T}\right] & \text{if } 1 \leq n_2 \leq n \\ 1 & \text{elsewhere} \end{cases}; \quad 0 \leq n_1 < L; \quad 0 \leq n_2 < N_T; \quad 1 \leq n < L$$
(10.27)

The link variables are uniform in the $n_2$ direction. All the plaquettes have no electric field except for the ones between $n_1 = 0$ and $n_1 = 1$ and between $n_1 = n$ and $n_1 = n + 1$. Between $n_1 = 0$ and $n_1 = 1$ there is a net flux of $2\pi$ and this corresponds to a localized instanton "ring" with charge $+1$. Between $n_1 = n$ and $n_1 = n + 1$ there is a net flux of $-2\pi$ and this corresponds to a localized anti-instanton "ring" with charge $-1$. The total flux is zero and this configuration has no topological charge. Continuum arguments will imply that the chiral determinant will decrease as $n$ increases as long as $n << L$: If the distance $n$ (and the size $L$) were infinite there would be two fermionic zero modes and the chiral determinant would vanish. For large but finite distances the absolute value of the chiral determinant can be estimated by computing the two smallest eigenvalues of $\mathbf{C}^\dagger \mathbf{C}$ by perturbation theory around the two zero modes. These eigenvalues are in turn determined by the amount of overlap of the two approximate zero modes localized at the instanton and at the anti-instanton. Simply put, there is an attractive potential between an instanton and an anti-instanton induced by the chiral fermion. In this particular case of finite temperature, the zero modes decay exponentially and the potential is expected to be linear with the constant of proportionality given by $\frac{\pi}{N_T}$. This can be seen by solving the effectively one–dimensional differential equation for the zero mode in the continuum. We wish to show that the overlap reproduces the fermion induced instanton – anti-instanton interactions correctly.

---

\* Throughout the remainder of this section we shall use the terms instanton and anti-instanton in a loose sense, meaning a configuration that has a relatively localized lump of topological charge density that integrates to unity. Away from the lump the configuration is trivial.

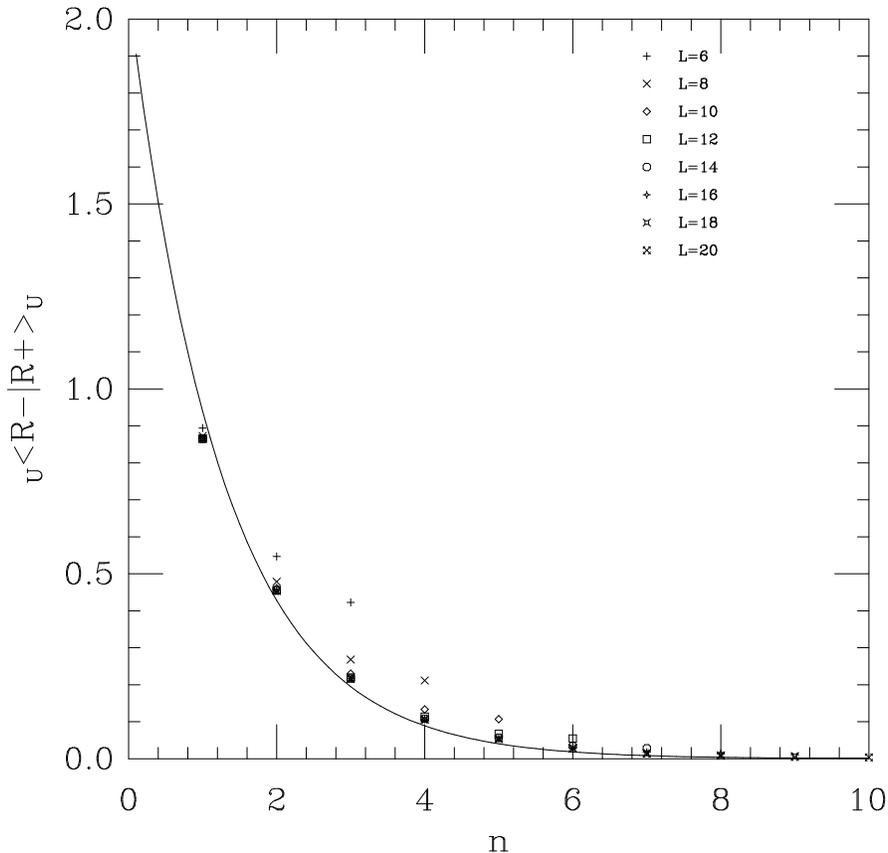

**Figure 10.7** Overlap in a background of an instanton "ring" separated spatially by $n$ lattice spacings from an anti-instanton "ring" in two dimensions at finite temperature.

To do this numerically we fix $N_T = 4$ and $m = 0.9$ in $\mathbf{H}^{\pm}$ and compute the determinant for the field configuration in (10.27) for a range of lattices with $L = 6, 8, 10, 12, 14, 16, 18, 20$. On each lattice we consider $n$ in the range $1 \leq n \leq L/2$. The results for the "chiral determinant", $_U\!< R-|R+>_U$, are plotted in Figure 10.7. For fixed $n$, we expect the chiral determinant to approach a limiting value as $L \to \infty$. This is seen in Figure 10.7. The limiting values are expected to fit an exponential of the form $C \exp(-\frac{\pi n}{4})$ and the curve shown in Figure 10.7 is one such exponential obtained by tuning $C$. The curve smoothly passes through the points for a wide range of $n$ showing that the overlap reproduces the expected behavior quantitatively.

*10.5 Fermion expectation values in an instanton–anti-instanton background*

Although the gauge configuration considered in the previous section has no topological charge, it is made up of one localized instanton of charge +1 and another localized instanton of charge −1. As mentioned in the previous subsection, formal arguments imply the existence of two approximate zero modes one localized at the site of the instanton and another at the site of the anti-instanton. This can be uncovered by looking at fermion expectation values in this background. The method to compute fermion expectation values was described in section 5. In this case we would like to compute the fermion propagator. Normally we would expect the fermion propagators to vanish at large distances. But in this gauge background we will get a significant contribution even at large distances if the pair of points are located at the sites of the instanton and anti-instanton. In particular, from subsection 10.3, we know that there is an excess of negative eigenvalues for $\mathbf{H}^+$ in a background with positive topological charge. Therefore we should insert a fermion creation operator at the site of the instanton and a fermion annihilation operator at the site of the anti-instanton and look at $_U{<}R-|a_{x\alpha}a^\dagger_{y\beta}|R+{>}_U$. If we keep $y$ fixed at the site of the instanton field ($y=0$) and vary $x$, we should see the propagator reach a peak at the site of the anti-instanton. Further, this behavior should remain the same, up to an overall multiplication by a factor, if one picks another $y$ in the vicinity of 0.

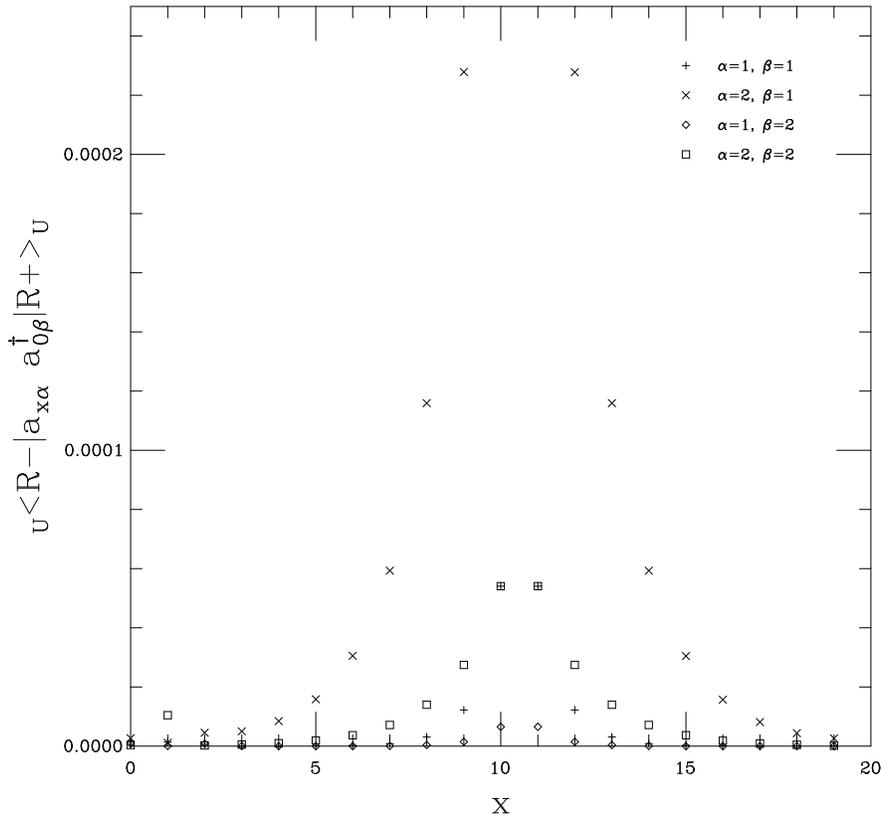

**Figure 10.8**   The fermion two point function in the instanton–anti-instanton background of Figure 10.7

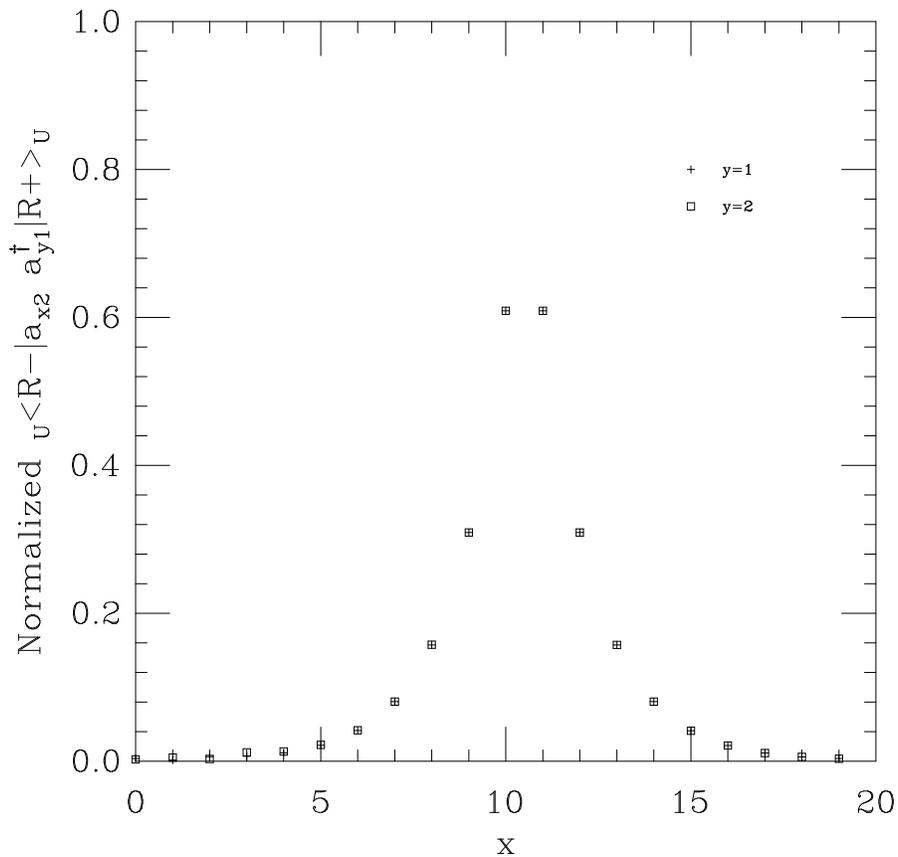

**Figure 10.9**  Two point function for $\alpha = 2, \beta = 1$ in Figure 10.8, normalized in $x$, plotted for two values of $y$.

We study the above behavior by using the configuration in the previous subsection. Here we fix $L = 20$ and $n = 10$ in addition to fixing $N_T = 4$. We compute ${}_U\!< R - |a_{x\alpha} a^\dagger_{y\beta}|R+ >_U$. As discussed in section 5, we are using Dirac operators to define expectation values of chiral fermions, but only half of the modes are physical. This is made very obvious numerically because it turns out that for one value of $(\alpha, \beta)$ one gets a matrix element that is significantly larger than the matrix elements corresponding to the three other possible $(\alpha, \beta)$ pairs. In Figure 10.8, we plot ${}_U\!< R - |a_{x\alpha} a^\dagger_{0\beta}|R+ >_U$ as a function of $x$ for all four values of the pairs $(\alpha, \beta)$. We see that out of the four only $\alpha = 2, \beta = 1$ contribute significantly. In particular, the complementary pair, $\alpha = 1, \beta = 2$ is completely suppressed. Focusing on the dominating matrix element we see that indeed the propagator peaks at $x = n$, the site of the anti-instanton.

In Figure 10.9, we plot the same propagator in a normalized form, where the square of the normalized propagator summed over all $x$ gives unity at fixed $y, \alpha, \beta$. We only plot the $\alpha = 2, \beta = 1$ component and display the results for two different values of $y$, namely $y = 1, 2$, both of which are near the site of the instanton. We see that indeed the shapes of the propagators are identical. This shows that the fermion propagator in the overlap regularization factorizes into two factors, one for the instanton and another for the anti-instanton as expected on the basis of continuum manipulations. The factorization is important because it shows how the 't Hooft vertex gets exponentiated in the dilute gas approximation.

*10.6 Fermion expectation values in non-zero topological charge*

The zero modes referred to in the previous two subsections can be computed explicitly in non-zero topological charge. Their shape should approximately determine the results we have obtained in the previous two subsections and this will provide another check on the selfconsistency of viewing the overlap as a discretization of the continuum chiral determinant. In particular, let us focus on a gauge configuration consisting of a single localized anti-instanton. The configuration has topological charge $-1$ and can be realized on an $N_T \times L$ lattice by the following set of link variables:

$$U_1(n_1, n_2) = \begin{cases} \exp\left[\frac{2\pi i n_2}{N_T}\right] & \text{if } n_1 = 1 \\ 1 & \text{elsewhere} \end{cases} ; \qquad U_2(n_1, n_2) = \begin{cases} \exp\left[-\frac{2\pi}{N_T}\right] & \text{if } n_1 = 1 \\ 0 & \text{elsewhere} \end{cases} \qquad (10.28)$$

The plaquettes between $n_1 = 0$ and $n_1 = 1$ carry identical amounts of electric field adding up to a total flux of $-2\pi$. All other plaquettes carry no electric field at all. The Wilson loop encircling the torus in the 1 direction at fixed $n_2$ is different from unity and winds around the complex unit circle when $n_2$ is taken around the torus in the direction 2. The nontrivial winding reflects the non-trivial topology of the gauge configuration. Apart from the winding, the configuration is very similar to one half of the instanton–anti-instanton configuration considered in the previous two subsections. The overlap in such a background is zero due to a mismatch and from subsection 10.3 we know that $\mathbf{H}^+$ will have one more positive eigenvalue than $\mathbf{H}^-$ and therefore the matrix element $_U\!< R - |a_{x\alpha}|R+ >_U$ should be non-zero. Further, the matrix element, when viewed as a function of $x$ can be interpreted as the zero mode in the background gauge field and should therefore peak at the site of the anti-instanton.

In order to establish the existence of the zero mode in the above background we fix $N_T = 4$, $L = 20$ and expect that one value of $\alpha$ should dominate. Based on the results of the previous subsection we know that the relevant $\alpha$ ought to be 2. We plot $_U\!< R - |a_{x\alpha}|R+ >_U$ for the two values of $\alpha$ as a function of $x$ in Figure 10.10 and see that the $\alpha = 2$ matrix element is indeed the dominant one and that it peaks at the site of the anti-instanton as it should.

As discussed in the previous subsection, the propagator in the presence of an instanton and anti-instanton factorizes and therefore the propagator plotted in Figure 10.8 should represent just the zero mode localized at the site of the anti-instanton. Also, the discussions in subsection 10.4 imply that the decay of the overlap in the presence of the instanton and anti-instanton is governed by the decay of the zero mode. The overlap representing the chiral determinant is non-zero only due to the nonvanishing overlap between the two zero modes. As one separates the instanton and anti-instanton this latter overlap decreases causing the chiral determinant to decrease. It is therefore clear that the rate of decay of the chiral determinant should be the same as the rate of decay of the zero modes. We can show that these considerations are obeyed by the lattice overlap by comparing the normalized zero mode from Figure 10.10 with the normalized propagator from Figure 10.9 and with the normalized decay of $_U\!< R - |R+ >_U$ from Figure 10.7. The three sets of points are plotted together in figure 10.11 and all neatly fall one on top of the other. With this we have tested that the real part of the overlap along with the definition of fermion expectation values have the right behavior for large but smooth gauge fields in the zero topological sector and also in the non-zero topological sectors.

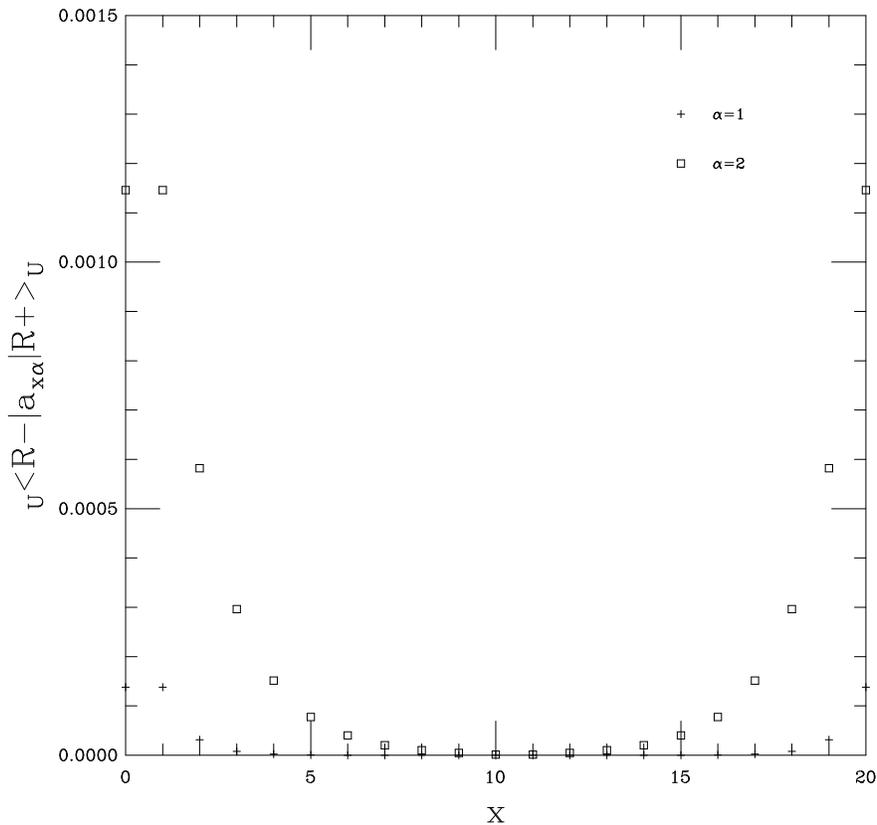

**Figure 10.10** Zero mode in an anti-instanton background.

*10.7 Axial anomaly in nonzero topological sectors*

We have tested the continuum formula for the anomaly in perturbative backgrounds. In this subsection we wish to test it in a background that is non-perturbative because it carries topological charge. We cannot do that from the overlap because it vanishes (as does the corresponding continuum chiral determinant). However, if we insert a local, gauge invariant operator that carries the right amount of anomalous charge we know that its expectation value will become non-zero. Suppose we change the fermions by a local gauge transformation corresponding to the anomalous (ungauged) group. Since the operator carries anomalous charge the expectation value will change even at the classical level in a definite way. At the quantum level, the anomaly will add an additional term reflecting the anomaly. We can pick the local inserted operator at a location $y$ say and carry out the anomalous gauge transformation in a localized way away from $y$. Then there is no classical variation at all, and all we should get is the anomaly.

A test of the anomaly in the unit topological sector will support the extension of the Wigner-Brillouin phase choice to this case (see section 5.2) and at the same time test the theory in some more detail in non-perturbative backgrounds.

Here we consider a two dimensional abelian vector theory (the matter is a charged Dirac fermion) in a background $U$ carrying unit topological charge. Let $\mathcal{O}$ be the local operator that has a nonzero vacuum matrix element in this background. The overlap now involves the states $|V\pm>_U$ and the definition of their phases is slightly more complicated, as explained in subsection 5.2. The expectation value of $\mathcal{O}$ in this background is

$$\frac{{}_1\!<V-|V->_U}{|{}_1\!<V-|V->_U|}{}_U\!<V-|\mathcal{O}|V+>_U \frac{\sum_{x\alpha}{}_U\!<V+|a^R_{x\alpha}[a^L_{x\alpha}]^\dagger|V+>_1}{\sum_{x\alpha}{}_U\!<V+|a^R_{x\alpha}[a^L_{x\alpha}]^\dagger|V+>_1} \tag{10.29}$$

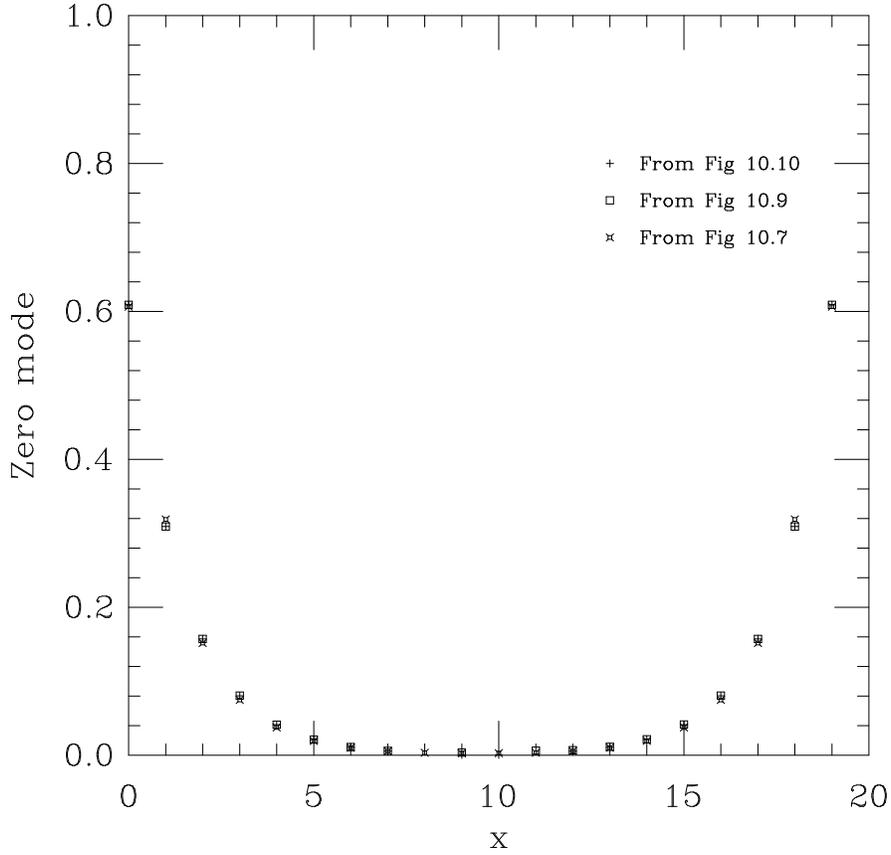

**Figure 10.11**  Checking that the zero mode of Figure 10.10 essentially determines the shape of Figures 10.7 and 10.9.

$|V\pm>_U = |R\pm>_U \otimes |L\pm>_U$ are the many body states for the vector theory. There are level crossings only in the positive side and therefore the modified definition of the phase is needed only there. $\sum_{x\alpha} a^R_{x\alpha}[a^L_{x\alpha}]^\dagger$ is the 't Hooft operator that has to be inserted for the overlap to be non-zero. It is a Lorentz scalar and is gauge invariant. Since there is an axial anomaly in this theory, this expectation value is not expected to be invariant under a local axial phase change made at a location far away from $\mathcal{O}$. The anomaly being local is still expected to be given by the gauge variation of (10.6), namely, by $-\frac{i}{2\pi}E(x)$ in the continuum. Further, the middle term in (10.29) is invariant under the above axial phase change since $\mathcal{O}$ is. (If it were not we would also get the classical contribution to the axial variation.) Thus the anomaly must come from the variation of the first and third terms.

As an example of a gauge field configuration with unit topological charge, we take the perturbative configuration defined in (10.15) with electric field (10.16) and add to that a configuration that has a uniform electric field of strength $\frac{2\pi}{l^2}$. If we now make an axial gauge variation of the type given in (10.19), the uniform field does not contribute to the anomaly. Therefore only the perturbative plane wave part contributes and the anomaly should be given by (10.20) with an overall factor of 2 to account for two chiralities. To implement this on the lattice, we multiply the links representing the perturbative configuration in (10.21) by the ones representing the uniform field configuration in (10.26). We choose the same parameters as in section 10.2 and again compute the gauge variations on several lattices. Like in the perturbative case the negative side is gauge invariant in the continuum limit and only the positive side has an anomaly. However, the approach to the continuum in the present case is qualitatively different from the perturbative case. An extrapolation to the continuum gives 0.0511(2) in good agreement with the continuum value $\frac{0.16}{\pi} = 0.0509296$ and the test is successful.

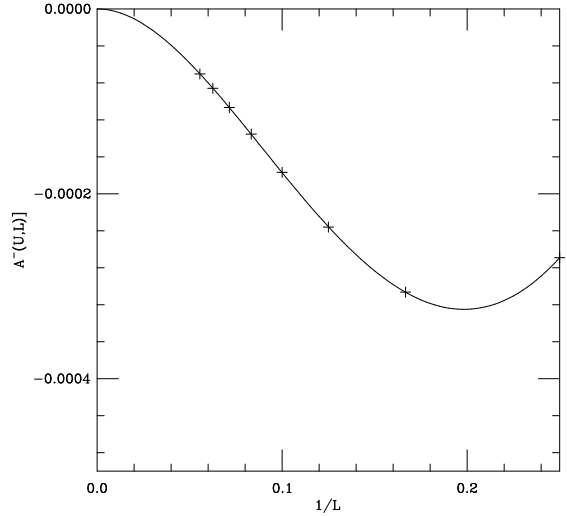

**Figure 10.12**  The contribution to the anomaly coming from the "inactive" side, $\mathcal{A}^-$ as a function of number of lattice points for a configuration with topological charge.

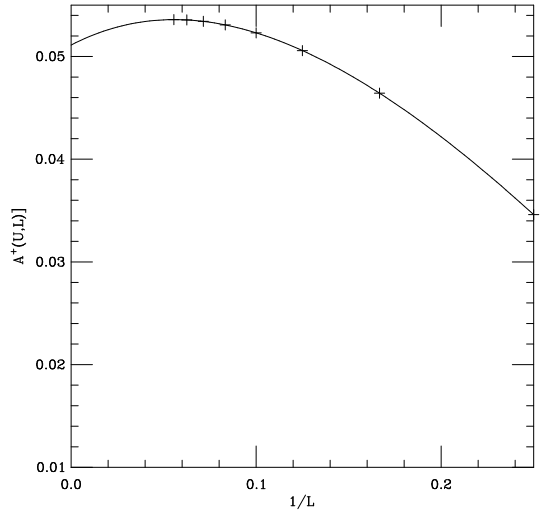

**Figure 10.13**  The contribution to the anomaly coming from the "inactive" side, $\mathcal{A}^+$ as a function of number of lattice points for a configuration with topological charge.

## 11. Tests in Four Dimensions.

In this section we check that the overlap in four dimensions reproduces the abelian anomaly and the rules of counting zero modes in abelian and non–abelian backgrounds. We use the transfer matrix formalism throughout. For the simple tests we have carried out the computational burden is not too bad and since future work with more ambitious goals will most likely focus on the Hamiltonian formalism it is useful to establish once and for all that the transfer matrix version has been also checked (at least to some extent). The basic logic of the checks is the same as in the previous section. We shall see that the overlap formalism passes both perturbative and non-perturbative tests successfully. In particular, some of the most crucial ingredients necessary in order to reproduce the baryon decay in the minimal standard model are seen to be incorporated in the lattice overlap.

*11.1 U(1) anomaly.*

We imagine a single righthanded fermion interacting with an abelian plane wave background. Under a gauge transformation the phase of the overlap has to change by the amount dictated by the famous triangle diagram. We shall show that the continuum result is reproduced and therefore conclude that all the results based on triangle diagrams are reproduced. Combining this with the obviously holding Zumino [24] consistency conditions we conclude that all perturbative four dimensional anomalies, including non–abelian ones, will be reproduced by the overlap.

The technical difficulty we have to overcome is that the matrices one has to diagonalize are larger in four dimensions. If the system is a lattice torus with equal sides $L$ one deals with matrices of size $4L^4 \times 4L^4$. From subsection 10.2 we expect to need to go to $L$'s of the order of 10 at least in order to be able to extract the continuum limit. This implies rather large matrices and we do not wish to get into ponderous numerics. We opt therefore to use a trick by devising a background that has an anomaly but also has translational invariance in two directions, thus reducing the matrices down to $L^2$ $4L^2 \times 4L^2$ blocks. Of course, it is much easier to diagonalize, say, 100 400×400 matrices rather than one 40000 × 40000 one. We need a configuration that is independent of two of the components $n_i$ of the lattice site coordinate $n$ but manages to produce magnetic and electric fluxes through a sufficiently large number of plaquettes to make a relatively coarse discretization reasonable. The configuration we choose is translational invariant in the 3 and 4 directions:

$$U_{n,4} = e^{i\frac{A_4}{L}\cos\frac{2\pi k_2 n_2}{L}}; \quad U_{n,3} = e^{i\frac{A_3}{L}\cos\frac{2\pi k_1 n_1}{L}}$$

$$U_{n,2} = e^{i\frac{\phi}{L}\left[\cos\frac{2\pi}{L}(k_1 n_1 + k_2 n_2 + k_2) - \cos\frac{2\pi}{L}(k_1 n_1 + k_2 n_2)\right]}$$

$$U_{n,1} = e^{i\frac{\phi}{L}\left[\cos\frac{2\pi}{L}(k_1 n_1 + k_2 n_2 + k_1) - \cos\frac{2\pi}{L}(k_1 n_1 + k_2 n_2)\right]} \quad (11.1)$$

To get the anomaly we elect to use a more direct approach (also a less efficient one) than the one in subsection 10.2. We do this because here, unlike in two dimensions, we shall not carry out any other computation that directly tests the imaginary part of the induced action. For this reason we put in a free parameter $\phi$ in the configuration. By varying $\phi$, one generates a family of gauge equivalent configurations. Using the continuum "consistent" value of the anomaly we conclude that we expect the following:

$$\lim_{L\to\infty} -\frac{i}{\pi} \frac{\partial \Gamma^R(U^\phi, L)}{\partial \phi}\bigg|_{\phi=0} = -\frac{1}{12\pi} A_3 A_4 k_1 k_2 \qquad (11.2)$$

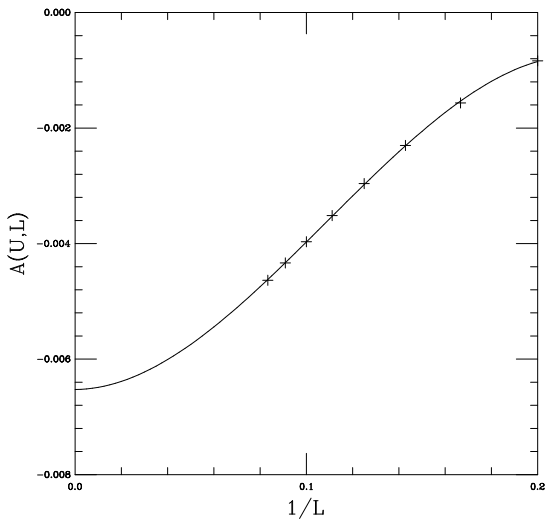

**Figure 11.1** The anomaly as a function of inverse linear lattice size for a four dimensional plane wave.

We set $k_1 = k_2 = 1$ and $A_3 = A_4 = 0.5$ and compute numerically, using the overlap and the Wigner–Brillouin phase choice, the left hand side of the above equation for $L = 5, 6, ..., 12$. Antiperiodic boundary conditions for the fermions are implicit and the parameter $m$ in the transfer matrices has been set to .5. Upon closer examination one realizes that there is more symmetry in the problem, related to "rotations" and significantly fewer matrices need to be actually diagonalized. Since this is quite technical we omit the details. Also, because of another symmetry, the $\phi$ derivative at zero can be computed by evaluating the phase at a single small positive value of $\phi$.

The results are plotted in Figure 11.1 and the extrapolation to $L = \infty$ is smooth yielding $-0.00663(1)$ for the anomaly in agreement with the continuum value of $-0.00663$.

*11.2. Topological charge.*

In this subsection we shall discuss a series of tests whose purpose is to ascertain that in four dimensions in the presence of nonabelian gauge backgrounds the insertion of the appropriate number and kind of creation/annihilation operators between the states making up the overlap renders an otherwise vanishing result nonzero. This is done by obtaining the number of eigenstates of the "active" transfer matrix whose eigenvalues are less than unity. Actually, the interesting number is the deficit or surplus relative to the free case, or, equivalently, relative to the "inactive" transfer matrix. Again, the main technical problem is to avoid dealing with too large matrices.

The gauge group is chosen to be $SU(2)$. Based on our two dimensional experience is is easy to guess that on a torus even abelian gauge configurations can carry topological charge. Such a configuration can be embedded in one of the $U(1)$'s existing inside $SU(2)$, for example the one generated by $T_3 = \frac{1}{2}\begin{pmatrix} 1 & 0 \\ 0 & -1 \end{pmatrix}$. Since the number of zero modes associated with $T_3 = 1/2$ is equal to the one associated with $T_3 = -1/2$ only even $SU(2)$ topological charge can be generated this way. However, in practice we can only look at $T_3 = 1/2$ say, and do not need to carry group indices at all. This reduces the size of the matrix by a factor of two. Clearly, it is wasteful to fully diagonalize the transfer matrix (we only consider the "active" one, denoted by "+" before); it suffices to track eigenvalues crossing unity. We wish to identify all those crossings, but do not need the other eigenvalues. In order to be able to focus on the crossings we need a trick to put the crossings at the edge of the spectrum of some operator, where they could be observed by projecting out an extremal eigenvalue. To achieve this, as explained in section 8, we construct $\mathcal{T}(U) = \frac{1}{2}\left[e^{\mathbf{H}_+(U)} + e^{-\mathbf{H}_+(U)}\right]$, where the Hamiltonian is the one associated with the transfer matrix formulation. Clearly, generic unity crossings in the spectrum of $e^{\mathbf{H}_+}$ appear as quadratic local minima of the lowest eigenvalue of $\mathcal{T}(U)$. The minima of interest occur at one, the lower bound of $\mathcal{T}(U)$. To obtain the lowest eigenvalue we start with a random unit norm vector and operate on it with $C - \mathcal{T}(U)$; we renormalize the result and iterate the procedure. $C$ is chosen slightly larger than the upper bound of $\mathcal{T}(U)$, so that the state we are interested in has the highest eigenvalue of $C - \mathcal{T}(U)$. After a few iterations the expectation value of $\mathcal{T}(U)$ in the state stabilizes and this is our estimate for the lowest eigenvalue of $\mathcal{T}(U)$ which we denote by $\frac{1}{2}(t_0 + \frac{1}{t_0})$. This method is very unsophisticated but suffices for our needs.

Using the two dimensional configuration 10.26 in all the 1–2 and all the 3–4 planes we construct a four dimensional abelian link configuration that has (in the continuum approximation) fluxes $F_{12} = F_{34} = \frac{2\pi q}{L^2}$ through every plaquette. $q$ will be varied continuously from zero to 3 and whenever it attains an integer value the lattice configuration can be thought of as representing a smooth $SU(2)$ configuration with topological charge $2q$ and uniform topological charge density.

We set $L = 8$ and plot in Figure 11.2 the lowest eigenvalue of $\mathcal{T}(U)$ for several $q$ values (represented by plus signs) between zero and three. The eigenvalues are connected by a solid line to guide the eye through the flow. One sees clear evidence of three crossings, one in each of the intervals connecting two consecutive integers. The number of "zero modes" of the $SU(2)$ fermions is twice the one obtained from the figure. The breaks in the line correspond to instances when a different eigenstate of $e^{\mathbf{H}_+}$, whose eigenvalue $t_m$ minimizes $t_n + \frac{1}{t_n}$ over $n$, where $n$ labels all eigenvectors of $e^{\mathbf{H}_+}$, takes over the dominating position. Since the mapping inverse to $t \to t + \frac{1}{t}$ is double valued the flow of $t + \frac{1}{t}$ need not be differentiable, the dominating eigenvector switching between the two branches.

Because $q$ appears in all the exponents of the link variables, and only there, the result for the higher $q$'s can be reinterpreted as evidence that the right number of zero modes will be generated for fermions in the representations $I = 1/2, 1, 3/2$ of the gauge group. In particular this includes the adjoint and hence the supersymmetric case. It is gratifying to get the right zero mode structure in the theory whose continuum limit is supposed to come out supersymmetric because instantons play a special role in the dynamics of supersymmetric theories.

The abelian check is insufficient because we have not tested the simplest case, with unit topological charge. Also, we have not yet done any test that is sensitive to the nonabelian group structure of the transfer matrices. Further, we wish to show that even localized topological density will lead to zero modes in the overlap formula. It is inconvenient to work on a torus at this point (we shall come back to this later) because the instanton is a solution on four space compactified to a sphere. We shall therefore use an open hypercube, i.e. our manifold will look like a piece of $R^4$ whose boundary has the topology of $S^3$ and the instanton on the boundary is almost a pure gauge. The associated gauge transformation maps the boundary into the group with unit three dimensional winding.

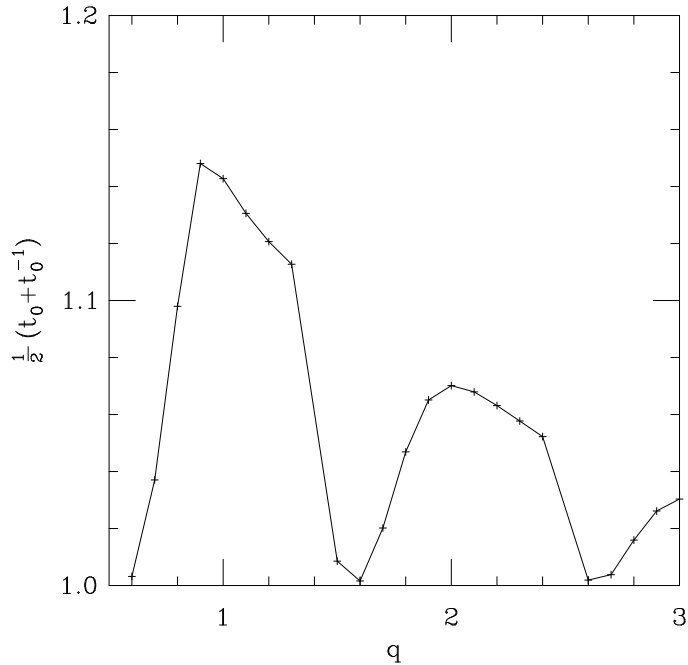

**Figure 11.2**  Lowest eigenvalue of $\mathcal{T}$ as a function of $q$ for the abelian background. Each time the graph touches the value one an eigenvalue of the "active" transfer matrix crosses unity.

We use antihermitian notation for the continuum gauge fields $A_\mu(x)$. We wish to latticize the following configuration:

$$A_\mu(x) = \frac{x^2}{x^2 + \rho^2} g^\dagger(x) \partial_\mu g(x) \quad g(x) = \frac{x_4 + i\vec{\sigma} \cdot \vec{x}}{\sqrt{x^2}} \tag{11.3}$$

In the above equation the $\vec{\sigma}$'s are the three Pauli matrices with the indices $1, 2, 3$ combined into symbols carrying a vector sign. For $R^2$ sufficiently larger than $\rho^2$, the instanton "size" squared, the topological charge inside the ball $|x| \leq R$ is close to one and it makes sense to expect such a configuration to act as if it carried topological charge exactly equal to unity. As a criterion we choose to require that $R/\rho$ be so large that more than two thirds of the classical instanton action in infinite $R^4$ be contained in the ball $|x| \leq R$. A simple computations shows that this roughly means that we wish the ball radius to be at least three times as large as the instanton size.

To accommodate the lattice we do not use a ball but rather a hypercube centered around $x = x_0$ defined by $|x - x_0| < l$ and consider an instanton whose size is $l/3$ or smaller. We embed a

hypercubic lattice with side $L$ into the continuous medium. We choose $L$ to be even and make the center be the point with coordinates $(\frac{L-1}{2}, \frac{L-1}{2}, \frac{L-1}{2}, \frac{L-1}{2})$, keeping the center off the lattice sites at a symmetrical location. The lattice sites $x$ have integer coordinates $0 \leq x_\mu \leq L - 1$. The link configuration is computed by evaluating the parallel transporters along the links of the embedded lattice associated with the continuum gauge field. We wish to compute the parallel transporters exactly. If we do this then a gauge transformed continuum configuration will yield a lattice gauge transformed link configuration. Since the eigenvalue problem for the transfer matrices is gauge invariant it does not matter which gauge we choose to represent the instanton configuration in and we can stick to the "regular" form in (11.3).

The evaluation of the path ordered exponentials giving the link variables is less formidable than one may think because of the symmetry of our configuration. Let us write the gauge field configuration more explicitly:

$$A_\mu(x) = \frac{i}{h(x)} \vec{a}_\mu(x) \cdot \vec{\sigma} \tag{11.4}$$

Denote $(x - x_0)_\mu$ by $\Delta_\mu$. Then $h(x)$ in the above equation is given by $h(x) = \Delta^2 + \rho^2$ and the $\vec{a}$'s are:

$$\begin{aligned} \vec{a}_1(x) &= (\Delta_4, \Delta_3, -\Delta_2), & \vec{a}_2(x) &= (-\Delta_3, \Delta_4, \Delta_1), \\ \vec{a}_3(x) &= (\Delta_2, -\Delta_1, \Delta_4), & \vec{a}_4(x) &= (-\Delta_1, -\Delta_2, -\Delta_3) \end{aligned} \tag{11.5}$$

Note that for a given $\mu$ the $\vec{a}_\mu(x)$ variable does not depend on $\Delta_\mu$. Since the link variables are defined by

$$U_\mu(x) = \mathcal{P} e^{\int_0^1 dt A_\mu(x + t\hat{\mu})}, \tag{11.6}$$

the integrand along the paths does not change its orientation in group space. Only $h$ has some $t$ dependence and the path ordering symbol can be dropped. The explicit form of the link variables is

$$U_\mu(x) = e^{i\vec{a}_\mu(x) \cdot \vec{\sigma} \Phi_\mu(x)}, \tag{11.7}$$

where the scalar quantities $\Phi_\mu(x)$ are given by

$$\Phi_\mu(x) = \int_0^1 \frac{dt}{h(x + t\hat{\mu})} = \frac{1}{\sqrt{\rho^2 + \sum_{\nu \neq \mu} x_\nu^2}} \tan^{-1} \frac{\sqrt{\rho^2 + \sum_{\nu \neq \mu} x_\nu^2}}{\rho^2 + \sum_\nu x_\nu^2 + x_\mu} \tag{11.8}$$

The exponents of the linear combinations of Pauli matrices in (11.7) have familiar closed expressions in terms of trigonometric functions.

In our numerical implementation we wish to work with the smallest volumes possible. It is reasonable to demand that we have at least of the order of two lattice spacings sampling the core of each instanton and since we want the size of the system to be be at least three times as big we take $L \geq 8$ (recall that $L$ has to be even and, since the hypercube is open, its linear extent is $L - 1$ lattice spacings). We test cases with $L = 8$, $L = 10$, $L = 12$, $L = 14$ and with $\rho$ values going from $\frac{L-1}{3}$ for the smallest volume to $\frac{L-1}{6}$ for the largest.

We do not wish to interpolate the nonabelian configuration to a trivial one this time. Instead we vary the mass parameter in the transfer matrix $\mathbf{H}_+(U)$ as explained in section 8. We need to consider only the "active" side, corresponding to negative mass values in our conventions. The number of zero modes one would associate with a given gauge configuration and with a given choice for the mass parameter $m$ is equal to the number of unity crossings when the mass parameter is varied from zero to its final value. This procedure not only tests more directly the method described in section 8, but also gives some indication of the dependence of the lattice topological charge definition on the extra regulator $m$. For each intermediate value of the mass parameter we construct $\mathcal{T}(U)$ and project out its lowest eigenvalue in the same manner as above. This time however we take the corresponding "eigenvector" and compute the expectation value of $e^{\mathbf{H}+(U)}$ in it. In this way the crossings are made more evident in Figure 11.3.

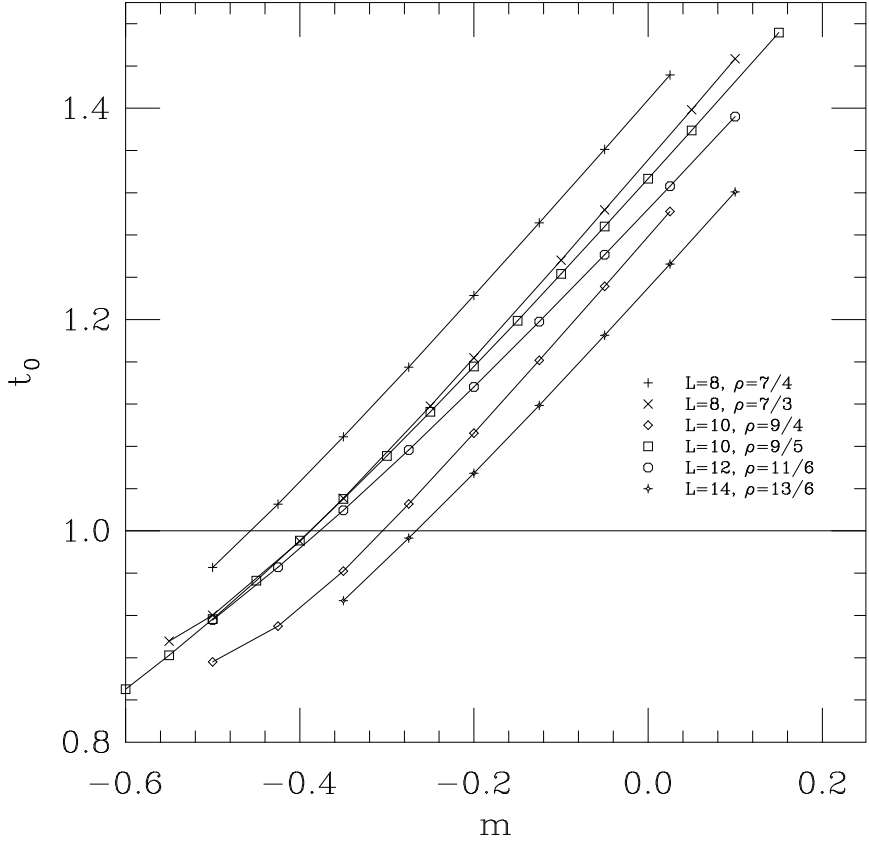

**Figure 11.3** Lowest eigenvalue of the transfer matrix as a function of $m$ for a background consisting of a nonabelian $SU(2)$ instanton on a hypercube with open boundary conditions. Different box sizes $L$ and instanton sizes $\rho$ are shown. Crossings of the horizontal line indicate the creation of an effective zero mode.

We note that one cannot make $m$ too small relative to the instanton size. We also see that tiny instantons can easily be missed. The minimal $m$ needed to detect the crossing decreases when we increase the fraction of the infinite-space instanton contained in our hypercube. All of these features make qualitative sense. Since the open hypercube can be embedded anywhere in a much larger system and the associated eigenvectors of the transfer matrix are more or less localized in it, we expect that the number of approximate zero modes in a dilute gas of instantons and anti-instantons will be correctly reflected by the overlap.

We still would like to see an instanton–like configuration on the torus performing correctly. We also would like to ascertain that the other, higher, eigenvalues of $\mathcal{T}(U)$ do not come down too and mess up the picture of the crossings. We might suspect that they just escaped detection until now. To address these concerns we take a configuration generated by the methods of [25] * using "supercooling" ("cooling" [26] with the help of a specially augmented lattice action designed to stabilize instanton-like configurations). We use anti–periodic boundary conditions for the fermions. We carry out almost the same analysis as before, only we substitute the power method used to extract the ground state of $\mathcal{T}(U)$ for each intermediate mass value there by a Lanczos algorithm which gives us several low states. We resort back to plotting the estimates for the eigenvalues of $\mathcal{T}(U)$ rather than of $\mathbf{H}_+(U)$ as we did in Figure 11.3. However, unlike in Figure 11.2 we do not include the factor $\frac{1}{2}$ in the definition of $\mathcal{T}(U)$ so the crossings occur now at 2. The result is shown in Figure 11.4 and we see that there is one crossing and that the crossing eigenvalue is clearly separated from the rest, as we would have wished.

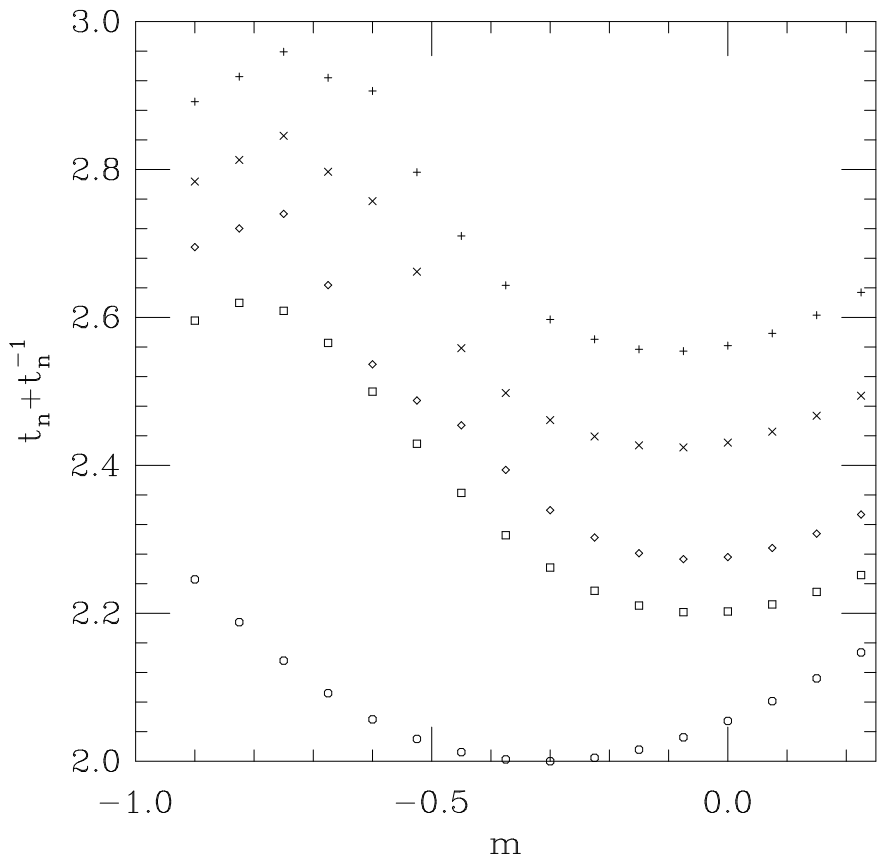

**Figure 11.4**  Several of the lowest eigenvalues of $2\mathcal{T}$ as a function of $m$ for a background consisting of a nonabelian $SU(2)$ instanton on a hypercube with toroidal boundary conditions. The lattice instanton has been generated by the method of "supercooling". When the bottom state hits the value 2 there is a crossinq of unity in the spectrum of the transfer matrix.

---

\*   We are grateful to M. G. Perez for supplying us with the configuration.

*11.3. Shape of zero mode.*

We have stressed before that it is important to see that the shape of the zero modes is reproduced on the lattice. The direct way to do this is to compute matrix elements of the appropriate creation or annihilation operators. We did this in two dimensions. A complete diagonalization is involved and would require a larger computational effort in four dimensions than we are able to invest at present. We opted for a more modest goal: we simply wish to see how the lattice states contributing to the lattice topological charge in (8.5) compare with the continuum states contributing to the continuum topological charge in (8.6). In the continuum, it is obvious that only the zero modes contribute. On the lattice, only the crossing state, evaluated at the crossing point contributes. Note that (8.5) holds both for the hamiltonian and for the transfer matrix formulation.

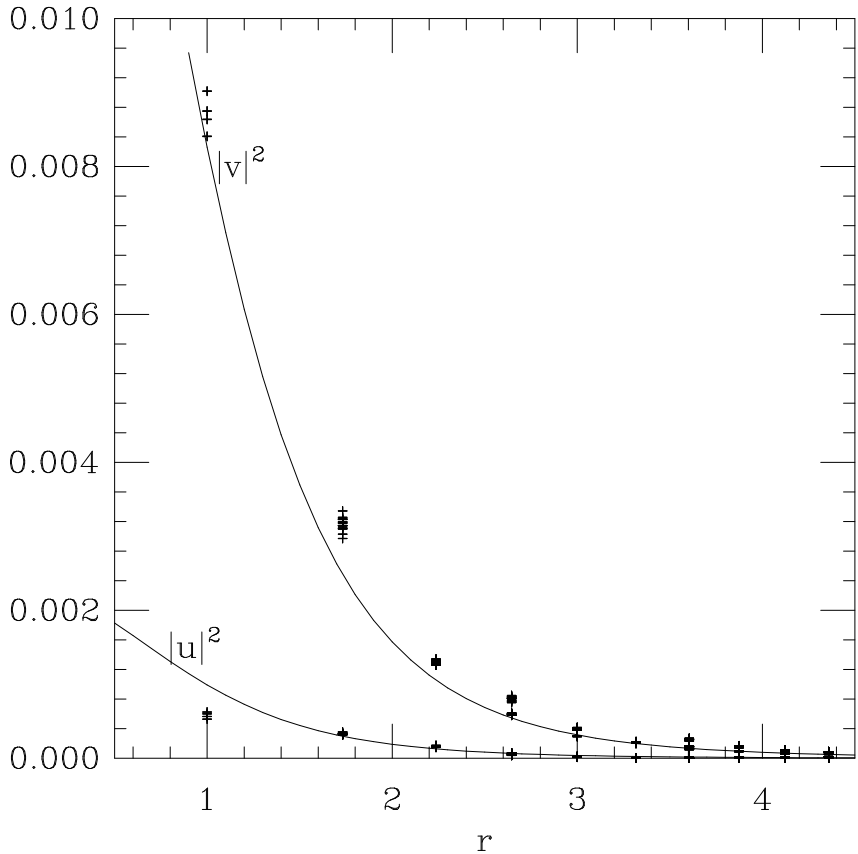

**Figure 11.5** Shape of crossing state for $L = 8$, $\rho = \frac{7}{4}$ on an open hypercube as a function of euclidean lattice distance. The solid lines are $\frac{C}{(x^2+\rho^2)^3}$ with $C$ determined by overall normalization.

In Figure 11.5 we plot the norms squared of the $u, v$ components (the notation is that of section 8) of the crossing state on an open hypercube of size $L = 8$ in the background of one instanton of size $\rho = \frac{7}{4}$. From Figure 11.3 we see that the crossing takes place at $m = -.45$. We take the mass parameter as $m = -0.5$. The plot is given as a function of the euclidean lattice distance $r \equiv \sqrt{x^2}$ and the lines shown are the well known continuum shapes $\frac{C}{(x^2+\rho^2)^3}$ with $C$ determined so that the integral in the continuum over infinite space is equal to the sum over the contributions of both of the $u, v$ components over the finite lattice. The ratio of the contributions of the two componets $u, v$ on the lattice is preserved between the $C$–constants corresponding to the two lines. There are many lattice points corresponding to a given $r^2$ and all the values are displayed. Their dispersion gives a feel for the amount of $O(4)$ invariance breaking due to the lattice. We see that the zero mode is reasonably symmetric and that it falls off with the correct power. We also see that most of it is concentrated in the $v$ component, reflecting the definite chirality in the continuum. The shape is unchanged as long as the crossing state is well separated energetically from the other eigenstates of the "active" transfer matrix.

## 12. Discussion

We claim that our work up to now shows that our regularization reproduces everything we know about the continuum field theories we are considering. With this we can declare victory on the problem of regularizing chiral gauge theories; all other questions about the theories can be answered only after actually carrying out the integrals we have defined. This point of view may not be widely accepted and we shall have more to say about this later. First, however, we accept the regularization and try to organize our thoughts about the dynamics by listing a few simple possibilities.

Consider an anomaly free gauge theory. The pure gauge action is given by the usual single plaquette term. The lattice gauge coupling is $\beta$ and to get to the continuum limit we have to take $\beta$ to positive infinity. The parameter $m$ in the fermionic sector is fixed and does not get tuned.

The most obvious question is whether there is a phase transition separating strong coupling (small $\beta$) from weak coupling. We would like not to have such a transition, at least in the vector–like case, but we think that it is more likely that such a transition occurs. The basic reason is related to instantons: For large $\beta$ we expect the expectation value of the 't Hooft vertex to be nonvanishing. However, for small $\beta$ the link variables fluctuate rapidly and could be approximated by zero; in that case, when $m$ is varied from zero to its chosen value, the gap of the "active" Hamiltonian never closes and eigenvalue flow across zero does not occur for almost all gauge configurations. Without an imbalance in the number of negative energy states the 't Hooft vertex has zero expectation value.

A possible way to investigate this more thoroughly at $\beta = 0$ is to expand everything in the link variables. To leading order the overlap is unity since the two Hamiltonians have identical negative energy subspaces. We may therefore work in a "quenched" approximation, in which the effect of the overlap on the distribution of the link variables is ignored. (Another way to justify the "quenched approximation" is to work in the limit of an infinite number of colors.) We are then interested in the properties of the two random Hamiltonians $H^{\pm}(U)$. A simple random variable that can exhibit transitions is the average level density defined (for the active side say) by $\rho(E, U) = \sum_i \delta(E - E_i(U))$ where $H^+(U)\phi_i = E_i(U)\phi_i$. The simplest observable is the average $\bar{\rho}(E) =< \rho(E, U) >_U$ and a possible question is whether $\bar{\rho}(E)$ has a gap or does not have a gap around $E = 0$. One can also consider the topological charge observable $q(U) = \int_{-\infty}^{\infty} dE \epsilon(E) \rho(E, U)$ where $\epsilon(E)$ is the sign function. At zero $\theta$ parameter $< q(U) >_U = 0$ but the topological susceptibility, $\frac{<q^2(U)>_U}{\text{volume}}$, can change discontinuously as $\beta$ varies. The question is whether the topological susceptibility is zero or non-zero.

To summarize, the question is whether a phase transition separating strong from weak coupling exists in the quenched approximation and/or in the full dynamical model. The suspected transition is somehow related to global anomalous symmetries. In exploratory numerical studies carried out in collaboration with Gyan Bhanot [27] we saw some indications from the $SU(2)$ lattice topological charge susceptibility that indeed such a transition takes place in the quenched approximation.

A deeper question relates to strictly chiral theories. If we go back for a moment to an action point of view we know that the overlap represents a strictly infinite number of charged fermionic lattice degrees of freedom. (In the vector–like case one does not need to be strict about the infinity, as long as one tolerates small masses for the fermions.) If we accept that the strict infinity of fermionic species is more than an opportunistic trick we may attach importance to the fact that Elitzur's famous theorem [28] about absence of spontaneous breakdown of local symmetries does not hold necessarily if the number of fields per unit volume is infinite. Do we have an indication here that local gauge symmetries can really be spontaneously broken in chiral gauge theories ?

Of course, before we speak about breaking the gauge symmetry spontaneously we must argue how it gets restored on the lattice when the target theory is anomaly free. It is the most important dynamical question facing our approach. Before we discuss this point in general terms, we focus on a simpler, specific question: For backgrounds that are trivial up to a gauge transformation, does one obtain a continuum like behavior in the abelian case, where there is no dependence on the gauge in the chiral determinant because the anomaly vanishes for zero field strength. On the lattice we might expect some residual gauge breaking terms that vanish only in the continuum limit. However, in our formalism the two dimensional gauge invariance on the trivial orbit is exact, just like in the continuum. When we try to go to the continuum limit we are sampling mainly the neighborhood of the trivial orbit in orbit space, so the above bodes well for the chances to obtain the right target theory. We now present the proof of our assertion:

Let $|R\pm>_{1^g}$ and $|L\pm>_{1^g}$ be the many body states for a 2-D U(1) gauge field configuration which is a gauge transformation of unity and let the states obey the Wigner-Brillouin phase choice. Using (7.8) and (7.14) we can write these states as

$$|R\pm>_{1^g} = \mathcal{G}^\dagger |R\pm>_1 \frac{{}_1<R\pm|\mathcal{G}|R\pm>_1}{|{}_1<R\pm|\mathcal{G}|R\pm>_1|}$$
$$|L\pm>_{1^g} = \mathcal{G}^\dagger |L\pm>_1 \frac{{}_1<L\pm|\mathcal{G}|L\pm>_1}{|{}_1<L\pm|\mathcal{G}|L\pm>_1|}$$
(12.1)

All the dependence on $g$ comes in through $\mathcal{G}$. We claim that the overlaps are, in fact, independent of $g$. That is,

$${}_{1^g}<R-|R+>_{1^g} = {}_1<R-|R+>_1; \quad {}_{1^g}<L-|L+>_{1^g} = {}_1<L-|L+>_1 .$$
(12.2)

To see this, we begin by noting that $G$ is a unitary matrix and from Lemma 4.0, it follows that

$$\det G = \frac{{}_1<R\pm|\mathcal{G}|R\pm>_1}{{}_1<L\pm|\mathcal{G}^\dagger|L\pm>_1}.$$
(12.3)

The above equation is valid for both $\pm$ independently. This is because the full set of single particle states that make up $|R\pm>_1$ and $|L\pm>_1$ form a complete set in which we can represent $G$. The numerator and denominator are simply the determinant of the two blocks that appear in Lemma 4.0. The explicit form of the denominator in the above equation is

$$_1< L \pm |\mathcal{G}^\dagger| L\pm >_1 = \det_{K,K'} \sum_{x\alpha} \left\{ \left[\psi_{K'}^{L\pm}(x\alpha;0)\right]^* g^*(x) \psi_K^{L\pm}(x\alpha;0) \right\} \tag{12.4}$$

There is no group index above because the group under consideration is U(1). We have used the notation $A = 0$ in the right hand side of the above equation to be consistent with the notation in section 4. Now we can use (4.24) to rewrite the above equation in terms of the right handed single particle states. We get

$$_1< L \pm |\mathcal{G}^\dagger| L\pm >_1 = \det_{K,K'} \sum_{x\alpha} \left\{ \psi_{K'}^{R\pm}(x\alpha;0) g^*(x) \left[\psi_K^{R\pm}(x\alpha;0)\right]^* \right\} \tag{12.5}$$

We have used the fact that $\gamma_1^2 = 1$. The phase factors in (4.24) cancel out in the determinant. The above equation simply says that

$$_1< L \pm |\mathcal{G}^\dagger| L\pm >_1 = {_1< R \pm |\mathcal{G}^\dagger| R\pm >_1} \tag{12.6}$$

(12.6) when used in conjunction with (12.3) says that the phases of $_1< L \pm |\mathcal{G}| L\pm >_1$ and $_1< R \pm |\mathcal{G}| R\pm >_1$ do not depend on the $\pm$ signs and are equal to half of the phase of $\det G$. Therefore when we compute the overlaps in (12.2) the phases from the positive and negative sides cancel and the result is independent of $g$.

Having addressed the very special case of the trivial orbit in two dimensional abelian models, we now turn to the general picture. The following presents our point of view in more detail, as promised in the opening paragraph.

Our solution to the problem of regularizing anomaly free chiral gauge theories is not perfect [29]. A perfect solution would elevate the algebraic conditions for anomaly cancelation to the non–perturbative level and show that they are a necessary and sufficient condition for writing down a completely gauge invariant well defined path integral. Whether such a perfect non–perturbative regularization exists or not is an interesting and potentially deep question.* However, it is logically possible that while gauge invariant continuous chiral gauge theories exist, perfect lattice regularizations do not; and even if they do, it is plausible that the desired limits will be approachable from imperfect schemes also.

We claim that our imperfect scheme is quite natural, almost naïve in the sense that it is all based on one simple viewpoint, exposed is section 2. Conceptually, it is much simpler than other approaches, in particular the ones based on strongly coupled lattice Yukawa models, where some non–trivial dynamics seems to be required and the ultimate goal appears to be a perfect regularization, although the role of anomalies is not made explicit.

To argue the simplicity of our approach in more detail we start by reviewing the situation in the continuum, at the level of the formal path integral [30]. The integration variables are some set of chiral fermions $\psi$ and some gauge fields $A$. The integrand does not depend on all of the integration variables: a gauge transformation $\psi \to \psi^g$, $A \to A^g$ leaves the integrand invariant. If we could do that in a natural and local way we would eliminate the redundant variables completely and replace the set of integration variables by gauge orbits and fermions. We cannot however do that and we are stuck with using more fields than we want. The special features of chiral gauge theories can be cast into a simple framework if we are willing to accept that the fermion integration measure depends on $A$. Fujikawa has shown that the anomalous dependence of the chiral determinant on the coordinates along the orbits in gauge field space can be usefully viewed in this way. Therefore, generically, the functional integral is not gauge invariant after all and we do not have a theory that describes fermions in interaction with gauge orbits. However, if the anomalies cancel, we do have such a theory.

---

\* In our formalism we are defining the phase of each chiral fermion multiplet separately whereas it is sufficient to define the overall phase of the combined many body states. It is conceivable that a natural gauge invariant definition can be found when the theory is anomaly free.

The clearest way to see the difference is to integrate out the fermions and obtain a functional of $A$, $e^{\Gamma(A)}$. When anomalies cancel $\Gamma(A)$ is invariant under $A \to A^g$ and when they do not we have $\Gamma(A^g) - \Gamma(A) = k\Gamma^{WZ}(A, g)$ where $k$ is some quantized non–vanishing constant and $\Gamma^{WZ}(A, g)$ is a universal, known, functional of both $A$ and $g$. While $\Gamma(A)$ is non–local and unknown $\Gamma^{WZ}(A, g)$ is both known and local. If anomalies do not cancel we need to deal with one more degree of freedom but, essentially, we know the exact dependence of the induced action on it.

Similarly to the gauge invariant case we cannot explicitly separate the degrees of freedom into orbits and coordinates along orbits in a local and natural way. We deal with this problem by adding a redundant variable, $g$, and having the integrand and measure invariant under a (new) gauge transformation $A \to A^h$, $g \to h^{-1}g$. Under this new transformation $\Gamma(A) + k\Gamma^{WZ}(A, g)(= \Gamma(A^g))$ is trivially invariant. We therefore change the bosonic action by adding the term $k\Gamma^{WZ}(A, g)$ to the original action.* We have not really altered the original functional integral because the new gauge invariance can be used to set the new integration variable $g$ to unity and then the added term vanishes. So up to an uninteresting overall factor, nothing has changed. However, in the new form we can deal with the coordinate along gauge orbits first: We can now try to integrate out $g$ because we know the exact dependence on it. After that, the new and old gauge invariances mean the same thing and we are left with something gauge invariant, i.e. the "true" orbit – orbit interaction. Unfortunately, the integration over $g$ changes the theory drastically, and in four dimensions, we cannot make sense of the result even perturbatively. Basically, the problem is that $g$ needs to be a dimension zero field to honestly play the role of a coordinate along the gauge orbit. In two dimensions there is no problem with this and therefore one can construct anomalous gauge theories perturbatively, but in four we do not know how to do that, except in the trivial case where there is no dependence on $g$ in the path integral at all. In four dimensions, when there are anomalies, one can either try to view $g$ as a dimension zero field similar to the nonlinear fields appearing in effective Lagrangians, but the theory then has a limited energy range of applicability (i.e. the theory is non–renormalizable), or, hope that somebody will discover a new way to quantize the degrees of freedom $g$ in the $\Gamma^{WZ}(A, g)$ action [31]. Anyhow, we cannot construct anomalous gauge theories around an ordinary gaussian fixed point as we do with all other relativistic field theories known to date.

When we go to the lattice, it is best to avoid having too explicit a Grassmann integration in mind, because then we cannot realize Fujikawa's gauge dependent measure [32]. The simplest is to start from the chiral determinant itself and require the lattice expression for it to be a faithful representation of the continuum $e^{\Gamma(A)}$. This is what we have done. When we compute on the lattice $\Gamma(A^g) - \Gamma(A)$ we do not get an answer exactly proportional to $\Gamma^{WZ}(A,g)$ but we know now that we do get something sufficiently close to it for slowly varying $A$ and $g$. For an anomalous theory we would need to integrate over $g$ the exponent of the Wess-Zumino action and we do not know what the answer would be. Unlike in the continuum, what we have to do is precisely defined, but the result, obviously, is still unknown. If the anomalies cancel in the continuum the cancelation on the lattice will not be exact. So, we shall still have to carry out the $g$ integral. The action is not special now (except being purely imaginary and obeying some discrete symmetries) and the dependence on $g$ is weak in the sense that the leading operator has cancelled out. It is therefore reasonable to expect the integration over $g$ to simply give the exponent of some generic, local, functional of $A$. By construction, this functional also has to be gauge invariant and if sufficiently local will only inconsequentially alter the tree level lattice bosonic action. That lattice gauge (invariant) theories are robust under not too large perturbations by gauge breaking terms has been known for a long time [33]. Essentially, the lattice, unlike the continuum, has no trouble dealing with dimension zero fields: they tend to decouple from the low energy physics if such exists. This is how we expect an anomaly free theory to emerge. We have seen that the lattice gauge dependent remnants in two dimensions are indeed numerically small for some trial configurations, so it is plausible that the breaking is of the inconsequential type. We feel that the real mystery is not that the anomaly free theory could come out of this, but what exactly happens to the anomalous theory on the lattice. For this we must understand the $g$–dynamics of the lattice Wess-Zumino action constructed via the overlap.

---

\* The result of integration over $g$ is the same as averaging the overlap over all gauge transforms of the link variables $U$.

To summarize, we claim that our approach will lead to continuum interacting anomaly free chiral gauge theories without any tuning beyond the need to take the bare gauge coupling to zero. To be sure, we have not proven this to happen. But we do claim that if indeed chiral anomaly free gauge theories exist our claim is very plausible. In addition, one can always add some explicit local gauge breaking terms from the beginning and fine tune them. This would be disappointing to us, but we should remember that it is done implicitly in ordinary perturbation theory (it cannot be done in the anomalous case because the locality requirement cannot be satisfied and without it the perturbative scheme breaks down).

In short, our basic premise is that a Euclidean regularization of a chiral gauge theory is successful if it is based on local objects (the operators $H^{\pm}(U)$ are local in $U$) and satisfies the following criteria:

1. It is bilinear in the fermions. This means that in a fixed gauge background all fermion correlations are given in terms of two point functions and zero modes.
2. It reproduces an acceptable continuum chiral determinant in weak, slowly varying external gauge potentials to any number of external legs attached to the fermionic bubble. By "acceptable" we mean that universal terms are exactly reproduced and non–universal ones can be written as a series in a basis of local operators of increasing irrelevance. Gauge breaking must be allowed to occur, but should be restricted strictly to the imaginary part of the induced action. Of course, anomalies must be reproduced in their consistent form.
3. It reproduces instanton effects in some detail, including non–vanishing expectation values for 't Hooft vertices and correct long distance behavior of fermion propagators.
4. The regularized chiral determinant is a naturally defined complex function over gauge field space that admits generic zeros in the sense of section 4. The regularized chiral determinant respects the symmetries under charge conjugation, parity, global gauge transformations, some selected Eulidean $O(4)$ transformations and chirality interchange as defined in section 4.

Not all of the requirements must be obeyed, but our construction shows that it is possible to satisfy them all. Conspicuously absent from the list is a requirement of exact gauge invariance. If our requirements are met we believe that gauge invariance takes care of itself in the anomaly free case.

## Acknowledgements

We would like to thank W. Bardeen, R. Levien and E. Witten for discussions. R. N.was supported in part by the DOE under grant # DE-FG02-90ER40542. H. N. was supported in part by the DOE under grant # DE-FG05-90ER40559 and by the Monell Foundation.